\pdfoutput=1
\documentclass[twocolumn]{aa} 
\usepackage{graphicx}
\usepackage{rotating}
\usepackage{txfonts}
\usepackage{caption}
\usepackage{natbib}
\bibpunct{(}{)}{;}{a}{}{,}
\usepackage[section] {placeins}

\begin{document}

   \title{Modelling the asymmetric wind of the luminous blue variable binary MWC~314}

   \authorrunning{A. Lobel et al.}
   \titlerunning{Wind Modelling of Massive Binary MWC~314}

    \author{A. Lobel$^1$,  J. H. Groh$^2$,  C. Martayan$^3$, Y. Fr\'{e}mat$^1$, K. Torres Dozinel$^4$,
            G. Raskin$^5$, H. Van Winckel$^5$, S. Prins$^5$, W. Pessemier$^5$, C. Waelkens$^5$,
            H. Hensberge$^1$, L. Dummortier$^1$, A. Jorissen$^6$, S. Van Eck$^6$,
            and H. Lehmann$^7$}

   \offprints{A. Lobel}

    \institute{Royal Observatory of Belgium, Ringlaan 3, B-1180 Brussels, Belgium \\
                   \email{alobel@sdf.org; Alex.Lobel@oma.be  }
    \and
    Geneva Observatory, Geneva University, Chemin des Maillettes 51, CH-1290 Sauverny, Switzerland
    \and
    European Southern Observatory, Alonso de Cordova 3107, Vitacura, Santiago, Chile
    \and
    University of S{\~a}o Jo{\~a}o Del Rei, CAP, 36420-000 Ouro Branco, MG, Brazil
    \and
    University of Leuven, Instituut voor Sterrenkunde, Celestijnenlaan 200 D, B-3001 Heverlee, Belgium
    \and
    Universit\'{e} Libre de Bruxelles, Boulevard du Triomphe, B-1050, Brussels, Belgium
    \and
    Th\"{u}ringer Landessternwarte, Sternwarte 5, D-07778 Tautenburg, Germany
     }

   \date{Received;}

 
  \abstract
   {}
   {We present a spectroscopic analysis of MWC~314, a luminous blue variable (LBV)
   candidate with an extended bipolar nebula. The detailed spectroscopic variability is investigated 
   to determine if MWC~314 is a massive binary system with a supersonically accelerating wind or a low-mass 
   B[e] star. We compare the spectrum and spectral energy distribution to other LBVs (such as P Cyg) 
   and find very similar physical wind properties, indicating strong kinship.  
    }
    {We combined long-term high-resolution optical spectroscopic monitoring and $V$-band photometric observations 
     to determine the orbital elements and stellar parameters and to investigate the spectral variability with 
     the orbital phases. We developed an advanced model of the large-scale wind-velocity and wind-density structure 
     with 3-D radiative transfer calculations that fit the orbitally modulated
     P~Cyg profile of He~{\sc i} $\lambda$5876, showing outflow velocities above 1000~$\rm km\,s^{-1}$.
    }
   {We find that MWC~314 is a massive semi-detached binary system of $\simeq$1.22 AU, observed at an inclination 
   angle of $i$=72.8$^{\circ}$ with an orbital period of 60.8 d and $e$=0.23. The primary star is a 
   low-$v$sin$i$ LBV candidate of $m_{1}$=39.6~$\rm M_{\odot}$ and $R_{1}$=86.8~$\rm R_{\odot}$. 
   The detailed radiative transfer fits show that the geometry of wind density is asymmetric 
   around the primary star with increased wind density by a factor of 3.3, which leads the orbit of the primary. 
   The variable orientation causes the orbital modulation that is observed in absorption portions of P Cyg wind lines. 
   Wind accretion in the system produces a circumbinary disc.
   }
   {MWC 314 is in a crucial evolutionary phase of close binary systems, when the massive primary star has its H
    envelope being stripped and is losing mass to a circumbinary disc. MWC 314 is a key system for studying the
    evolutionary consequences of these effects.}


   \maketitle
%

\section{Introduction}

 There is currently little known about the wind physics of the most massive 
 binary stars. Although massive binary systems have been investigated for decades,
 it is notoriously difficult to infer important wind properties, such as 
 accurate mass-loss rates, wind driving, and wind structuring mechanisms from traditional 
 spectroscopic and photometric observations. Ultraviolet spectroscopy is one of the main 
 sources for information about massive star winds but is currently available in only 
 a small number of operational space-based instruments. Optical and near-IR long-term 
 spectroscopic monitoring with ground-based spectrographs is an important alternative. 
 In recent years the quality and efficiency of these high-resolution spectroscopic 
 observations have dramatically improved, offering new 
 opportunities for investigating detailed spectral line variability in the 
 most massive binaries \citep[e.g.,][]{2011AJ....142..201R}.          
 
 The star MWC~314 (V1492~Aql; BD$+$14$^{\circ}$3887; $V$=$9^{\rm m}$.9) is a candidate LBV 
 and has been proposed as one of the most luminous 
 stars in the Galaxy by \citet{1998A&AS..131..469M} 
 with log(${L}_{\star}/{L}_{\odot}$)$\simeq$6.1$\pm$0.3, $T_{\rm eff}$$\simeq$20 to 30 kK, and
 $\dot{M}$$\sim$ $\rm 3\,10^{-5}$ $\rm M_{\odot}\,{yr}^{-1}$ (assuming a single star). 
 More recently, \citet{2008A&A...487..637M}
 have found that the star is a (possible) binary system with an orbital period of $\sim$30 d, however, 
 without determining other orbital parameters. \citet{2008A&A...477..193M} demonstrated
 the presence of an extended double-lobe H$\alpha$-emission (bipolar) nebula or a large east-west bipolar 
 feature $>$13 pc (end to end) around the star. They suggest that the large size 
 corresponds to a nebula age ($>$$10^{5}$ yr) that is greater than for LBV nebula in general.
 The lifetime of the LBV phase is, however, quite uncertain and may be much longer 
  {    \citep[2 - 5 $10^{5}$ yr;][]{1994PASP..106.1025H}} than previously assumed. 
 The LBV phase can be the evolutionary phase, when most of the mass loss of the most massive stars
 occurs to form the enigmatic Wolf-Rayet stars.
 {    In addition, recent evolutionary and radiative transfer calculations have shown that low-luminosity 
 LBVs can be the end stage of massive star evolution \citep{GrohAAsubm}.}
 
 A persisting problem for analysing and interpreting the spectrum 
 of MWC~314 is that only a small number of stellar absorption features have been observed and identified 
 thus far. 
 The optical spectrum is, however, dominated by strong and single-peaked H Balmer emission and He~{\sc i} lines.
 Numerous metallic Fe~{\sc ii}, [Fe~{\sc ii}], Cr~{\sc ii}, 
 and Ti~{\sc ii} emission lines were identified by \citet{1943ApJ....97..226S}. The double-peaked shape 
 of the metallic emission lines signal formation in a rotating disc and could indicate a 
 B[e] star in lieu of an LBV. \citet{2008A&A...487..637M} therefore conjectured that the disc is produced
 by low-velocity outflow, as in B[e] supergiants, instead of an accretion mechanism, although they do not 
 rule out accretion disc formation. Except for the H$\alpha$
 equivalent width variability observed by \citet{2006PASP..118..820W}, there is no evidence 
 in the literature of significant spectral line variability that can confirm 
 an LBV nature for MWC~314.
 
 { P Cygni-type optical line profiles due to expanding wind opacity 
 have  not been observed thus far around MWC 314
 \citep[see also Sect. 6.2 of][]{2008A&A...477..193M}. 
 In this paper, we show that the optical He~{\sc i} 
 lines of MWC 314 are P Cyg profiles, with detailed shape variability 
 that correlates to the orbital phases. We show that MWC 314 is spectroscopically, 
 and in its overall spectral energy distribution, almost identical 
 to the LBV P Cygni. The bipolar nebula observed around 
 MWC 314, combined with its LBV-like spectroscopic properties, 
 signal a unique massive binary system that can possibly contain 
 an LBV in a dormant state. The candidate LBVs 
 MWC 349A and Sher 25 have comparable physical properties. 
 MWC 349A was recently proposed to be an LBV by 
 \citet{2010A&A...519A..33G} and Sher 25 earlier by \citet{1997ApJ...475L..45B}. 
 Interestingly, six of the nine known Galactic LBVs 
 are also surrounded by bipolar nebulae. 

 There are currently 
 two proposed LBV binaries, $\eta$ Car \citep{2008MNRAS.386.2330D} and 
 HD 5980. { The binary system $\eta$ Car is well-studied, currently involving sophisticated 
 hydrodynamic wind modelling efforts by \citet{2012MNRAS.420.2064M} and \citet{2008MNRAS.388L..39O}}. 
 HD 5980 is an eclipsing OB+WR binary system 
 \citep{1978PASP...90..101H} with an orbital period of $\simeq$19.3 d. More 
 recently, it has been identified as an LBV/WR binary (containing a third star)
 with stellar and orbital parameters that have been
 relatively well determined by \citet{2011AJ....142..191G} 
 and \citet{2008RMxAA..44....3F}.
 {    The third (and possibly fourth) components appear to be a pair of fairly normal, 
  understudied O-type stars either coincident to the line of sight or in a very long 
  orbit with HD 5980AB. MWC~314 is also very similar to massive 
 binary HDE 269128 (RMC~81) in the LMC with $P$=74.655 d 
 \citep{2002A&A...389..931T}. RMC~81 is an eclipsing 
 close binary ($e$=0.57) with an cLBV primary 
 star of $T_{\rm eff}$=19500 K, $R_{\star}$=96~$\rm R_\odot$, 
 and $K_{1}$=35~$\rm km\,s^{-1}$. 

 More recently, \citet{2012ASPC..464..293M} have
 imaged a companion near the cLBV HD~168625 for the first time with Very Large Telescope/NACO
 (Nasmyth Adaptive Optics System). 
 The system is known to be surrounded by several rings similar to those of SN1987A.}
 We note, however, that the high $V$-brightness
 changes typical of bona fide LBVs, presumably due to recurrent eruption 
 activity, have so far not been observed in MWC 314.
 It is therefore technically considered a cLBV 
 without signs of S Dor variability as for AG Car and HR Car. 
 We find in this paper that the slow rotation of MWC 314 fits well in 
 the scheme in which fast-rotating LBVs show S Dor variability, while 
 slowly rotating ones do not \citep{2009ApJ...705L..25G}. } 

 We spectroscopically monitored MWC~314 for two years to test for binarity \citep{2012ASPC..465..358L}.
 We present the observations in Sect. 2. In Sect. 3, we discuss 
 new H$\alpha$ imaging observations of inner envelope emission in MWC~314. Section 4 presents
 the high-resolution optical and near-IR spectra with radial velocity measurements for determining the 
 binary period and other orbital elements. The spectra are combined with an analysis of ASAS $V$-band 
 photometry that also provides accurate stellar parameters. 
 Detailed spectroscopic analysis reveals strong correlations and clear dependence of 
 permitted and forbidden emission line variability on the orbital phases, which 
 we observe, for example, 
 in variable H$\alpha$, H$\beta$, and metallic emission lines. We also present a detailed 
 comparison to the spectral energy distribution of P Cygni using broadband photometry 
 combined with IR {\it Spitzer} and ISO spectra. The IR SEDs of
 MWC~314 and P Cyg are practically identical. In Sect. 5, we present a quantitative model 
 of the large-scale asymmetric wind structure around MWC~314 that is computed with 3-D radiative 
 transfer by modelling  the orbitally modulated P Cyg line profile variability in detail. 
 We discuss and summarise the results in Sect. 6. We show that 
 MWC~314 is a massive binary system and accurately determine its orbital and stellar parameters.
 We find that MWC 314 can contain an LBV star with wind properties 
 comparable to P Cygni and put strong constraints on the stellar masses and the size 
 of the system.

\section{Observations}

\subsection{Echelle spectroscopy}

We observed 15 high-resolution spectra ($R$$\simeq$85,000) of MWC~314 between Sep 2009 and Aug 2011 
with the HERMES instrument of the 1.2-m Mercator Telescope at the Roque de los Muchachos
Observatory on La Palma Island (Canary Islands, Spain). 
HERMES is a high-efficiency bench-mounted {\'e}chelle spectrograph 
that observes the complete wavelength range from 420 nm to 900 nm in a single exposure \citep{2011A&A...526A..69R}. 
The HERMES spectra of MWC~314 have large signal-to-noise ratio (S/N)$\sim$100 
required for accurate radial
velocity measurements and detailed studies of line profile variability. 
Three subsequent exposures of $\sim$1800 s each are added to minimise 
the amount of CCD cosmic hits. The spectra are calibrated with the 
latest version of the HERMES pipeline (release 4.0) developed at the Royal Observatory of 
Belgium (ROB) in collaboration with the HERMES Consortium. 
The typical {\'e}chelle calibration steps are 
performed, which include spectral order tracing and extraction, average flat-fielding, 
Th-Ar lamp wavelength calibration, and hot pixel removal using cross-order profiling. 
The wavelength scale is corrected to the heliocentric rest frame, which for most 
observations includes a wavelength rebinning of the sum spectra.
An overview of the HERMES observations of MWC~314 is provided in Table 1.
On 5 \& 10 Sep 2009 and on 18 \& 21 Mar 2011, we observe two spectra
within 5 d to investigate short-period spectroscopic variability in 
MWC~314. 

One FEROS spectrum of MWC~314 is retrieved from the ESO Science Archive Facility. 
The FEROS spectrum of 5 June 2009 precedes the start of our HERMES monitoring campaign 
by only $\sim$3 m and is very useful for testing the orbital solution for MWC~314 that we discuss 
in Sect. 4.1. The FEROS instrument on the MPG/ESO 2.2-m telescope is a bench-mounted fiber-linked 
high-resolution ($R$=48,000) {\'e}chelle spectrograph with 
high efficiency ($\sim$20\% between 5000~\AA\, and 6000~\AA), covering 
the wavelength range from 356~nm to 921~nm with 39 orders in one exposure. 
For example, a $V$=12$^{\rm m}$.0 star can be observed with a S/N 
of 105 in two hours \citep{1999Msngr..95....8K}. 
The spectrum of MWC~314, observed for 900 s, is calibrated in {\sc eso-midas} with 
the FEROS-Data Reduction System (DRS). The common {\'e}chelle pipeline reduction steps are 
performed starting with bias subtraction, optimal order extraction, 
and division of the science exposures with an averaged flat-field frame, 
which is followed by the order wavelength calibration using a Th-Ar lamp spectrum.       
The flat-fielding step in the FEROS-DRS pipeline removes the 
blaze-function similar to the HERMES calibration pipeline.

\subsection{Imaging and photometry}
We observed MWC~314 on 10 Mar 2011 with the MEROPE instrument on Mercator:              
MEROPE is a direct camera in the Cassegrain focal station of the telescope.
The camera chip is a 2k $\times$ 2k back-side illuminated device 
cooled with liquid nitrogen to an operational temperature of $-$130 $^{\circ}$C. 
The total field of view is 6.6\arcmin $\times$ 6.6\arcmin\, (standard setting) with a 
sampling of 0.193\arcsec per pixel \citep{MeropeMan...2008}.
The inner portions of the extended bipolar H$\alpha$ nebula 
of MWC~314 were observed with three subsequent MEROPE exposures.
The star was first observed during 300 s through a special narrow-band H$\alpha$ 
filter. {    The filter's transmission function has a full width at half
maximum (FWHM) $\sim$50~\AA\, with a
peak transmittance of $\sim$90\% at H$\alpha$. The passband is very
narrow with transmittance levels, decreasing below 10\% shortward of
6520~\AA\, and longward of 6610~\AA.} Next, a broad passband optical continuum 
filter, which excludes H$\alpha$, is used for comparison during 200 s. 
{    The continuum filter has a transmission function FWHM$\sim$100~\AA\, 
centred around 6365~\AA.} 

Finally, a second H$\alpha$ image is 
observed for 414 s to increase the combined S/N.
Master flat-field images are observed in both filters before and after the target sequence.
The image calibration steps include bias level measurement of the mean value 
in the overscan areas. The mean bias level is subtracted from the combined 
flat-field frames and the science exposures before flat division.
The comparison of the H$\alpha$ and continuum MEROPE images of MWC~314
is discussed in Sect. 3. It also provides a comparison
of the inner H$\alpha$ nebula surrounding MWC~314 and its extended 
bipolar nebula observed by \citet{2008A&A...477..193M}.  
   
We retrieved photometric fluxes of MWC~314 in the $V$-band 
from the All Sky Automated Survey (ASAS) project public 
database \citep{2002AcA....52..397P}; ASAS-3 observations of MWC~314 
were performed at Las Campanas Observatory (Chile) between 
Mar 2003 and Oct 2009. The entire facility is housed in 
an automated enclosure and consists of four telescopes,
equipped with a telelens and a 2k $\times$2k 
CCD camera that observes wide and narrow fields of the sky 
through $V$- and $I$-band filters. The exposure time 
is 180 s per frame and 372 $V$-band observations (quality 
grades A to D) are offered of which we select 280 with 
the highest quality (grade A only). The typical $V$-error 
is $0^{\rm m}.03$ - $0^{\rm m}.04$. It is of note 
that the FEROS spectrum of June 2009 and the 
HERMES observations of Sep 2009 are coeval with the
ASAS-3 observations of MWC~314. The comparison 
of the radial velocity values (RV) and the corresponding 
$V$-magnitudes rules out the $\sim$1 m orbital period 
proposed in \citet{2008A&A...487..637M}. The $V$-brightness
curve consists of two minima over a variability period of 
$\sim$2 m. In Sect. 4.2, we further discuss the ASAS $V$-curve
in relation to the RV monitoring of MWC~314.  
   
\section{H$\alpha$ imaging}

Figure 1 shows the MEROPE narrow band H$\alpha$ image
that we observed around MWC~314 ({\em inset image}). The outer flux image is
adapted from Fig. 1 of \citet{2008A&A...477..193M}. The
MEROPE image is contoured in the linear intensity scale
to enhance the faint H$\alpha$ nebulosity. 
The intensity contour image is very practical 
for cross-comparison with the continuum passband image in 
the right-hand panel of Fig. 2. The faint east-west H$\alpha$ emission
is not detected in the broadband continuum image. With the absence of 
other similarly strong emission lines in the optical spectrum, we find 
that MWC~314's bipolar nebula primarily consists of an 
extended loop-like filament structure of ionised hydrogen gas. 
We observe very strong and broad H$\alpha$ emission 
with line peak fluxes more than 20 times above the continuum flux level
and equivalent line widths of 120 - 150 \AA\,({\em see Sect. 4.4.2}). 
The aperture size of our spectroscopic H$\alpha$ observations is limited 
to $\simeq$2\arcsec, and MEROPE imaging is therefore useful for 
detecting possible spatial asymmetry of H$\alpha$ emission 
that is related to the extended E-W lobe structures. 

We devised various image calibration
methods but were unable to detect any 
H$\alpha$ flux asymmetry within 20\arcsec\, from the central
source. The left-hand panel of Fig. 3 shows a narrow-band 
H$\alpha$ subimage of 526 $\times$ 526 pixels for a field of view
of 100\arcsec\, by 100\arcsec. The continuum subimage is shown in the
right-hand panel. The angle between two
tick marks is 100 pixels or 100 $\times$ 0.19\arcsec = 19\arcsec.       
The inner (nearly Gaussian) intensity distribution is practically 
circular and symmetric in both contour images, having a HWHM of 7.0$\pm$0.5
pixels or 7$\times$0.19\arcsec=1.33\arcsec. The innermost contour 
is drawn at a level of 1.82\% of the central 
intensity maximum. The outermost contour is drawn at 0.92\% ({\em dots}). 
The comparison of these intensity levels clearly reveals the larger
spatial extension of the H$\alpha$ emission envelope around 
MWC~314. The contours of the stars towards the S and SW
have the same (almost circular) shape in both images. Only at distances 
well beyond 20\arcsec\, from the central star, where H$\alpha$ flux levels
rapidly decreases to below 0.5\% of the central peak intensity, 
do we observe patchy regions of faint H$\alpha$ emission. Our 
imaging observations therefore do not indicate a preferred 
direction of (wind) outflow that may (have) produce(d) the extended
bipolar H$\alpha$ nebula. The images do reveal that the 
inner H$\alpha$ envelope is directly linked to MWC~314 and 
associated with very bright H$\alpha$ line emission emerging 
from the central binary system.

\section{Spectroscopic and photometric variability}
\subsection{Orbital elements and stellar parameters}
We investigate a number of unblended {\em absorption} lines 
for accurate RV measurements. Three lines S~{\sc ii} $\lambda$5454.8, $\lambda$5473.6,
\& $\lambda$5647.0, and one line Ne~{\sc i} $\lambda$6402.2
are selected, because they show well-defined flux minima
(for SNR$>$100) with respect to the neighbouring continuum 
flux level. We measure the RV of each line
from the line bisector at half flux minimum after local continuum
normalisation. The spectra are converted to the
heliocentric rest frame, and we estimate the rms error 
of the RV-values based on the noise
levels of fluxes observed inside these lines. For the strongest line, 
the RV mean errors stay below $\pm$3-7 $\rm km\,s^{-1}$, which 
can nearly double for the weaker absorption lines. 

We use the {\sc Phoebe} code of \citet{2005ApJ...628..426P} for computing the best fit 
to the RV-curve and the ASAS $V$-brightness curve provided in heliocentric Julian days (HJD).
Figure 4 shows the 16 RV-values that we measure ({\em see Table 1}) from the spectra
with the best fit solutions from {\sc Phoebe} ({\em curved solid lines in top 
and bottom panels}). {\sc Phoebe} calculates the best fit orbital period ($P$) 
by varying the values of the eccentricity ($e$), maximum distance between both stars 
($a$), systemic velocity ($\gamma$), orbital inclination angle ($i$), 
and longitude of periastron ($\omega$). The best fit solution to
the combined RV- and $V$-curve also varies the mass ratio of both stars.
The code computes the stellar masses and radii with corresponding surface 
gravities. It also gives the surface brightness ratio, employing a logarithmic 
limb darkening law for both stars. We use the {\sc Phoebe} internal fitting method 
of differential corrections and obtain the lowest $\chi^{2}$-values 
of 0.01248 for the 280 $V$-points and of 0.01180 for the 16 RV-points.  

Table 2 lists the best fit orbital elements and stellar parameters of MWC~314.
The error values of the orbital elements are computed from the errors
of the RV and $V$ input data. The RV rms errors are listed in Table 1,
while the $\sigma$-errors of $V$ stay typically below $\pm$0$^{\rm m}$.05, 
as shown in the bottom panel of Fig. 5. The top panel of Fig. 5 
shows the best fit RV-curve as a function of orbital phase ($\phi$).
An eccentric orbit with $e$=0.23 best fits the RV observations, 
while a circular orbit ($e$=0) does not. We find that the systemic velocity of 
$\simeq$$+$28~$\rm km\,s^{-1}$ is indeed very close to the bisector positions 
of many static metal emission lines ({\em see Sect. 4.4.3}).

The ASAS $V$-curve covers $\sim$38 orbital periods (or cycles). 
There is substantial scatter that can be due to a long-term 
trend observed towards fainter magnitudes in 2008-2009, when compared 
to 2003-2006. It can be due to multiple 
periodicity and/or cycle-to-cycle variations, since it 
substantially decreases when using only cycles 20 to 38. 
This is even more apparent when using data from cycles 30 to 38 only. 
In the middle panel of Fig. 5, we fold the $V$ data with $P$=60.8 d. 
The scatter in $V$ over three time intervals is shown with three 
different symbols. The $V$-curve shows two brightness minima with 
the orbital phases: the main minimum is around $\phi$=0.55-0.65, and a 
shallower minimum is around $\phi$=0.95-0.05. 
The mean amplitude does not exceed $0^{\rm m}.1$, but the 
cycle-to-cycle scatter between $V$-extrema is $\simeq$$0^{\rm m}.15$. 
The unequal brightness minima are due to an eclipsing binary system. 
The main minimum is due to partial eclipses by the companion star 
of the more luminous primary, while the companion is partly occulted 
by the primary star in the shallower $V$-minimum.

It is important to point out that the orbital and stellar
parameters in Table 2 are computed with {\sc Phoebe} 
using $T_{\rm eff}$=18\,000 K for the primary star. We determine
the latter value from detailed non-LTE fits to three Si~{\sc iii}
absorption lines in Sect. 4.2. By varying $T_{\rm eff}$ of the secondary 
star, {\sc Phoebe} computes the minimum $\chi^{2}$-values by simultaneously 
best fitting the RV- and $V$-observations. We obtain $T_{\rm eff}$=6227 K   
for the secondary star. It yields a surface brightness ratio of $\sim$15.
The slow $V$-brightness variability modelled with the code results 
from the deformation from spherical shape of the primary's surface
in a semi-detached binary system. The primary fills its Roche lobe
and alters the projected surface area (against the sky) during its 
orbit. Figure 6 shows the increase in projected surface area
at $\phi$=0.8 around maximum $V$ brightness ({\em bottom left-hand panel}). 
The surface deformation is smaller at $\phi$=0.3 ({\em top right-hand panel}), 
because the distance between the stars further increases towards 
apastron around $\phi$=0.6 and the partial eclipse by the secondary star. 

During periastron passage, the shape deformation of the primary is 
strongest but mostly behind the star for $i$=72$^{\circ}$, while 
the occultation of the secondary causes a stronger brightness decrease of $V$.
We do not incorporate reflections or spots in the 
{\sc Phoebe} model of the $V$-curve. {    The best fit signals a yellow 
supergiant secondary star with a small radius. Interestingly, 
the binary system HDE 326823 also has a large mass-function with 
a secondary that is likely obscured by a thick torus \citep{2011AJ....142..201R}. 
The best $V$-fit with {\sc Phoebe} of MWC~314 is sensitive to $i$, 
and the brightness minima result from the partial eclipses. In the case there
is a thick torus around the secondary, the torus' optical thickness 
in $V$ changes the duration of the brightness minima.}

We compute an rms errorbar of $\pm$13$^{\circ}$ 
to the best fit value of $i$  with {\sc Phoebe}. We use the error value for 
computing errors to the provided stellar masses.
Figure 7 shows $m_{2}$ as a function of $m_{1}$ for 
$i$=72.8$^{\circ}$.
We compute
\begin{equation}
\frac{m^{3}_{2}}{m^{2}} {\rm sin}^{3}i = \frac{P}{2\,\pi\,G} {K_{1}}^{3} \,,
\end{equation}  
where $m$=$m_{1}$+$m_{2}$=65.92$\pm$9~$\rm M_{\odot}$ and 
the semi-amplitude $K_{1}$-velocity of $\simeq$84.5~$\rm km\,s^{-1}$ 
from the RV-curve. The solid drawn
vertical and horizontal lines mark the 
best fit values of $m_{1}$=39.66~$\rm M_{\odot}$ and $m_{2}$=26.26~$\rm M_{\odot}$. 
The dashed drawn lines mark the computed error boundaries of 30.7~$\rm M_{\odot}$ $\leq$ 
$m_{1}$ $\leq$ 44.2~$\rm M_{\odot}$ and 25.3~$\rm M_{\odot}$ $\leq$
$m_{2}$ $\leq$ 30.7~$\rm M_{\odot}$ for $i\pm$13$^{\circ}$. The mass 
we compute for MWC 314 signals a massive binary system of $m$$>$65~$\rm M_{\odot}$
with stellar mass errors below $\pm$9~$\rm M_{\odot}$.     
The total bolometric luminosity $L_{\star}$ of the system is 
7.1~$10^{5}$~$\rm L_{\odot}$. The latter value is lower 
than previous estimates \citep[i.e.,][]{1998A&AS..131..469M}, but
appears to be a normal trend for stars in the upper portion of the H-R diagram 
(for example also in the Pistol star). The $L_{\star}$ of MWC 314  
is close to the Eddington luminosity limit with $\Gamma_{\rm Edd}$ $>$0.5 for 
electron scattering. It is much higher than in other B-type hypergiants 
\citep[such as Cyg OB2 No. 12, BP~Cru, and HD 152236, see][]{2012A&A...541A.145C},
placing MWC 314 in the ballpark of an LBV.

Figure 8 depicts the orbit of the primary star in MWC 314. After 
the main $V$-brightness minimum, the primary approaches the observer 
with the largest RV (e.g., maximum blueshift of the absorption lines) 
at around $\phi$$\simeq$0.75-0.8. At the largest elongation, the total 
brightness is at a maximum. Since we compute $\omega$=289$^{\circ}$ ({\it Table 2}), 
periastron occurs around $\phi$=0.12. The dashed line in Fig. 8 
is the primary orbit computed for $e$=0.23 around the centre-of-mass 
of the binary system. The orbital phases that we observe are marked. 
We compute the semi-major axis 
of $a_{1}$sin$i$=98.7$\pm$6~$\rm R_{\odot}$ with $K_{1}$ 
and $P$. The periastron distance, computed with $a_{1}$(1-$e$) and 
$i$=72.8$^{\circ}$$\pm$13$^{\circ}$, is 79.5$\pm$8~$\rm R_{\odot}$. 
The apastron distance is 127.0$\pm$13~$\rm R_{\odot}$. The long 
axis of the primary's orbit is 79.5$+$127.0~$\rm R_{\odot}$ 
= 206.5~$\rm R_{\odot}$, or very close to 1 AU. 

\subsection{Absorption lines and atmospheric parameters}

We investigated the continuum normalised line profiles of three absorption lines 
of MWC~314 in Fig. 9. The bottom panels show flux plots evenly
stacked according to the orbital phase to avoid overcrossing lines.
The spectrum numbers are labelled in the bottom left-hand panel. 
The top panels show the corresponding dynamic spectra in the heliocentric velocity scale. 
We observe that the depth and width of the S~{\sc ii} and Ne~{\sc i} 
absorption lines do not substantially vary over the orbit, although the lines become 
deeper ($\sim$15\%) for $\phi$=0.65-0.85. We performed
detailed comparisons of the absorption line profiles and do not find 
indications that these high-excitation lines are influenced by emission 
contributions with orbital phase. It signals that the RV shifts 
result from orbital motion of the entire absorption line formation region 
and that Doppler effects of possible atmospheric pulsations
stay small. This is also supported by the high total RV amplitude 
of 169~$\rm km\,s^{-1}$, which with the short period of 
$P$=60.8~d, rules out RV variability caused by stellar pulsation. 

\citet{1998A&AS..131..469M} identify a good number of 
S~{\sc ii} (3), N~{\sc ii} (2), Ne~{\sc i} (4), and Al~{\sc iii} (2) 
absorption lines in the optical spectrum of MWC~314 (see their Table 5).
We search for other doubly ionised absorption lines,
such as C~{\sc iii} $\lambda$4647.42 and $\lambda$4650.25,
but do not find any evidence of them. We neither detect 
the N~{\sc iii} absorption lines at $\lambda$6478 and
$\lambda$6487. We observe the Si~{\sc iii} absorption 
line at $\lambda$5739.7, and the Si~{\sc iii} triplet lines at $\lambda$4552.62,
$\lambda$4567.82, and $\lambda$4574.76. The last line is a weak absorption line
blended in the red wing due to veiling from neighbouring emission lines.
We do not find any evidence of Si~{\sc iv} absorption lines
(i.e., $\lambda$4116, $\lambda$4631, and $\lambda$4654).
A study of the Si~{\sc ii}/Si~{\sc iii}-ionisation balance 
in MWC~314 turns out to be complicated in that 
many of the Si~{\sc ii} lines are filled in by (static) emission above 
the continuum flux level. The photospheric Si~{\sc ii} absorption 
line becomes noticeable during some orbital phases 
({\em see Sect. 4.4.4 for a discussion of Si~{\sc ii} $\lambda$6347}), 
but the emission contributions severely compromise the Si~{\sc ii} lines 
for reliable depth and equivalent width (EW) measurements.   

We performed detailed radiative transfer fits to three Si~{\sc iii}
lines using the {\sc Multi} code \citep{1986UppOR..33.....C} for determining 
atmospheric parameters of the primary. The $\lambda$4552.62,
$\lambda$4567.82, and $\lambda$5739.7 lines are fairly unblended
for modelling detailed shapes and EWs. 
The Si~{\sc iii} $\lambda$4574.76 line is weak (the weakest line of the $3s4p$ triplet lines) 
in MWC~314, and we do not attempt to fit this line. The $\lambda$4552.62 and
$\lambda$4567.82 triplet ($3S$-$3P^{\circ}$)  lines belong to the $3s4s$-$3s4p$ Si~{\sc iii} 
atomic configuration with both sharing the lower energy level of $\chi_{\rm low}$=19.01 eV (with $\chi_{\rm
up}$=21.7 eV). 
The $\lambda$5739.7 line is a singlet ($1S$-$1P^{\circ}$) transition with slightly larger
$\chi_{\rm low}$=19.72 eV (and $\chi_{\rm up}$=21.88 eV). The latter line is also 
weakest of the three lines, and we adopt log($gf$)=$-$0.096 from the NIST database.
We also adopt log($gf$)=0.292 for $\lambda$4552.62, and log($gf$)=0.068 for 
$\lambda$4567.82, because the NIST values are provided with the $B^{+}$
accuracy flag. We developed a Si~{\sc iii} atom model with 12 energy levels, which range from 0 eV to 21.88 eV, 
and a (Si~{\sc iv}) continuum level. The upper levels of the triplet lines 
are treated as separate levels, while the three levels of $3s3p,$ $3p^{2}$, and $3s3d$ are
combined into superlevels. We solve the detailed balance for
29 selected lines (and 12 continua), of which some can directly (de-)excite 
the upper and lower energy levels of the observed Si~{\sc iii} triplet and singlet lines. 
The (non-LTE) lambda iterations to detailed balance are performed, with the temperature of 
the stellar continuum radiation field ($T_{\rm rad}$) set equal 
to 18 kK, until variations in the line source functions decrease to below 
$\sim$1~\%. For example, our Si~{\sc iii} atom model
also includes the $3p^{2}$ $^{1}S$ and $2s3d$ $^{1}D$ energy 
levels between the upper and lower levels of the triplet lines.
The collisional and radiative equilibrium calculations therefore 
also include the $\lambda$4338.5 line, which has the upper level in common with the 
singlet line. We observe a weak $\lambda$4338.5 line in the spectrum of MWC~314, but 
it is too noisy for detailed fits. The $\lambda$9323.89 and $\lambda$10525.5
lines are also included (connecting $^{1}D$-$^{1}P^{\circ}$ and $^{1}D$-$^{3}P^{\circ}$ levels),
but these lines are outside our observed wavelength range. We also 
compute the $\lambda$4325.888 ($^{3}S$-$^{1}P^{\circ}$)
and $\lambda$6173.61 ($^{1}S$-$^{3}P^{\circ}$) lines but do not detect them 
in the observed spectra; the log($gf$)-values are below $-$2.6. 
   
We employ a grid of plane-parallel atmosphere Kurucz models 
computed with the opacity distribution functions of \citet{2004astro.ph..5087C}.
{ The Kurucz atmosphere models are utilised
in {\sc multi} for performing the detailed Si~{\sc iii} line profile calculations. 
{ The \sc multi} code was developed for non-LTE modelling of detailed
line profiles using input multi-level atom models that can
be modified for energy levels, continua, selected line transitions, 
branching ratios, cross sections, etc. The transfer code
is very useful for testing the effects of various atomic input
parameters on the theoretical line profiles. We find it particularly 
advantageous for constructing atom models of the
higher ions Si~{\sc ii} and Si~{\sc iii}, because the input atomic data are
collected and combined from various databases.}

We adopt mean solar metallicity [M/H]=0.0 Kurucz atmosphere models and use solar abundance values 
for silicon. The model $T_{\rm eff}$ ranges from 17 kK to 21 kK in steps of 1000 K.
The model surface gravity values, log $g$, range from 2.5 to 4.0 in steps of 0.5 dex.
We compute the Si~{\sc iii} lines for two projected microturbulence velocity ($\zeta_{\mu}$) 
values of 5 and 10 $\rm km\,s^{-1}$. Next, the computed lines are broadened
with an instrumental profile of $R$=80,000, and the projected rotational velocity 
$V_{\rm rot}$sin$i$ range from 40 to 100~$\rm km\,s^{-1}$ in steps of 5~$\rm km\,s^{-1}$.
A least squares fit to the observed profiles of the three Si~{\sc iii} lines
reveals that $V_{\rm rot}$sin$i$ does not exceed 50~$\rm km\,s^{-1}$. We obtain best fits 
for atmosphere models of ($T_{\rm eff}$, log $g$) = (17 kK, 2.2), (18 kK, 2.6),
and (19 kK, 3.0) in the case that $\zeta_{\mu}$=5~$\rm km\,s^{-1}$. Figure 10 shows 
the observed Si~{\sc iii} line profiles overplotted
with the computed profiles for 17 kK, 18 kK, and 19 kK.
The boldly drawn lines show profiles computed for log $g$=2.5, while the thin drawn 
lines for log $g$=3.0. Models with $T_{\rm eff}$ above 19 kK yield EW values
incompatible with the observed values. This is also the case for higher 
$\zeta_{\mu}$=10~$\rm km\,s^{-1}$, requiring $T_{\rm eff}$ values 
well above 20 kK. Models with $T_{\rm eff}$$\geq$20~kK are not supported, 
because we do not observe the optical Si~{\sc iv} lines. 
Note that the wavelength regions of the Si~{\sc iv} lines are void 
of strong emission lines, so that these absorption lines ought to be observed 
in the case that $T_{\rm eff}$ of MWC~314 would exceed 20 kK. Our modelling of the 
Si~{\sc iii} absorption spectrum of MWC~314 indicates a minimum 
log $g$$\simeq$2.2 dex. Smaller log $g$-values 
would require $T_{\rm eff}$ below 17 kK, which is not compatible
with several high-excitation Ne~{\sc i} absorption lines that we also 
observe in the spectrum.

\subsection{Spectral energy distribution}
We compare the spectral energy distributions of MWC~314 and the (first discovered) 
LBV P Cygni. The comparison is justified by very similar optical
spectra \citep{1992iue..prop.4281P}. 
\citet{2001ASPC..233..215D} provides an overview of $T_{\rm eff}$-values for P Cyg published over the past 80 years.
For 2000, he adopts $T_{\rm eff}$=18600~K with an estimated uncertainty of a few 100 K.
The effective acceleration ($g_{\rm eff}$) in the atmosphere and 
wind is dominated by radiative forces (more than 10 times the gravity acceleration), 
and values for log $g_{\rm eff}$ ranging from 1.2 to 2.05 are derived from 
spectral lines that form in different parts of the wind of P Cyg. These values are 
comparable to log $g$=2.2-2.6 and $T_{\rm eff}$=17-18 kK that we compute for MWC~314 in Sect. 4.2 
using non-LTE fits of three photospheric absorption lines. 

Figure 11 shows the flux density of broadband (BB) photometric observations
of MWC~314 ({\em left-hand panel}) and P Cyg ({\em right-hand panel})
that we collect from the literature. Solid black dots mark published flux values
(we convert to flux density) observed in the {\em Johnson} 
(\citet{2001KFNT...17..409K}; \citet{1978A&AS...34..477M}; 
\citet{2007PASP..119.1083R}; \citet{2006yCat.2168....0M}; \citet{2008PASP..120.1128O};
\citet{2002yCat.2237....0D}), {\em Str\"{o}mgren} \citep{1998A&AS..129..431H}, 
and {\em Vilnius} \citep{1989BICDS..37..179S} photometric systems
and in various 2MASS \citep{2003yCat.2246....0C}, TD~1 \citep{1976A&A....49..389H}, 
OAO~2 \citep{1980ApJS...43..501C}, Tycho~2 \citep{2000A&A...355L..27H}, 
Hipparcos \citep{1997ESASP1200.....P}, and SDSS \citep{2008ApJS..175..297A} 
photometric bands. The BB flux density values are plotted at the 
effective wavelength $\lambda_{\rm eff}$ of the passbands. 
We observe the SED maximum around $\sim$2~$\mu$m in MWC~314, 
longward of the SED maximum of P Cyg. The latter maximum occurs 
shortward of 1~$\mu$m. This was also observed by \citet{1996A&A...312..941M} 
with $UBVRIJHK$ photometry ({\em see his Fig. 1}). 
He computes  $A_{\rm V}$=$5^{\rm m}.5$ from model 
optical SEDs with $T_{\star}$=20 kK and log $g$=2.5 for MWC~314. 
The $E_{B-V}$=$1^{\rm m}.83$ signals a very reddened 
luminous B-type star with intrinsic colours close 
to P Cyg. 

For a more detailed study of the SEDs longward of 3~$\mu$m 
we complement Fig. 11 with photometric data from 
the AKARI ({\em solid boxed symbols}), MSX ({\em open boxed symbols}), 
WISE ({\em solid triangles}), and {\it Spitzer}-IRAC missions ({\em open pentagonal symbols 
longward of 3~$\mu$m in MWC~314}). The (linear) absolute IR flux densities and errors 
are listed in Table 3. 

We investigate the SED slopes 
of both stars with NASA  Spitzer Space Telescope and  
ESA Infrared Space Observatory (ISO) spectra. The {\it Spitzer}-IRSX 
spectrum of MWC~314 is shown in Fig. 11. 
It was observed between 5.21~$\mu$m and 34.98~$\mu$m on 6 Nov
2008\footnote{sha.ipac.caltech.edu/applications/Spitzer/SHA}.   
\citet{2010ApJS..191..301A} note the presence of the broad silicate dust 
absorption feature at 10~$\mu$m that can originate in the 
interstellar medium. The feature is weak in MWC~314 
compared to other LBVs and LBV candidates. LBV candidate 
Cyg OB2 No. 12 (B5 Ie) has a very strong silicate absorption 
feature ({\em see their Fig. 5A}), indicating its large dust extinction
may be of circumstellar nature. The right-hand
panel of Fig. 11 shows the ISO flux density spectrum of 
P Cyg ({\em solid drawn line with spikes}) between 2.36~$\mu$m and 12~$\mu$m.
It was observed with ISO-SWS on 17 Oct
1996\footnote{iso.esac.esa.int/ida}.
The ISO spectrum longward of 12~$\mu$m is not shown, because the absolute 
flux errors rapidly exceed 1 Jy. We overplot the {\it Spitzer}-IRSX
spectrum of P Cyg that is observed between 14.0~$\mu$m and 39.72~$\mu$m
(on 17 Dec 2008 for SIRTF IRS Calibration Program), 
which have small flux errors below 0.05 Jy shortward 
of 37~$\mu$m. The {\it Spitzer} and ISO spectra show that the 
IR SED slope of P Cyg is indeed only slightly steeper than the {\it Spitzer} 
spectrum of MWC~314 that we also observe with AKARI, MSX, and WISE
photometry. 

The sub-panel in the right-hand panel of Fig. 11 compares 
the {\it Spitzer} spectrum of MWC 314 and the ISO and {\it Spitzer} 
spectrum of P Cyg. The spectrum flux values are plotted 
with the flux errors, provided by the ISO and {\it Spitzer} 
calibration pipelines. The P Cyg spectrum is shifted down by 
0.9 for detailed comparison to the spectrum slope in MWC 
314. The IR spectrum of MWC 314 can be well modelled by 
$\lambda^{-0.68}$ up to 22~$\mu$m ({\em dash-dotted line}). 
On the other hand, the IR spectrum of P Cyg is best fit with a 
steeper slope. We find that 
changes in the (inverse lambda) power of $\sim$$\pm$0.05 
significantly alter the model slope. The best fit to the 
continuum flux level that we observe in both spectra yields $\lambda^{-0.68\pm0.05}$
in MWC 314 and $\lambda^{-1.00\pm0.10}$ for P Cyg.

The IRAS 12~$\mu$m, 25~$\mu$m, and 60~$\mu$m 
fluxes in Fig. 11 are high-quality observations
and reveal IR SED slopes that are almost identical to the {\it Spitzer} or/and ISO spectra. 
In P Cyg, however, the 60~$\mu$m BB flux exceeds the $\lambda^{-1.0}$ flux 
density dependence, and the {\it Spitzer} spectrum reveals a 
decrease of the gradient in flux density towards longer 
wavelengths. The IR SED slope, connecting the IRAS 60~$\mu$m and 
the JCMT-Scuba submillimetre ({\em open circle shown at 859~$\mu$m}) 
flux values \citep{2008ApJS..175..277D} in P Cyg, yield 
a far-IR slope of $\lambda^{-0.73}$ ({\em dotted drawn line}).    
 
In Fig. 11, we plot the photospheric continuum flux distribution 
of both stars. We compute photospheric fluxes
using Kurucz models ({\em see Sect. 4.2}) of $T_{\rm eff}$=18 kK and
log $g$=2.5 for MWC~314 and $T_{\rm eff}$=19 kK and log $g$=2.5 for P Cyg.
The photospheric spectra are almost identical longward of the 
SED maxima, following a Rayleigh-Jeans tail ($\lambda^{-2}$). 
Between 2~$\mu$m and 100~$\mu$m, the SEDs however substantially exceed the photospheric
model flux values. The excess flux is attributed to free-free 
(continuum) radiation emerging from the ionised and expanding winds
of massive hot stars with emission lines (provided scattering by dust grains is 
negligible). \citet{1999isw..book.....L} show that the predicted energy distribution 
is a power law of $\lambda^{-0.61}$ for radio free-free emission
from an isothermal wind with constant velocity ({\em see their Eq. 2.20}).
The shape of the SED shortward of 100~$\mu$m depends on the gas
kinetic temperature and density in the layers where the 
wind velocity increases outwards. In P Cyg, the radiative wind driving 
mechanism yields a `beta'-power law with $\beta$$\simeq$1 
{ \citep[i.e.,][\citet{2001ASPC..233..133N}]{1997A&A...326.1117N}}. 
Since we observe comparable near-IR SED slopes for $T_{\rm eff}$-values of 
P Cyg and MWC~314 that are nearly identical and compute integrated total 
excess fluxes that largely exceed the photospheric flux by similar factors, 
we can adopt the $\beta$=1 wind velocity structure for MWC~314 
(We further discuss the values of $\beta$ for 3-D RT modelling in Sect. 5.3.) 
The conservation of wind momentum requires $\rho(r)$=$\rho_{0}$/$r^{2}$ 
for the smooth wind density structure of MWC~314.
However, the spherical wind geometry is not supported 
since we find that MWC~314 is a short-period binary system, and the 
values of $\dot{M}$ and $v_{\infty}$ observed in P Cyg cannot 
readily be assumed for MWC~314. In Sect. 5, we measure $v_{\infty}$ from 
detailed radiative transfer fits to He~{\sc i} P Cyg-type line profiles
in MWC~314. The 3-D model of the radial wind velocity structure employs a modified 
$\beta$-law that accounts for the orbital motion of the primary star observed 
at different orbital phases in the rest frame of the observer. 
 
\subsection{Emission line variability}
\subsubsection{He~{\sc i} P Cyg lines}
The strong He~{\sc i} $\lambda$4471, $\lambda$5876, 
$\lambda$6678 emission lines show remarkable variability in the spectrum 
of MWC~314. The normalised line fluxes and dynamic spectra are shown with $\phi$ in the heliocentric 
radial velocity scale in Fig. 12. The He~{\sc i} lines are always single-peaked with emission flux 
maxima around the systemic $\gamma$-velocity. The high-resolution spectra reveal variable 
blue-shifted absorption which becomes strongest during $\phi$=0.65-0.85. 
The He~{\sc i} lines show variable P Cyg-type line profiles. The P Cyg profile is always 
observed in He~{\sc i} $\lambda$5876. 
In Fig. 13, we compare the He~{\sc i} $\lambda$5876
profiles of three LBVs, $\eta$ Car, P Cyg, and AG Car, to the line in MWC~314. 
The top panel plots two spectra of MWC~314 ({\em solid thick and thin drawn 
lines}) to 1300~$\rm km\,s^{-1}$. We observe the blue-shifted wing absorption well
beyond 1000~$\rm km\,s^{-1}$ in MWC~314, while it is observed to $\sim$500~$\rm km\,s^{-1}$
in the other LBVs. \cite{2007A&A...465..993G} find a similar behaviour in the
He~{\sc i} $\lambda$10830 line of MWC~314 with the absorption wing
extending to 1800~$\rm km\,s^{-1}$ at certain epochs.  {    \citet{2010AJ....139.1534R}
discuss P Cyg type absorption near $-$500~$\rm km\,s^{-1}$ in H$\alpha$ of $\eta$ Car
which could be possibly due to binary interaction around periastron passage.}
The bottom panel of Fig. 13 shows strong absorption 
variability in He~{\sc i} $\lambda$5876 of MWC~314 for wind velocities below 
500~$\rm km\,s^{-1}$ ({\em solid thick and thin drawn lines}). We observe similar 
blue-shifted absorption variability in the higher H Balmer emission lines, 
such as H$\delta$ ({\em solid thick and thin drawn lines}).      

The blue-shifted absorption observed in the He~{\sc i} lines
of MWC~314 is produced in a fast-expanding warm wind surrounding the 
primary, a massive hot supergiant. The expanding 
wind opacity becomes best visible in high-temperature (i.e., He~{\sc i}) 
lines because the absorption line formation region is located relatively close to the 
primary within a few $R_{\star}$ above the stellar photosphere ({\em see Sect. 5}). 
The He~{\sc i} line emission emerges from a wind region close to the supergiant 
and does not considerably fill in the blue-shifted wind absorption. 
The strong emission portions of the lower H Balmer series lines (that form at lower excitation 
temperatures), on the other hand, emerge from a much more extended circumbinary wind 
envelope that completely squelches blue-shifted absorption. 
The variable blue-shifted P Cyg absorption in the He~{\sc i} lines results from
the increase in line-of-sight wind opacity in front of the hot star
during orbital phases of fastest approach to the observer.
During $\phi$=0.65-0.85, the radial velocity of the blue-shifted wind absorption is 
most separated from the velocity of the emission line, and hence 
the overall P Cyg line shape becomes also best visible. The orbitally modulated 
absorption that we observe in the He~{\sc i} lines signals an asymmetrically 
extended wind envelope with enhanced density, which is oriented towards the wind region 
that leads the orbit of the primary. {    Interestingly, similar behaviour is discussed 
by \citep{2002A&A...389..931T} for RMC 81 with higher $v_{\infty}$ at orbital phases when 
the secondary is in front of the primary, which is observed in the higher Balmer lines and 
metallic P Cyg profiles. They attribute this to an increase in the mass-loss rate 
in the direction of the secondary. This is caused by tidal interaction with the 
stellar wind, which can result in a focused wind \citep{1982ApJ...261..293F}.}
In Sect. 5, we develop a 3-D model of the binary wind structure in MWC~314, 
which fits the observed P Cyg He~{\sc i} line variability.      
 
\subsubsection{H Balmer and Paschen emission lines}

We also investigate the hydrogen Balmer and Paschen emission lines
in relation to the spectroscopic orbital variability of MWC~314. 
The normalised H$\alpha$ and H$\beta$ line fluxes are plotted 
according to the orbital phases in the top panels of Fig. 14. 
Both H lines are the strongest emission lines in the optical and near-IR spectrum. 
The FWHM of the H$\alpha$ line is $\simeq$200~$\rm km\,s^{-1}$, while 
$\simeq$150~$\rm km\,s^{-1}$ in H$\beta$.   
The FWHM values gradually decrease towards the higher Balmer series lines,
while the unblended H Pa14 emission line has a 
FWHM$\simeq$100~$\rm km\,s^{-1}$ ({\em bottom panels}). 
The line wings of H$\alpha$ extend well beyond $\pm$300~$\rm km\,s^{-1}$.
The high-resolution profiles reveal that
the violet emission line wing is always more extended
than the red one. Towards the higher H Balmer lines, stronger 
{\em absorption} extending to $\sim$300~$\rm km\,s^{-1}$ becomes more apparent in 
the violet line wing. The H$\delta$ and Pa14 lines in the 
bottom panels of Fig. 14 exhibit enhanced blue-shifted absorption during $\phi$=0.65-0.85,
similar to the He~{\sc i} lines in Fig. 12. The P Cyg-type line
shape of the higher Balmer and Paschen series lines is also noticeable, 
although it is less pronounced than in the He~{\sc i} lines.     
  
We find clear evidence that the continuum normalised
emission line maxima of H$\alpha$ and H$\beta$ are variable and 
orbitally modulated. Figure 15 shows the upper portion of the continuum 
normalised H$\alpha$ ({\em bottom panel}) and H$\beta$ ({\em top panel})
emission lines. The line flux maxima are shifted to the orbital
phase $\phi$ and marked with the spectrum number.  
The normalised H$\beta$ line flux maxima vary between 6.4 and 7.3, 
while H$\alpha$ varies between 22.8 and 25.6 ({\em Table 1}). 
Both lines show largest flux maxima for $\phi$$\simeq$0.65 (spectrum Nos. 6 and 16),
and smallest flux maxima for $\phi$$\simeq$0.32 (Nos. 13, 15, and 7).
The $\phi$-dependent H$\alpha$ ({\em open symbols}) and 
H$\beta$ ({\em filled symbols}) flux maxima are shown scaled 
in the top panel of Fig. 16. For comparison purposes, the normalised H$\alpha$ 
line flux maxima are divided by 24, and the H$\beta$ maxima by 6.75.
We measure emission line EW values of 120 - 150 \AA\,
for H$\alpha$ and 35 - 45 \AA\, for H$\beta$. The H$\alpha$
line is however mutilated by strong telluric absorption lines 
at some orbital phases ({\em see the bottom panel of Fig. 15}), which 
can slightly perturb the EW-orbital phase dependence.

We also find a clear correlation between the continuum normalised emission 
flux maxima observed in the H$\alpha$ \& H$\beta$ lines and the
$V$-brightness curve of MWC~314. The $V$-band brightness
maximum around $\phi$=0.3 in the bottom panel of Fig. 5 yields
H$\alpha$ and H$\beta$ emission line maxima that become apparently weaker
after continuum flux normalisation compared to orbital
phases when the $V$-continuum brightness is dimmest ($\phi$=0.65).
The orbital variability of about $\pm$7\% in the continuum normalised emission 
flux maxima of both lines ({\em top panel of Fig. 16}) is fully accounted 
by the photometric $V$-continuum flux changes of about 
$\pm$$0^{\rm m}.05$ that we observe in the ASAS light curve of MWC~314.
The product of the visual flux (we compute from $V$ with
orbital phase) and the continuum normalised emission line
maxima is almost invariable.
The near constancy of the {\em absolute} H$\alpha$ and H$\beta$ emission line 
fluxes therefore indicates that the H emission line formation regions are not 
substantially influenced by orbital motion. It signals a 
very extended formation region for the H$\alpha$ and H$\beta$  
emission lines. We think therefore that both H emission lines form in a very extended 
circumbinary envelope around MWC~314 in which the bulk of the line emission 
regions are not substantially modulated by the orbital motion.  

\subsubsection{Orbital modulation of permitted and forbidden emission lines}

The optical spectrum of MWC~314 shows many prominent metal emission lines of 
Fe~{\sc ii}, Ti~{\sc ii}, and Si~{\sc ii}.
Figure 17 shows a portion of the spectrum with the permitted Fe~{\sc ii} $\lambda$6416.9, 
$\lambda$6432.7, and $\lambda$6456.4 emission lines. The normalised 
flux spectra are shifted upwards according to the orbital phase ($\phi$ and spectrum No.
are marked). The Doppler shift of the Ne~{\sc i} $\lambda$6402.2 absorption line with $\phi$ 
is shown by the vertically plotted curved line ({\em dotted}). The three Fe~{\sc ii} lines 
mostly show double and occasionally triple flux maxima. It is very remarkable to 
observe that the detailed shapes of the emission lines almost perfectly depend on $\phi$.
The flux maxima of the blue and red emission peaks vary with $\phi$; the red
maximum exceeds the blue peak for $\phi$=0.95 to $\sim$0.5, while the blue peak
exceeds the red one for $\phi$=0.5 to 0.95. The bottom panel of Fig. 16
shows the ratio of blue/red (B/R) emission flux maxima with $\phi$ for 
Fe~{\sc ii} $\lambda$6433 ({\em filled triangles}) and Fe~{\sc ii} $\lambda$6456
({\em open boxed symbols}). {    The B/R flux ratios are provided in the last two columns of 
Table 1}. Notice the strong correlation to the (normalised) intensity variations in 
the H$\alpha$ and H$\beta$ emission line 
maxima in the top panel of Fig. 16. The B/R flux ratio variations 
of about $\pm$10~\% that we observe, however, result from intrinsic 
changes in the detailed profile shape of the Fe~{\sc ii} emission lines with $\phi$. 

We also observe orbital modulation over short $\phi$-intervals. For example, 
the Fe~{\sc ii} lines from $\phi$=0.405 to
0.462 reveal {\em three} separate flux maxima in the central
emission cores (see spectrum Nos. 12, 1 , 8, and 4). We compare the spectra 
of 5 June 2009 (No. 1; $\phi$=0.440) and 4 Oct 2010 (No. 8; $\phi$=0.444)
and observe nearly identical emission line shapes. Both spectra are however 
observed 16 m apart or separated in time by $\sim$8 orbital periods.        
It is of note that very weak and stable spectral features close to the stellar 
continuum almost exactly repeat over time (i.e., observed around 6445 \AA\, in Fig. 17),
signaling that orbital modulation strongly determines the spectral 
line variability that we observe in MWC~314. {    The triple-peaked metal 
emission lines of MWC 314 are very remarkable spectral features. 
They are clearly linked to the spectral variability with orbital phases 
and will be investigated in the future. The similarity to the 
phenomenon in the spectra of Be-stars is also striking and may signal 
a common emission line formation mechanism, which is due to the circumbinary 
disc of a close binary system. For example, \citet{2010AJ....140.1838S} mention 
the triple emission line maxima can be caused by a one-armed 
oscillation or precessing disc in Be-star $\zeta$ Tau.}

We investigated the detailed shapes and FWHM widths of
various permitted and forbidden emission lines in the high-resolution 
HERMES spectra of MWC 314. Very striking are the double-horned shape 
of many permitted and forbidden emission lines. For example,
the wings of Fe~{\sc ii} $\lambda$6433 in the right-hand 
bottom panel of Fig. 18 are very steep, while the line core 
is almost flat around the central flux minimum. Similar 
detailed line shapes are observed in [Fe~{\sc ii}] $\lambda$7155 and
[Fe~{\sc ii}] $\lambda$8616 in the top and bottom 
right-hand panels of Fig. 18. We do not measure any significant 
Doppler shift in the position of the steep emission line wings 
with $\phi$. The line wings are static within $\pm$3~$\rm km\,s^{-1}$. 

Three [N~{\sc ii}] emission lines 
$\lambda$6583, $\lambda$5754, and $\lambda$6548 are observed.
The first two are shown with $\phi$ in the top and bottom 
left-hand panels of Fig. 19. 
The [N~{\sc ii}] lines are single peaked with almost flat-topped 
emission maxima that can occasionally become double peaked 
(i.e., [N~{\sc ii}] $\lambda$5754).
{    The [N~{\sc ii}] emission lines are formed in nebulae 
around many massive stars, signaling CNO processed material. 
The lines are observed in a number of LBVs with $T_{\rm eff}$ 
that are very similar to MWC 314 like P Cyg, AG Car, HR Car, and HD 168625. 
Cooler hypergiants such as HR~8752 (currently of A-type), 
also reveal the [N~{\sc ii}] lines \citep{2013ASPC..470..167L}.} 
We searched for other forbidden emission lines in the spectrum
but failed to detect the [O~{\sc i}] $\lambda$5577 and 
$\lambda$6300 lines reported by \citet{2005A&AT...24..317K}.
There is no indication of stellar emission at these line rest wavelengths.    
The permitted O~{\sc i} triplet $\lambda$7771-5 lines, on the other hand,
reveal a strong and broad P Cyg shape, while O~{\sc i} $\lambda$8446
is a prominent double-peaked emission line. We also 
observe [Ca~{\sc ii}] $\lambda$7291.506 and $\lambda$7323
emission lines, although they are heavily mutilated by 
overlying narrow telluric $\rm O_{2}$ doublet lines.
The Ca~{\sc ii} H \& K lines also show P Cyg profiles, 
while the near-IR Ca~{\sc ii} triplet lines are prominently 
double- or multiple-peaked emission lines. The S~{\sc ii} $\lambda$6347
line shows a shallow and extended P Cyg line shape. We also search for forbidden metal 
emission lines of [Ti~{\sc ii}] and [Cr~{\sc ii}] but 
fail to identify them. We employ the [Ti~{\sc ii}] line list 
in Table 2 of \citet{2005MNRAS.361..206H}
(we convert to air wavelengths) but do not detect
any [Ti~{\sc ii}] emission lines in MWC~314.

Table 4 provides a list of emission lines that we investigate in
the HERMES spectrum of MWC 314 of Sept. 5 2009 (spectrum 
No. 2 in Table 1), when the radial velocity is close to
the $\gamma$-velocity. It also lists the
heliocentric radial velocities of the blue and red emission maxima of
double-peaked lines.

\subsubsection{Emission line formation}

\citet{2008A&A...487..637M} find that the double-peaked metal 
emission lines in MWC 314 are consistent with line formation 
in a rotating disc. They measure the blue and red component 
velocities of permitted and forbidden Fe~{\sc ii} 
(Cr~{\sc ii} and Ti~{\sc ii}) double-peaked emission lines 
and compute the detailed line shape with a model of a projected 
Keplerian rotating disc. Interestingly, they also compute 
that the Fe~{\sc ii} emission 
occurs from 2$\pm$1 times up to 15$\pm$2 $R_{\star}$. 
The [Fe~{\sc ii}] lines form 
in the same disc at distances to the star between 12$\pm$2 to 
15$\pm$2~$R_{\star}$. 

We observe that the permitted and forbidden Fe~{\sc ii} lines 
have almost identical FWHM of $\simeq$110~$\rm km\,s^{-1}$ 
and comparable differences between blue-and red-peak (radial) velocities of 
$\sim$65~$\rm km\,s^{-1}$. The strongest and broadest double-peaked 
lines in the spectrum of MWC 314 are the near-IR H Paschen emission lines. 
All Pa lines reveal substructures in the emission flux maxima 
(i.e., H Pa14 in Fig. 14) with FWHM$\sim$100 to 140~$\rm km\,s^{-1}$. 
In contrast, we observe much smaller FWHM values for the 
[N~{\sc ii}] emission lines that do not exceed 50~$\rm km\,s^{-1}$, 
which are considerably less than the Pa and Fe~{\sc ii} lines. 

The FWHM values of the emission lines in MWC~314 are 
a measure of the (projected) rotation velocity in the line formation 
regions of a Keplerian disc. \citet{2011AJ....142..201R} 
discuss the formation of double-peaked Fe~{\sc ii}, Si~{\sc ii}, and Ne~{\sc ii} 
lines in a circumbinary disc of the massive WR$+$O binary 
HD~326823. They observe B/R variability in these double-peaked 
emission lines, which are very similar to the orbitally modulated 
B/R variability that we observe in MWC 314. In HD 326823, 
the Fe~{\sc ii} emission lines form in a circumbinary disc consistent 
with Keplerian motion required to explain the double-peaked 
profiles. \citet{2011AJ....142..201R} provide an expression 
for the projected Keplerian rotational velocity in the 
circumbinary disc gas as a function of distance $r$ from the 
binary. It follows a $r^{-1/2}$-dependence with highest emission 
speed at the inner disc boundary. We think therefore that 
the Pa lines in MWC~314 form closer to the binary centre
of gravity and at higher gas temperatures, than the metallic 
emission lines, while the [N~{\sc ii}] lines form at longer 
distances in the outer circumbinary disc regions. 

The semi-amplitude $K_{1}$-velocity of $\simeq$84.5~$\rm km\,s^{-1}$ that 
we measure in the RV-curve of absorption lines ({\em see Sect. 4.1}) is 
larger than the HWHM and/or half of the difference in B/R-peak 
velocities of double-peaked emission lines: $\sim$65~$\rm km\,s^{-1}$ 
in Pa, $\sim$55~$\rm km\,s^{-1}$ in Fe~{\sc ii}, and $\sim$25~$\rm km\,s^{-1}$ 
in [N~{\sc ii}] lines. It indicates that the radius of the primary's 
orbit (producing the photospheric absorption lines) is smaller than 
the mean line formation region distances of the emission flux maxima in a 
Keplerian rotating disc.

The orbitally modulated B/R variability that we observe in 
MWC 314 can result from two separate line formation regions. 
The top and bottom left-hand panels of Fig. 18 show 
the continuum normalised Si~{\sc ii}
$\lambda$6347 line variability with $\phi$. The single-peaked Si~{\sc ii}
emission line is static around the 
$\gamma$-velocity. The detailed shape of this Si~{\sc ii} emission line is however 
periodically altered by Doppler displaced photospheric 
absorption (following an S-shape) and yields the double-peaked 
Si~{\sc ii} emission line around $\phi$=0.9-0.1. 

We also subtract the average flux profile from the Fe~{\sc ii} 
$\lambda$6456 and $\lambda$6433 lines shown in Fig. 18. It yields residual 
flux spectra with a weak (and sometimes P Cyg-type) absorption 
line that Doppler shifts and causes the B/R peak variability 
({\em bottom panel of Fig. 15}). The triple-peaked emission 
profiles that we observe around $\phi$=0.4-0.5 result from rapid blue-shift 
in the photospheric line becoming noticeable for
RV-values around the $\gamma$-velocity between the blue and red
emission line maxima. We find that the orbital modulation of the 
B/R variability observed in double-peaked metal emission 
lines of MWC 314 can result from photospheric absorption 
Doppler displacing below the (static) emission line formed 
in the circumbinary disc. A comparable analysis of the orbitally 
modulated B/R variability observed in Fe~{\sc ii}
emission lines of HD~326823 is presented in \citet{2011AJ....142..201R}.

\subsection{Spectroscopic signatures of the companion star?}

The high S/N ratios and wavelength coverage of the HERMES 
spectra from 420 nm to 900 nm are useful for searching for possible 
spectral signatures of the companion star. Interestingly, \citet{2010ApJS..191..301A}
find a pattern of emission lines in the
{\it Spitzer} spectrum ($R$$\sim$90) of MWC~314. They mention strong He~{\sc i} 
and He~{\sc ii} but weak metallic lines, which are similar to early-type O stars, such as 
the O5f$+$ star HD~14947. Their Figs. 8 and 14 mark two weak 
He~{\sc ii} lines at $\simeq$22.5 $\mu$m and $\simeq$27.7 $\mu$m. 
However, we do not observe the He~{\sc ii} $\lambda$4685.7 line in the HERMES spectra. 
The spectral region around 4686~\AA\, is almost continuous (flat) without 
significant emission or absorption features. 

The bottom spectrum of Fig. 20 shows the {\it Spitzer} spectrum
of MWC~314 ({\em thick drawn line}) between 5 $\mu$m and 
33~$\mu$m. The top spectrum ({\em thin line}) shows
the theoretical model spectrum computed with {\sc cmfgen}. The vertically 
drawn lines mark the positions of H~{\sc i} and He~{\sc i} lines in 
the synthetic spectrum calculations.
We compute strong H emission lines (from energy levels 11-9 and 
13-10) at 22.38 $\mu$m and at 27.8 $\mu$m (from levels 9-8). 
These H lines are also observed in {\it Spitzer} spectra of B5 supergiants, 
which do not show the He~{\sc ii} $\lambda$4685.7 emission line at mid-B 
spectral type. Our synthetic spectrum modelling shows that the {\it Spitzer} 
spectrum of MWC~314 is consistent with $T_{\rm eff}$$\simeq$17-19 kK. 
We do not observe optical or mid-IR spectral signatures of
an O-type star in MWC~314. 

\section{3-D binary wind modelling} 

We find several indications of a non-spherical geometry 
of the wind and circumstellar environment of MWC 314. First, MWC 314 
contains a hot supergiant primary star in a semi-detached massive binary 
system. The orbital modulation of P Cyg absorption 
provides important new information about the asymmetry of wind
geometry around the primary. It is supported by the deformation of 
the surface of the primary from a spherical shape due to filling of its
Roche lobe ({\em see Sect. 4.1}). Second, the prominently double-peaked
emission line spectrum indicates a circumbinary disc that 
can result from wind accretion in MWC 314. Therefore, it is not 
appropriate to employ one-dimensional radiative transfer models to 
analyse the P Cygni-type lines of MWC 314.

In this section, we develop a quantitative 3-D model of the large-scale 
wind structure around the primary star in MWC~314. The wind velocity and 
density structure are computed with detailed 3-D radiative transfer (RT) 
fits to the variable P Cyg profiles observed in the He~{\sc i} lines 
of Figs. 12 and 13. Our goal is to obtain insights that help to constrain 
the physical parameters (e.g., wind density contrast) of the accretion 
flow in the system.

Figure 21 provides a schematic 3-D representation of MWC 314. 
The primary star is drawn at apastron passage ({\em small blue concentric spheres}) 
in orbit ({\em dashed line}) around the centre of gravity ({\em central dot}). 
The circumbinary disc is drawn in the equatorial plane of the binary 
({\em marked with dotted concentric lines}). The inner accretion 
disc radius ({\em thick drawn ellipse}) is drawn around 
periastron distance of the primary. We investigate wind accretion 
in MWC~314 combined with the results of the 3-D wind modelling 
in Sect. 6.

\subsection{Wind model velocity structure}

The 3-D wind model that we develop does not incorporate a model of 
an accretion disc in MWC 314. Detailed RT modelling of the 
variable shape of double-peaked emission lines is outside the 
scope of this paper. We concentrate here on variable absorption
observed in He~{\sc i} P Cyg profiles that form in the supersonically 
expanding wind of the primary. The stellar wind is driven by the 
strong radiation field of the hot supergiant to distances much 
larger than the size of its orbit. The line RT calculations
are performed with the {\sc Wind3D} code \citep{2008ApJ...678..408L}, 
using a maximum wind model radius of 30~$R_{1}$ around the binary 
centre of gravity. We adopt a $\beta$=1 radial velocity law $v(r)$ 
for the wind and a 'smooth' wind density structure $\rho(r)$=$\rho_{0}/r^{2}$ 
similar to P Cygni ({\em we justify in Sect. 4.3}). The density 
at the base of the wind $\rho_{0}$ is set equal to $\dot{m_{1}}/(4\,\pi\,R^{2}_{1}\,v_{0})$,
where $\dot{m_{1}}$ is the primary's wind mass-loss rate of 
$\sim$10$^{-5}$~$\rm M_{\odot}$\,yr$^{-1}$ and $v_{0}$$\simeq$10~$\rm km\,s^{-1}$ 
is the wind velocity around $r$=$R_{1}$. 

First, we perform a
Galilean transformation of $v$ from the primary's rest frame to the observer's
frame. Our 3-D RT calculations utilise a Cartesian model grid. 
The wind velocity is $ v$=($v_{x}$,$v_{y}$,$v_{z}$) in the frame
of the primary. The  wind velocity component $v_{x}$ is transformed 
to the observer's frame by $v^{'}_{x}$=$v_{x}$ + $V_{r}$, where the
last term is the observed radial velocity depending on $\phi$. 
The observer's line of sight is in the (x,z)-plane of the model grid ({\em i.e., the vertical frontside 
plane marked in Fig. 21}), and we incline the orbital plane by 18$^{\circ}$ from the (x,y)-pane. 
The transformation causes Doppler effects due to 
the primary's orbital motion for the calculation of detailed line profiles 
in the frame of the observer. The effect of $V_{r}$ on $v^{'}_{x}$ is 
largest in wind regions of $r<2$~$R_{1}$, where $\left| V_{r}\right|$$\simeq$$\left| v_{x}\right|$ 
(Note that $v_{x}<0$ in the front hemisphere.) At a longer distance from 
the primary, the wind rapidly accelerates and approaches $v_{\infty}$=1200~$\rm km\,s^{-1}$.
This linear (orthogonal) transformation however yields unphysical 
$v^{'}_{\infty}$-values that differ by 2 $\times$ $V_{r}$
for line-of-sight directions in front and behind the primary. For example, 
the terminal wind velocity in the front hemisphere
increases by $V_{r}$ during fastest approach to the observer, 
while it {\em decreases} in the far hemisphere
(due to the opposite direction of $v_{x}$ behind the primary). 
However, $v^{'}_{\infty}$ must 
approach the same velocity in all (radial) directions around the 
binary system at a longer distance from the stellar surface ($r>$10~$R_{1}$).
At a longer distance, the acceleration of the wind mainly 
results from the primary's strong radiation field. Particles escaping at 
the wind base will gain mechanical momentum from the orbital motion 
(measured in the observer's frame), which becomes negligibly small 
when compared to the radiative driving by the stellar radiation field at 
a longer distance in the wind. 

We compute a wind crossing time $t(r_{w})$=$\int_{R_{1}}^{r_{w}} \! \mathrm{d} r / v(r)$ 
of $t(10)$$\simeq$10 d with $\beta$=1  for particles travelling
from $R_{1}$ to 10~$R_{1}$. At the latter distance, the wind velocity is
$v$$\simeq$0.9 $\times$ $v_{\infty}$ = 900~$\rm km\,s^{-1}$. The wind particles travel 
another $\sim$12 d ($t(30)$=22 d) before reaching 30~$R_{1}$ in the model 
where $v$$\simeq$$v_{\infty}$. Hence, the dynamical timescale of the wind 
outflow in our model is sufficiently long compared to $P$=60.8 d 
for the radiative driving to homogenise $v^{'}_{\infty}$ to the same 
velocity in all directions around the binary system. We therefore adopt a 
modified $\beta$-law providing equal $v^{'}_{\infty}$-values at distances 
$r>10~R_{1}$. The 3-D wind velocity structure is given by
\footnotesize
\begin{equation}
v^{'}_{x}(r,\theta,\psi)=-\left( v_{0}
+(v^{'}_{\infty}(\phi)-v_{0})\left(1-\frac{R_{1}}{r}\right)^{\beta^{'}}\right)
{\rm sin}(\theta)\,{\rm cos}(\psi) + V_{r}(\phi)\,, 
\end{equation}
\normalsize
where $\theta$ and $\psi$ are the polar and azimuthal angles of a point in the wind at distance 
$r>R_{1}$ above the stellar surface. We choose the x-axis of the model
along the line-of-sight to the star. The wind speed at the photosphere
($r=R_{1}$) is $v_{0}$. The minus sign of the first term in
Eq. (2)
follows from the convention for the sign of radial velocity $V_{r}$,
depending on the last term on $\phi$. 
Note that $V_{r}(\phi)$ = $V^{\rm orb}_{x}(\phi)$ = $V^{\rm
orb}(\phi)$\,{\rm sin}($i$)\,{\rm cos}($\omega$), where $V^{\rm orb}$ is the orbital velocity. Likewise,
$V^{\rm orb}_{y}(\phi)$ = $V^{\rm orb}(\phi)$\,{\rm sin}($i$)\,{\rm sin}($\omega$) and $V^{\rm orb}_{z}(\phi)$ =
$V^{\rm orb}(\phi)$\,{\rm cos}($i$) are computed with $i$, $\omega$, and the orbital 
velocity $V^{\rm orb}(\phi)$ using the orbital elements in Table 2. We denote in Eq. (2):
\begin{equation}
v^{'}_{\infty}(\phi)=v_{\infty} + V_{r}(\phi)\,{\rm sin}(\theta)\,{\rm cos}(\psi) \,,
\end{equation}
which modifies $v_{\infty}$ in the frame of the primary by the x-axis projected radial velocity.
Inserting Eq. (3) in Eq. (2) for $r$$\rightarrow$$\infty$ 
yields $v^{'}_{x}$($\theta$=$\frac{\pi}{2}$, $\psi$=0)=$-$$v_{\infty}$ 
and $v^{'}_{x}$($\theta$=$\frac{\pi}{2}$, $\psi$=$\pi$)=$v_{\infty}$. 
Hence, the non-linear base transformation provides the same terminal wind speed 
in front and behind the primary in the frame of the observer by 
adjusting $v_{\infty}$ for the (projected) orbital velocity at $\phi$.  
The $v^{'}_{y}$- and $v^{'}_{z}$-components are given by:
\footnotesize
\begin{equation}
v^{'}_{y}(r,\theta,\psi)=-\left( v_{0} +(v^{'}_{\infty}(\phi)-v_{0})\left(1-\frac{R_{1}}{r}\right)^{\beta^{'}}
\right)
{\rm sin}(\theta)\,{\rm sin}(\psi) + V^{\rm orb}_{y}(\phi)\,, 
\end{equation}
\normalsize
and
\begin{equation}
v^{'}_{z}(r,\theta)=-\left( v_{0} +(v^{'}_{\infty}(\phi)-v_{0})\left(1-\frac{R_{1}}{r}\right)^{\beta^{'}} \right)
{\rm cos}(\theta) + V^{\rm orb}_{z}(\phi)\,,
\end{equation}   
where the last term of Eq. (3) is replaced by 
$V^{\rm orb}_{y}(\phi)\,{\rm sin}(\theta)\,{\rm sin}(\psi)$ and 
$V^{\rm orb}_{z}(\phi)\,{\rm cos}(\theta)$,
respectively.
We also modify the wind acceleration law in the base transformation. 
The value of $\beta^{'}$ in Eqs. (2), (4), and (5) denotes a small 
adjustment of $\beta$ required to sustain the same wind acceleration 
in all directions around the binary. It results from the product
of $v^{'}_{\infty}$ and $(1-R_{1}/r)^{\beta^{'}}$. If $\beta^{'}$=$\beta$ for constant $\beta$, 
the multiplication yields slightly variable wind acceleration over ($\theta$, $\psi$)-directions, 
because Eq. (2) modifies $v^{'}_{\infty}$ with the projected 
orbital velocity. Using an angle-dependent adjustment for $\beta$ of the form
 $\beta^{'}$ = $\beta$ + $p$ \,${\rm sin}(\theta)\,{\rm cos}(\psi)$,
we compute  almost identical wind accelerations 
in all directions in the observer's rest frame with $p$$\simeq$0.07
for $\beta$=1, $v_{\infty}$=1200~$\rm km\,s^{-1}$, 
and $\left| V_{r} \right|$ $\leq$100~$\rm km\,s^{-1}$. 

It is important to point out that the 3-D binary wind geometry we develop
for MWC~314 $ v^{'}$=($v^{'}_{x}$,$v^{'}_{y}$,$v^{'}_{z}$) is not spherically 
symmetric around the centre of gravity of the binary system. 
At the base of the wind, the outflow (for inertial observers) is centred around 
the primary orbiting the binary centre of gravity, while the wind flow 
at a long distance assumes constant terminal velocity in all directions around the system.
The 3-D wind geometry is therefore asymmetric in the vicinity of the primary's orbit
and directly depends on $\phi$, while the wind flow at a long distance becomes
symmetric and independent of $\phi$. The left-hand top panel of Fig. 22
shows the change in wind velocity $v^{'}$ $-$ $v$ in the equatorial plane around the primary
($\theta$=$\pi$/2) due to the orbital motion. We compute 
$v^{'}$ = $\sqrt{v^{' 2}_{x}+v^{' 2}_{y}}$ with Eqs. (2), (3), and (4) 
for $V_{r}$=$-$85~$\rm km\,s^{-1}$, $V^{\rm orb}_{y}$=0~$\rm km\,s^{-1}$, 
$v_{0}$=10~$\rm km\,s^{-1}$, and $p$=0.07. For given $\phi$, the primary moves in the direction towards the 
observer south in the panel. The wind velocity in the observer's frame 
$v$ is computed for $V_{r}$=0~$\rm km\,s^{-1}$ 
and $\beta^{'}$=$\beta$=1 ($p$=0). In front of the primary, 
the wind velocity {\em difference} in the observer's frame decreases from $-$85~$\rm km\,s^{-1}$ 
at the surface ($r=R_{1}$) to above $-$10~$\rm km\,s^{-1}$ for $r>$3 $R_{1}$. 
Behind the star, it increases from $-$85~$\rm km\,s^{-1}$ at
the surface to below 10~$\rm km\,s^{-1}$ for $r>$3 $R_{1}$. At longer distances
from the surface (shown in Fig. 22 to $r$=10 $R_{1}$), the difference of
the radial wind velocity and the stationary wind (when the primary were static for 
the observer) is less than 10~$\rm km\,s^{-1}$ in all directions around the star.
The bottom left-hand panel of Fig. 22 shows $v^{'}(r)$ for $\psi$=0, $\pi$/3, 
2$\pi$/3, and $\pi$. For clarity, the four curves are shifted 
to the right by $+$1~$R_{1}$. The dashed lines in the upper panel of Fig. 22 
mark contours of equal $v^{'}$ $-$ $v$ that are denser in front 
of the primary than behind it. It results from the orbital velocity, which 
slightly increases the radial wind velocity gradient in the direction 
of the star's orbital motion, while decreasing it in the opposite direction.          
The influence of orbital velocity on the overall wind velocity
structure, however, rapidly becomes negligibly small ($<$10~$\rm km\,s^{-1}$) 
due to the asymmetric base transformation presented. It allows us 
to adequately measure the effect of (small) $\phi$-dependent density enhancements    
in the wind of the primary using 3-D RT modelling of selected spectral lines. 

\subsection{Wind model density structure}  

We parametrise the wind density structure around the primary of MWC~314 with
\footnotesize
\begin{equation}
\rho(r,\theta,\psi) = \rho_{s}(r) \, \left( 1+ \frac{f}{r} \,{\rm exp}\left(- \left(\frac{\psi-
\Upsilon_{0}}{\varsigma_{0}}\right)^{2}   \right) 
\,{\rm exp}\left(- \left(\frac{\theta-\Theta_{0}}{\vartheta_{0}}\right)^{2}   \right)
\right) \,,
\end{equation}
\normalsize
where $\rho_{s}$ = $\rho_{0}$/$r^{2}$ is the smooth wind density and $f$$\geq$0 
a free parameter that increases $\rho_{s}$ in the ($\Theta_{0}$, $\Upsilon_{0}$)-direction.
We adopt the $r^{-3}$-dependence for the increase in wind density, hence
$\rho$ $\rightarrow$ $\rho_{s}$ for $r$ $\rightarrow$ $\infty$. 
The constant values (in units of angle) $\varsigma_{0}$ and $\vartheta_{0}$ 
determine the azimuthal and polar density distributions around the primary star. 
Smaller values provide wind density structures that are more asymmetric
and oriented towards the ($\Theta_{0}$, $\Upsilon_{0}$)-direction. The top
right-hand panel of Fig. 22 shows the density contrast $\rho$/$\rho_{s}$
in the equatorial plane ($\theta$=$\pi$/2) to $r$=10~$R_{1}$.
We set $\Upsilon_{0}$=0 and $\Theta_{0}$=$\pi$/2,
providing enhanced wind density in front of the star moving south. 
We set the azimuthal density distribution to $\varsigma_{0}$=$\pi$/4,
which is equivalent to a mean opening angle of $\sim$2$\varsigma_{0}$=$\pi$/2 for
the larger density contrast in the wind. The bottom right-hand panel plots
$\rho/\rho_{s}$ for $\psi$=0 ({\em solid drawn 
line}), $\pi$/3 ({\em dash-dotted line}), 2$\pi$/3 ({\em dashed line}), 
and $\pi$ ({\em short-dashed line}). For $f$=1,
the wind density increases by 100~\% at the surface in front of the 
star but rapidly decreases to below 50~\% at $r$=2~$R_{1}$ and $<$$10$~\% 
for $r$$>$10~$R_{1}$. The percentages evidently depend on the 
$f/r$-factor adopted in Eq. (6). 
The RT modelling presented in Sect. 5.3 shows that the latter 
factor, combined with the polar and azimuthal Gaussian distributions
in Eq. (6), adequately parametrizes the large-scale asymmetric 
geometry of wind density in MWC~314.

\subsection{3-D radiative transfer modelling}

The {\sc Wind3D} code calculates the transport of radiation 
in spectral lines in a 3-D Cartesian grid. The code solves the
line transfer problem for arbitrary 3-D wind velocity $v_{ijk}$($r,\theta,\psi$)- and 
density $\rho_{ijk}$($r,\theta,\psi$)-structures around a spherical 
source radiating continuum flux. The line source function 
iterations employ an accelerated exact lambda iteration 
scheme for computing the mean intensities $\bar{J}_{ijk}$ in 
$101^{3}$ gridpoints. The RT scheme is fully parallelized 
with excellent load-balancing. We compute  the 
detailed profile of He~{\sc i} $\lambda$5876 in two steps
using 36 CPU cores on the ROB's central compute 
server `Plato'. It addresses 192 64-bit SGI shared-memory 
cores at 1.7 GHz with batch queue processing. First, 
the 3-D line source function is converged starting from a 
Sobolev approximation to local intensity fluctuations below 1~\%. 
The source function iterations are the most time-consuming, 
because they require integration over 80 angles 
to account for random line opacity- and velocity-distributions 
in the 3-D wind model that irradiates each gridpoint from 
all directions. Convergence can considerably slow down in non-uniform 
$v$- and $\rho$-structures, although we implement the
Cartesian upwind discretisation method \citep{1990A&A...240..541A}
to avoid non-convergence. This finite element method 
is unconditionally stable but requires large memory allocations. 
Since the wind model is symmetric about the orbital plane,
we initially compute the source function above 
the plane, followed by a fast convergence test for 
all gridpoints below the plane. This about halves the 
computation time of the first step. The second  step 
interpolates the source function on a finer grid of $801^{3}$ 
points and calculates the transfer equation for user-defined 
radial lines of sight around the model. It integrates 
the specific intensities over the front plane 
perpendicular to the line(s) of sight, hence 
providing the 3-D emergent line fluxes. 

{\sc Wind3D} solves the non-LTE transfer problem in scattering-dominated 
spectral lines formed in the extended winds of 
massive stars. We compute the dynamic He~{\sc i} line profile for 60 
equal time intervals in 0$\leq$$\phi$$<$1 over the orbit of the primary. 
This corresponds to uniform steps of 6$^{\circ}$ 
for 60 directions around the star observed in orbit 
at a constant inclination angle of $i$=72.8$^{\circ}$. 
The detailed line profile is computed at 100 wavelength 
points between $-$1300 $\rm km\,s^{-1}$ and 1300~$\rm km\,s^{-1}$.
Note that the computation time per line profile is considerably 
shorter ($\sim$10 min) than for the line source function 
iterations of $\sim$3-5 h. The dynamic spectrum 
calculations require total computation
times of $\sim$9-11 h per input model (using 36 cores), 
which is still acceptable for exploring a broad range of wind model parameters.  
Finer meshes require more CPU time and even larger
memory allocation. The grid resolution of $101^{3}$ points 
covers $\pm$30~$R_{1}$, which is adequate for modelling 
the effects of small velocity perturbations and density enhancements 
in the parametrized binary wind model developed in Sect. 5.1 \& 5.2.  
Similar grid resolutions of $71^{3}$ were used 
in \citet{2008ApJ...678..408L} for the detailed modelling of discrete absorption 
components observed in P Cyg 
profiles of luminous hot stars with 3-D hydrodynamic 
wind models that incorporate co-rotating interaction regions. 

We compute the He~{\sc i} $\lambda$5876 line opacity 
$\chi_{ijk}^{l}$ discussed in Appendix B of \citet{2008ApJ...678..408L}
with the parametrized binary wind model of $\rho_{ijk}$ and $v_{ijk}$. 
We use the smooth wind opacity parameters $\alpha_{1}$=0.1, 
$\alpha_{2}$=0.8, and $T_{\rm tot}$=0.18. The smooth wind acceleration 
parameters are $\beta$=1, $v_{\infty}$=1200~$\rm km\,s^{-1}$, 
and $v_{0}$=10~$\rm km\,s^{-1}$. The thermal line broadening in 
the Gaussian line profile function is $v_{\rm th}$=11 $\rm km\,s^{-1}$, which is
appropriate for kinetic gas temperatures at $\sim$20 kK in 
the absorption line formation region of unsaturated He~{\sc i} lines.
The He~{\sc i} $\lambda$5876 line is a $D_{3}$ transition between triplet states
$1s2p^{3}P^{\circ}$ - $1s3d^{3}D$ having $\chi_{\rm low}$=20.9 eV, which is 
comparable to the Si~{\sc ii} triplet modelled in Sect. 4.2 with 1-D 
RT in the stellar photosphere. {\sc Wind3D} computes 
the 3-D RT in the wind in the 2-level atom 
approximation. The $T_{\rm tot}$ parameter determines 
the total wind line opacity, depending on the 
line oscillator strength and the He abundance in the wind.   
The projected (micro-)turbulent velocity in the profile 
function is $\zeta_{\mu}$=5~$\rm km\,s^{-1}$ ({\em see Sect. 4.2}). 
We compute the 3-D structured wind velocity with 
Eqs. (2)-(5) using the best fit RV-curve $V_{r}$($\phi$)
for the 16 RV observations ({\em upper panel of Fig. 5}). 
We model the large-scale 3-D wind density around the primary 
with Eq. (6) for $f$-values between 1 and 10.
We set $\Upsilon_{0}$($\phi$)=2\,$\pi$\,$\phi$ $+$ $\phi_{0}$ 
and $\Theta_{0}$=$\pi$/2. Hence,  the density 
enhancement in the wind model for $\phi_{0}$=0 is always oriented with $\phi$ 
towards the leading side of the primary's orbit in its orbital plane. 

The left-hand top panel of Fig. 23 shows the 
dynamic spectrum of He~{\sc i} computed    
for $f$=10 and $\varsigma_{0}$=$\vartheta_{0}$=$\pi$/4.
The absorption portion of the P Cyg profile
is variable and strengthens for $\phi$=0.55-0.95.
The increase in absorption results from the wind 
density enhancement in the model in our line of sight 
to the primary star. During these orbital phases, the line opacity
exceeds the smooth wind opacity by a factor of
$\rho$/$\rho_{\rm s}$. The value of $\varsigma_{0}$
determines the duration of higher absorption,
while $\phi_{0}$ sets the orbital 
phase when the increase occurs. 
The lower panels of Fig. 23 show the normalised line fluxes
at 12 orbital phases shifted upwards between 0 and 1. 
For comparison, the middle panel shows the dynamic spectrum 
computed with $\varsigma_{0}$=$\vartheta_{0}$=$\pi$/3. 
The increase in line absorption occurs 
for $\phi$=0.45-1.0 due to the increased  
opening angle for higher wind density
around the star. We also increase the $V_{r}$-amplitude by a 
factor of four to amplify the effects of orbital motion on the emergent 
line fluxes. The primary's orbital velocity causes periodical 
Doppler shifts in portions of the line profile  
with the radial velocity forming near
the wind base. The S-wave becomes clearly
noticeable in the red-shifted emission line wing. 
On the other hand, the blue-shifted absorption 
around $v$$\simeq$1200~$\rm km\,s^{-1}$ is
stationary due to the constant $v_{\infty}$ in 
our wind model in all directions around the binary system
({\em see Sect. 5.1}).
 
The right-hand panels of Fig. 23 show the best fit 
dynamic spectrum to the He~{\sc i} $\lambda$5876 line we 
observe in MWC~314. The parameter values 
of $f$=3.3 and $\varsigma_{0}$=$\vartheta_{0}$=$\pi$/4
provide the best fit of the absorption variability  with $\phi$
in the P Cyg profile. The absorption line variability computed 
explains the periodic changes observed with $\phi$ in the He~{\sc i} 
lines of Fig. 12. Figure 24 shows the detailed comparison of 
observed and theoretical He~{\sc i} $\lambda$5876 line profiles.
The solid drawn lines are spectra observed at $\phi$=0.444 ({\em thin drawn line}) 
and at $\phi$=0.795 ({\em thick line}). The theoretical spectra are 
overplotted with dashed lines. At $\phi$=0.795, the He~{\sc i} absorption 
increases by $\sim$15~\%  due to the increase of wind density 
in our line of sight by a factor of $\sim$3.3 for 
distances of $r$$<$2~$R_{1}$ in the wind of the primary. 
It causes a variable He~{\sc i} $\lambda$5876 line opacity in 
wind regions of $v$$<$600~$\rm km\,s^{-1}$ due to the 
orbitally modulated orientation of enhanced wind density 
in front of the supergiant. For example,  
the higher wind density leading the primary's orbit for $\phi$=0.1-0.5 
is partially blocked from view by the star's visible surface 
hence diminishing the line opacity contributions to the 
He~{\sc i} absorption. During these orbital phases, the 
absorption at low wind velocity also Doppler shifts to longer wavelengths
and becomes more filled in by the He~{\sc i} emission line,
yielding the decrease to weaker blue-shifted line absorption. 
The He~{\sc i} absorption computed for $f$=10 is significantly 
too strong at $\phi$=0.795 ({\em dotted line}).
In the modelling, we tested $\beta$-wind values between 1 and
3. We compute a slope in the violet He~{\sc i} 
emission line wing considerably too shallow below 200~$\rm km\,s^{-1}$ for $\beta$$\geq$1.5.
$\beta$=1 yields the best fit to the He~{\sc i} line wing, providing 
a wind velocity structure with sufficiently fast acceleration to
$v_{\infty}$=1200~$\rm km\,s^{-1}$ .

We emphasise that the $f$-factor determined
in Eq. (6) for an asymmetric geometry of wind density 
around the primary star of MWC 314 is limited to 3.3 
using 3-D RT modelling of orbitally modulated He~{\sc i} P Cyg profiles. 
At a distance of 1~$R_{1}$ above the stellar surface ($r$=2~$R_{1}$), 
the highest wind density in the direction of $\psi$=$\Upsilon_{0}$ 
is $\rho$($r$=2)=$\rho_{0}$\,(1$+$3.3/2)/4 $\simeq$2\,$\rho_{0}$/3
for $\rho_{\rm s}$($r$)=$\rho_{0}$/$r^{2}$ and
$\rho_{0}$ the density at the wind base of $r$=1~$R_{1}$. 
Hence, we find that the wind density in
front of the primary rapidly decreases to below 
$\sim$$\rho_{0}$ within 1~$R_{1}$ above the stellar surface.
The asymmetric geometry of wind density computed 
is therefore limited to a volume around the primary within 
the periastron distance. It indicates that the primary's orbital 
motion, besides periodically modulating the He~{\sc i} $\lambda$5876 
line formation region, also produces the asymmetric geometry of wind 
density around the supergiant that can result from its 
appreciable surface deformation from a sphere ({\em see Sect. 4.1}).
This directional increase of wind density towards the front side 
of the primary's orbital path can also support a wind 
accretion model in MWC~314 due to enhanced 
wind mass-loss (partly), forming a circumbinary disc in 
the system's orbital plane.

\section{Discussion}

MWC~314 is a single-lined spectroscopic binary 
with a luminous early B-type supergiant. Its overall wind and 
SED properties are very comparable to that of LBVs, 
such as P Cyg. This is not surprising, given the co-location 
of MWC 314 with other LBVs in the H-R diagram, as shown in Fig. 25. 
MWC 314 has an average luminosity of an LBV, which is very similar to P Cygni 
and W243 \citep{2009A&A...507.1597R}. MWC 314 is surrounded 
by a symmetric inner H$\alpha$ nebula to $\sim$20$\arcsec$, however 
showing a distinct bipolar shape farther out. 
Some important aspects of the spectroscopic monitoring and 
modelling of the binary system in this paper deserve a more 
careful discussion.

We propose that the primary star of MWC 314 is an LBV
in a dormant state. It is very similar to the cLBV RMC~81 
that has H Balmer and Paschen series lines showing strong 
P Cyg profiles. \citet{2000ASPC..204...43W} 
 present a `working' model of RMC~81 (see their Fig. 7)
 with an interaction region around the periastron passage between 
 the primary and the unseen companion. \citet{2006ApJ...638L..93M}
 remarks however that the optical spectrum of RMC~81 contains 
 He~{\sc i} lines without a strong emission component. 
 The He~{\sc i} $\lambda$7065 line in RMC~81 shows 
 weak emission. We observe considerably stronger He~{\sc i} 
 $\lambda$7065 emission in MWC~314 with normalised emission
 maxima $\sim$3-4 times above the continuum flux level, 
 which is comparable to the line in P Cyg. Its He~{\sc i} profiles are almost 
identical to the P Cyg profiles with weak blue-shifted absorption 
that is observed in the LBV He 3-591 by \citet{1994A&A...281..833S} 
(see their Figs. 2b and 8). P Cyg and He 3-519 are 
surrounded by nearly spherical nebulae found in other LBVs, 
although both stars currently lack evidence of a close binary 
system. \citet{2010A&A...519A..33G} 
report IR observations of an arc-like nebula attached to RMC 81. 
It is also interesting that \citet{2002A&A...395..891H} 
favour the binary hypothesis for MWC 349A to explain very 
strong H Balmer (and He~{\sc i}) emission and a non-spherical
distribution of circumstellar matter. MWC 349A shows a clear 
bipolar nebula and is surrounded by a circumstellar disc
observed almost edge-on. Its spectral properties are 
consistent with a B[e] supergiant, such as the strong He~{\sc i} 
emission lines that we also observe in MWC~314. \citet{2012A&A...541A...7G} 
have recently suggested that MWC 349A is likely to be an LBV. 
It prompts us to ask the question if the extended bipolar nebula of
MWC~314 has been produced by the close binary system. 
A binary-driven wind model shaping the bipolar geometry of the
material ejected from the system during outbursts appears to be 
a possible scenario. We note that the companion could plow 
through the outer layers of the primary if the primary star 
develops S Dor variability (i.e., a rapid increase of $R_{1}$). 
This scenario has been advocated by 
\citet{2009NewA...14...11K} to explain the Giant Eruption 
of $\eta$ Car. \citet{2009ApJ...698.1698G} presented an 
interesting analysis of these radius changes in the LBV 
AG Car related to changes of $L_{\star}$ in 2001-2003. \\

The unequal $V$ brightness minima observed in MWC 314 
are due to an eclipsing binary with a relatively small 
size of the primary's orbit ($\simeq$1 AU). We observe a 
semi-detached binary system with the primary star that fills 
its critical Roche lobe and strong tidal interactions 
that deform the stellar surface from the spherical shape. 
Our 3-D RT modelling reveals asymmetric wind outflow 
of increased wind density in front of the primary's 
orbit, which modulates the He~{\sc i} P Cyg absorption line 
portions around $\phi$=0.65-0.85.
The 3-D wind geometry computed is likely more
complicated due to important effects of tidal and orbital 
motion around an inner wind region where strong mass-loss can 
accrete and accumulate in an extended circumbinary disc. 
The Coriolis force and the radiative wind driving in 
the orbital plane would direct the wind flow of increased 
density from the primary away from the line through 
$L_{1}$ (connecting the stellar centres) towards the 
region leading the primary's orbital path. 
The transfer of wind mass towards the circumbinary 
disc efficiently converts wind momentum to angular 
momentum at an inner disc radius around the primary's 
periastron distance. There are two 
important findings in our analysis of MWC 314 related to this. 
First, the projected rotational velocity $V_{\rm rot}$sin$i$$\simeq$50~$\rm km\,s^{-1}$ 
in photospheric absorption lines ({\em see Sect. 4.2}) is smaller 
or comparable to the maximum kinematic velocity that we observe in prominent 
emission lines emerging from the circumbinary disc. The largest HWHM 
peak separation of double-peaked H Paschen emission lines is 
$\sim$65~$\rm km\,s^{-1}$ ({\em see Table 4}). The primary's (low) 
rotation rate is therefore synchronous with the disc rotation rate 
at the inner boundary where wind mass is accreted. The nearly synchronous 
rates can result from the steady transfer of wind mass to the disc 
possibly due to critical Roche overflow at $L_{1}$.    
Next, we can estimate the amount of disc accretion from stellar winds  
with Bondi-Holye-Lyttleton accretion \citep{2004NewAR..48..843E}. 
Literature generally provides the BHL accretion rate of wind 
mass from $m_{1}$ into a disc around $m_{2}$ by 
\begin{equation}
 \dot{m}_{d} = \frac{4\,\pi\,G^{2}\,m^{2}_{2}\,\rho_{a}}{v_{a}}\,,
\end{equation}
where $v_{a}$ is the wind velocity at the accretion radius $r_{a}$=2$G$$m_{2}$/$v^{2}_{a}$ \citep{1952MNRAS.112..195B}.
Wind outflow from $m_{1}$ is captured within $r_{a}$ around $m_{2}$. We use $r_{a}$$\simeq$$a_{1}$ 
with $a_{1}$$\simeq$$R_{1}$ in MWC~314. Hence, $a_{1}$ is the distance $r$=2~$R_{1}$ from the centre
of $m_{1}$, where the maximum wind density $\rho_{a}$($r$=2)$\simeq$2\,$\rho_{0}$/3 for $f$=3.3 in Eq. (6).
The radiatively driven wind velocity is $v_{a}$($r$=2)=$v_{0}$ $+$ $v_{\infty}$/{2$^{\beta}$}, 
or $v_{a}$=$v_{\infty}$/2, since $v_{0}$$\ll$$v_{\infty}$ and $\beta$=1. The wind mass-loss rate at the surface of
$m_{1}$ 
is $\dot{m}_{1}$=4\,$\pi$\,$R^{2}_{1}$\,$\rho_{0}$\,$v_{0}$, which yields with Eq. (7):
\begin{equation}
 \dot{m}_{d} =  \frac{16}{3}\frac{G^{2}\,{m^{2}_{2}}}{R^{2}_{1}\,v^{3}_{\infty}\,v_{0}}\,\dot{m}_{1}.
\end{equation} 
We calculate $\dot{m}_{d}$= 1.03\, $\times$ $\dot{m}_{1}$  for $m_{2}$=26.26~$\rm M_{\odot}$, $R_{1}$=86.8~$\rm
R_{\odot}$ ({\em Table 2}), 
$v_{\infty}$=1200~$\rm km\,s^{-1}$, and $v_{0}$=10~$\rm km\,s^{-1}$. The factor in front of
$\dot{m}_{1}$ in Eq. (8) scales with $m^{2}_{2}$ and ranges from 0.95 to 1.40
for 25.3~$\rm M_{\odot}$$\leq$$m_{2}$$\leq$\rm 30.7~$M_{\odot}$.
We estimate that the disc accretion rate in MWC~314 is about the same order of magnitude 
as the primary's wind mass-loss rate $\dot{m}_{1}$. 
Our estimate of $\dot{m}_{d}$ reveals that a massive companion star in MWC~314 can accumulate
a considerable fraction of the wind mass shed by the primary star by disc accretion.
For a typical LBV, smooth wind mass-loss rates are of $\sim$$10^{-5}$~$\rm M_{\odot}\,yr^{-1}$, 
this implies that $\sim$1 $\rm M_{\odot}$ can amass in the circumbinary disc during the 
LBV evolutionary phase of $\sim$$10^{5}$ yr. This mass-loss scenario yields a massive 
binary system supporting a large (and heavy) circumbinary disc. The inner boundary 
becomes dynamically more unstable with time due to strong tidal 
perturbations by the massive companion stars. A large amount of angular momentum stored 
in the disc would escape the binary during unstable events that disrupt the inner 
boundary structure on 
much shorter timescales. We think therefore that a break-up mechanism of the inner disc boundary
can cause short-lived outburst events with very strong mass-loss from MWC~314.
Such an outburst may have produced the large bipolar nebulae observed around MWC~314.
{    Note that we compute  a fraction $\dot{m}_{d}/\dot{m}_{1}$ exceeding 0.1 with Eq. (8). 
Over longer periods of 2-5 $\rm 10^{5}$ yr, at least 0.2 to 1 $\times$ $\rm M_{\odot}$ 
is stored in the disc, which considerably increases the total disc mass, thereby rendering it 
dynamically unstable in the massive binary system. This may cause recurring outburst events during which 
disc mass becomes partly ejected from the system, or drive punctuated events with very high 
mass-loss rates. BHL disc accretion provides an interesting mechanism for expelling accumulated
wind material very far from a massive binary system in (bipolar) directional outflow.}

One may also wonder what the evolutionary stage of MWC 314 currently is 
and what would be its ultimate fate. Is the massive primary star perhaps 
on the verge of core-collapse? 
If so, one would expect that the supernova shock wave would interact with the potentially massive 
H-rich circumstellar disc around MWC 314. In this case, a SN IIn would occur, or namely, a SN 
with narrow H lines in the spectrum. Perhaps MWC 314 is in an earlier stage of evolution 
of systems like SN 2009ip, which has shown multiple outbursts and possibly has gone under
core-collapse \citep{2013ApJ...767....1P}. It is also interesting to note that 
\citet{2012arXiv1211.4577L} find evidence of a circumstellar disc in SN 2009ip, 
resembling the geometry that we found in MWC 314. Alternatively, the primary of MWC 314 is perhaps
still far from core collapse. In that case, it will eventually lose its entire H envelope, becoming 
a W-R star and exploding as a SN Ibc.

\section{Concluding remarks}

We summarise the results for MWC~314 as follows:

\begin{enumerate}

     \item MWC~314 is a massive binary with $P$=60.8 d and $e$=0.23. The orbital 
           elements are accurately determined from best fits of the combined 
           radial velocity and $V$ curves. We observe the system for $i$=72.8$^{\circ}$$\pm$13$^{\circ}$. 
           The maximum separation of both stars is 1.22 AU.
 
     \item The primary star is a hot supergiant with $T_{\rm eff}$=18 kK and
           30.7~$M_{\odot}$$\leq$$m_{1}$$\leq$44.2~$M_{\odot}$.
           It is a slowly rotating massive supergiant with $V_{\rm rot}$sin$i$$\simeq$50~$\rm km\,s^{-1}$ and
           $R_{1}$=86.8~$R_{\odot}$ that fills it Roche volume in a semi-detached binary system.

     \item Long-term high-resolution spectroscopic monitoring during 2 yr reveals that the 
           optical emission line spectrum is orbitally modulated. The H Balmer and Paschen emission lines, 
           the He~{\sc i} P Cyg line profiles, and the
           double-peaked permitted metal emission lines all show variability linked to the orbital period. 

     \item A detailed comparison of the SEDs of MWC~314 and LBV P Cyg reveal nearly identical IR slopes
           for free-free emission with wavelength. It signals a radiatively accelerating wind 
           in MWC~314 with a mean $r^{-2}$ density structure of the smooth wind, which is similar to P Cyg.  

    \item  We develop a modified $\beta$-law for modelling the large-scale wind velocity (and density) 
           structure for a massive star observed in a binary orbit. 
           The 3-D wind velocity structure accounts for the orbital motion in a radiatively 
           driven wind with constant terminal velocity in all directions around the binary system.   

    \item  Detailed 3-D radiative transfer modelling of He~{\sc i} P Cyg line profiles with orbital phases 
           reveals that the geometry of wind density around the primary is asymmetric. It causes 
           orbitally modulated variability of P Cyg absorption that forms in the primary's extended wind. 
           We observe wind outflow above 1000~$\rm km\,s^{-1}$ in He~{\sc i} $\lambda$5876. 
           The RT modelling requires a density enhancement factor of 3.3 above the smooth wind density. 
           The asymmetric increase of wind density leads the orbital path of the primary.
     
    \item  The double-peaked emission lines reveal a significant decrease
           in peak velocity separation from $\sim$65~$\rm km\,s^{-1}$ 
           in H Paschen, $\sim$55~$\rm km\,s^{-1}$ in Fe~{\sc ii}, and $\sim$25~$\rm km\,s^{-1}$ 
           in [N~{\sc ii}] lines. It can signal a Keplerian rotating disc in MWC~314 
           with lower rotation velocities in emission line formation regions  
           farther from the binary centre of gravity.

   \item   We do not observe He~{\sc ii} lines in MWC~314. Radiative transfer modelling of the IR {\it Spitzer}
           spectrum rules out the spectral O-type for the secondary star. The radial velocity
           and brightness curves indicate a less-luminous cool giant of
           25.3~$\rm M_{\odot}$$\leq$$m_{2}$$\leq$\rm 30.7~$M_{\odot}$. We estimate a wind accretion 
           rate for the circumbinary disc of this massive binary system around the primary's wind 
           mass-loss rate.

    \item  We observe variable continuum normalised maximum fluxes in strong H$\alpha$ 
           and H$\beta$ emission lines that clearly correlate to 
           the orbital phases. The $V$-brightness curve is due to an eclipsing 
           binary system with periodical $V$-continuum flux changes 
           that cause H$\alpha$ and H$\beta$ line variability. Strong H Balmer emission 
           emerges from an inner H$\alpha$ nebulae that we observe symmetrically around the 
           binary to $\sim$20$\arcsec$. At larger angular 
           distances, weaker H$\alpha$ emission emerges from a bipolar nebula that was previously 
           observed around MWC~314.

\end{enumerate}

Long-term spectroscopic monitoring shows that MWC~314 
is an excellent target for studying the physics of winds in massive 
binaries. The binary period of only $\simeq$2 m and $e$=0.23 offers 
a unique laboratory for investigating orbitally modulated spectral 
variability in one of the most massive hot stars. Our asymmetric 
wind model of MWC 314 contributes to the ongoing efforts 
for unraveling the driving mechanisms of LBV winds with 
punctuated mass-loss. An important question is how these outburst 
events can eject large amounts of hydrogen gas and shape 
structured nebulae around LBVs before they evolve to the Wolf-
Rayet phase. It also poses the more fundamental question whether binarity 
or multiplicity is a prerequisite for S Dor and Hubble-Sandage 
stars. The system $\eta$ Car is a long-period ($\sim$5.5 yr) massive binary 
with a compact and, possibly, massive companion 
\citep[i.e.,][]{2000ApJ...528L.101D}. The system HD~5980 has a massive W-R companion 
star with an orbital period of $\sim$20 d. We find that the 
secondary of MWC~314 is also a massive star with an orbital 
period $\sim$3 times longer. The number of known binary hypergiants, 
showing the orbitally modulated P Cyg line variability 
that we observe in MWC 314, is however still very limited. We are in 
the course of spectroscopically monitoring other LBVs and 
cLBVs, to search for similar line profile variability in examples, 
such as Cyg OB2 12, HD~168607, the Pistol star, and P Cygni.
The large-scale 3-D wind geometry in massive binary systems 
should be further explored and compared to detailed radiative 
transfer modelling of high-resolution spectroscopic observations. 
  
\begin{acknowledgements}
A.L. acknowledges funding from the ESA/Belgian Federal
Science Policy in the framework of the PRODEX programme.
The HERMES project and team acknowledge support from the Fund for
Scientific Research of Flanders (FWO), Belgium, support from the Research
Council of K.U.Leuven (Belgium), support from the Fonds National Recherches
Scientific, Belgium (FNRS), from the Royal Observatory of Belgium and from
the Landessternwarte Tautenburg (Germany). J.H.G. is supported by an Ambizione 
Fellowship of the Swiss National Science Foundation. We thank Drs. N. Gorlova and 
N. Cox of Univ. of Leuven, Belgium, for Mercator-MEROPE observations of 
MWC~314 in March 2011. {    We thank the referee for helping to improve
the presentation of the paper.}

\end{acknowledgements}

\bibliographystyle{aa} 

\newpage
\newpage
\begin{table*}
\caption{High-resolution Mercator-Hermes (H) and ESO-Feros (F) {\'e}chelle spectroscopic observations of MWC~314. 
The radial velocity $V_{r}$-value is given in the heliocentric rest frame. $\phi$ denotes the orbital phase.
Em. max. is the continuum normalised line flux maximum. B/R max. denotes the ratio of Blue and Red
normalised maximum emission fluxes of double-peaked lines.}    
\label{table:1}      
\centering   
\begin{tabular}{c c c c c c l c c c c}        
\hline\hline                 
No. & Obs. date & HJD        & Instr. & $V_{r}$                 & $\sigma(V_{r})$              & $\phi$ &  H$\alpha$   
& H$\beta$  & Fe~{\sc ii} $\lambda$6433    & Fe~{\sc ii} $\lambda$6456 \\   
    &           &            &        & [$\rm km\,s^{-1}$] & [$\rm km\,s^{-1}$]     &      &  Em. max.  & Em. max. &
B/R max. & B/R max. \\
\hline                        
   1 & 05-06-2009 &     2454987.254 & F   &  $+$62  & 7 & 0.440 &   24.969 &  6.602 & 0.934   & 0.960   \\
   2 & 05-09-2009 &     2455080.426 & H   &  $+$36  & 5 & 0.963 &   23.353 &  6.627 & 0.992   & 0.973   \\
   3 & 10-09-2009 &     2455084.459 & H   &  $+$69  & 4 & 0.039 &   24.206 &  6.589 & 0.967   & 0.926   \\
   4 & 06-04-2010 &     2455292.659 & H   &  $+$50  & 5 & 0.462 &   24.448 &  6.583 & 0.947   & 0.959   \\
   5 & 29-06-2010 &     2455376.635 & H   &  $-$70  & 5 & 0.795 &   24.462 &  7.171 & 1.090   & 1.011   \\
   6 & 17-08-2010 &     2455426.461 & H   &  $-$36  & 5 & 0.651 &   25.394 &  7.355 & 1.143   & 1.033   \\
   7 & 27-09-2010 &     2455467.383 & H   &  $+$82  & 4 & 0.325 &   22.843 &  6.392 & 0.885   & 0.910   \\
   8 & 04-10-2010 &     2455474.362 & H   &  $+$58  & 3 & 0.444 &   23.070 &  6.325 & 0.948   & 0.968   \\
   9 & 29-10-2010 &     2455499.325 & H   &  $-$44  & 3 & 0.852 &   23.706 &  6.647 & 1.039   & 0.992   \\
  10 & 18-03-2011 &     2455638.764 & H   &  $+$93  & 4 & 0.157 &   23.215 &  6.420 & 0.981   & 0.955   \\
  11 & 21-03-2011 &     2455641.764 & H   &  $+$99  & 6 & 0.208 &   23.421 &  6.392 & 0.942   & 0.939   \\
  12 & 02-04-2011 &     2455653.681 & H   &  $+$74  & 2 & 0.405 &   23.730 &  6.487 & 0.947   & 0.946   \\
  13 & 26-05-2011 &     2455708.691 & H   &  $+$89  & 6 & 0.310 &   22.974 &  6.196 & 0.873   & 0.864   \\
  14 & 05-07-2011 &     2455748.442 & H   &  $+$30  & 3 & 0.957 &   23.912 &  6.503 & 0.971   & 0.986   \\
  15 & 27-07-2011 &     2455769.542 & H   &  $+$85  & 2 & 0.318 &   23.667 &  6.442 & 0.894   & 0.901   \\
  16 & 17-08-2011 &     2455791.546 & H   &  $-$38  & 3 & 0.656 &   25.621 &  7.127 & 1.105   & 1.023   \\
\hline                                   
\end{tabular}
\end{table*}

\newpage
\newpage
\begin{table}
\caption{Orbital elements and stellar parameters of the stars in MWC~314 obtained from the 
simultaneous best fit to the RV curve and ASAS $V$-magnitude curve with the 
{\sc Phoebe} code; $a$=$a_{1}+a_{2}$ is the maximum distance between both stars.}             
\label{table:2}      
\centering                          
\begin{tabular}{l r }        
\hline\hline                 
Element & Value  \\    
\hline                        
   $P$ [d]              & 60.799977$\pm$0.000014   \\      
   $T$ (HJD)            &  2454959.76$\pm$0.56       \\
   $e$                  & 0.235$\pm$0.003      \\
   $\omega$ [deg]       & 289$\pm$1      \\
   $\gamma$ [$\rm km\,s^{-1}$]   & 28.44$\pm$0.17    \\
   $a$  [$\rm R_{\odot}$]    & 262.58$\pm$19.52 \\
   $i$ [deg]                     & 72.79$\pm$13.05  \\
\hline
   Primary star & \\
   $T_{\rm eff}$ [K]  & 18000 \\
   log $g$ [dex]      & 2.26 \\
   $m_{1}$ [$\rm M_{\odot}$] & 39.66 \\
   $R_{1}$ [$\rm R_{\odot}$] & 86.80  \\
   $K_{1}$ [$\rm km\,s^{-1}$] & 84.5   \\
\hline
   Secondary star & \\
   $T_{\rm eff}$ [K] & 6227 \\  
   log $g$ [dex]     & 3.55 \\
   $m_{2}$ [$\rm M_{\odot}$] & 26.26 \\
   $R_{2}$ [$\rm R_{\odot}$] & 20.41 \\
\hline                                   
\end{tabular}
\end{table}

\newpage
\newpage
\begin{table*}
\caption{Broadband space-based IR flux density observations of MWC~314 and P Cygni from
five missions.}             
\label{table:3}      
\centering          
\begin{tabular}{l l r r l r l l}    
\hline\hline      
  & &  & \multicolumn{2}{c}{MWC~314}{} & \multicolumn{2}{c}{P Cygni}{} \\ 
Mission       & Band      & $\lambda_{\rm eff}$ & \multicolumn{2}{c}{$F_{\nu}$}{$\sigma(F_{\nu})$} &
\multicolumn{3}{c}{$F_{\nu}$}{$\sigma(F_{\nu})$}{ Ref.} \\
      &       & [$\mu$m]    &  \multicolumn{5}{c}{$\times$10$^{-23}$ [erg s$^{-1}$ cm$^{-2}$ Hz$^{-1}$]}{}  \\
\hline                    
   IRAS & F12  & 11.0357  & 2.2800 & 0.1596 & 6.9800  & 0.4188 & 1 \\  
        & F25  & 23.0724  & 1.3700 & 0.1507 & 0.1624  & 4.0600 & 1 \\ 
        & F60  & 58.1904  &  -     &  -     & 0.3850  & 2.7500 & 1 \\
  AKARI & S9W  & 8.85203  & 1.9930 & 0.0276 & 9.3020  & 0.0930 & 2 \\
        & L18W & 18.9186  & 1.2380 & 0.0432 & 0.0673  & 4.2900 & 2 \\
   MSX  & B1   & 4.29352  & -      & -      & 16.7399 & 2.4105 & 3 \\
        & A    & 8.48032  & 2.2140 & 0.0907 & 8.7429  & 0.3584 & 3 \\
        & C    & 12.1516  & 1.8070 & 0.1084 & 6.1250  & 0.3123 & 3 \\
        & D    & 14.6783  & 1.7620 & 0.1145 & 4.6300  & 0.2824 & 3 \\
        & E    & 21.5008  & -      & -      & 3.3230  & 0.2126 & 3 \\
   IRAC & 36   & 3.54391  & 1.3075 & 0.1459 & -       & -      & 4 \\

        & 45   & 4.48406  & 2.5730 & 0.1630 & -       & -      & 4 \\
        & 58   & 5.71638  & 2.6970 & 0.0678 & -       & -      & 4 \\
        & 80   & 7.82825  & 1.9870 & 0.0668 & -       & -      & 4 \\
   WISE & W1   & 3.34597  & 4.1277 & 0.2813 & 20.5167 & 2.9667 & 5 \\
        & W2   & 4.59522  & 4.7897 & 0.2470 & 26.0801 & 0.2608 & 5 \\
        & W3   & 11.5485  & 1.3569 & 0.0150 & 5.7990  & 0.0580 & 5 \\
        & W4   & 22.0789  & 1.0545 & 0.0223 & 3.8570  & 0.0386 & 5 \\
\hline
\end{tabular}
\tablebib{
(1)~\citet{1988iras....7.....H}; (2) \citet{2010AA...514A...1I};
(3) \citet{2003yCat.5114....0E}; (4) \citet{Glimpse...2008};
(5) \citet{2012yCat.2311....0C}.
}
\end{table*}

\newpage
\newpage
\begin{table}
\caption{Emission lines in the high-resolution HERMES spectrum of 
MWC~314. FWHM values are provided for single- and double-peaked 
lines. Columns 4 and 5 list heliocentric radial velocities of the blue and 
red emission maxima of double-peaked lines. $\lambda_{0}$ is the laboratory rest 
wavelength of the line in air.}             
\label{table:4}      
\centering                          
\begin{tabular}{l l r r r}        
\hline\hline                 
 Line & $\lambda_{0}$ & FWHM & $V_{\rm blue-peak}$ & $V_{\rm red-peak}$  \\    
  & [\AA] & [$\rm km\,s^{-1}$] & [$\rm km\,s^{-1}$] & [$\rm km\,s^{-1}$] \\ 
\hline                        
H$\alpha$ & 6562.797 & 188 & - & - \\
H$\beta$ & 4861.323 & 135 & - & -  \\ 
H$\gamma$ & 4340.462 & 132 & - & - \\
H$\delta$ & 4101.734 & 117 & - & - \\
H Pa14 & 8598.396 & 134 & $-$7 & $+$58  \\
He~{\sc i} & 4471.479 & 91 & - & - \\
He~{\sc i} & 5875.790 & 87 & - & - \\
He~{\sc i} & 6678.149 & 82 & - & - \\
O~{\sc i} triplet & 7772-5 & - & - & - \\
O~{\sc i} & 8446.359 & 111 & $-$5 & $+$56  \\
Fe~{\sc ii} & 6416.905 & 123 & $-$6 & $+$53  \\
Fe~{\sc ii} & 6432.654 & 118 & $-$7 & $+$62  \\
Fe~{\sc ii} & 6456.376 & 131 & $-$5 & $+$54  \\
Si~{\sc ii} & 6347.10  & 149  &$-$9 & $+$85  \\
Si~{\sc ii} & 6717.04 & 143 & $-$8 & $+$81  \\
$[$N~{\sc ii}$]$ & 5754.59 & 49 & $+$18 & $+$39  \\
$[$N~{\sc ii}$]$ & 6548.05 & 45 & - & - \\
$[$N~{\sc ii}$]$ & 6583.45 & 47 & - & - \\
$[$Fe~{\sc ii}$]$ & 7155.16 & 93 & $-$8  & $+$59  \\
$[$Fe~{\sc ii}$]$ & 8616.95 & 96 & $-$10  & $+$58  \\
$[$Ca~{\sc ii}$]$ blends $\rm O_{2}$ & 7291.47 & 46 & - & - \\
$[$Ca~{\sc ii}$]$ blends $\rm O_{2}$ & 7323.89 & 40 & - & - \\
\hline                                   
\end{tabular}
\end{table}

\onecolumn
\begin{sidewaysfigure}
\vspace*{18cm}
\centering
\includegraphics[width=19cm,angle=0]{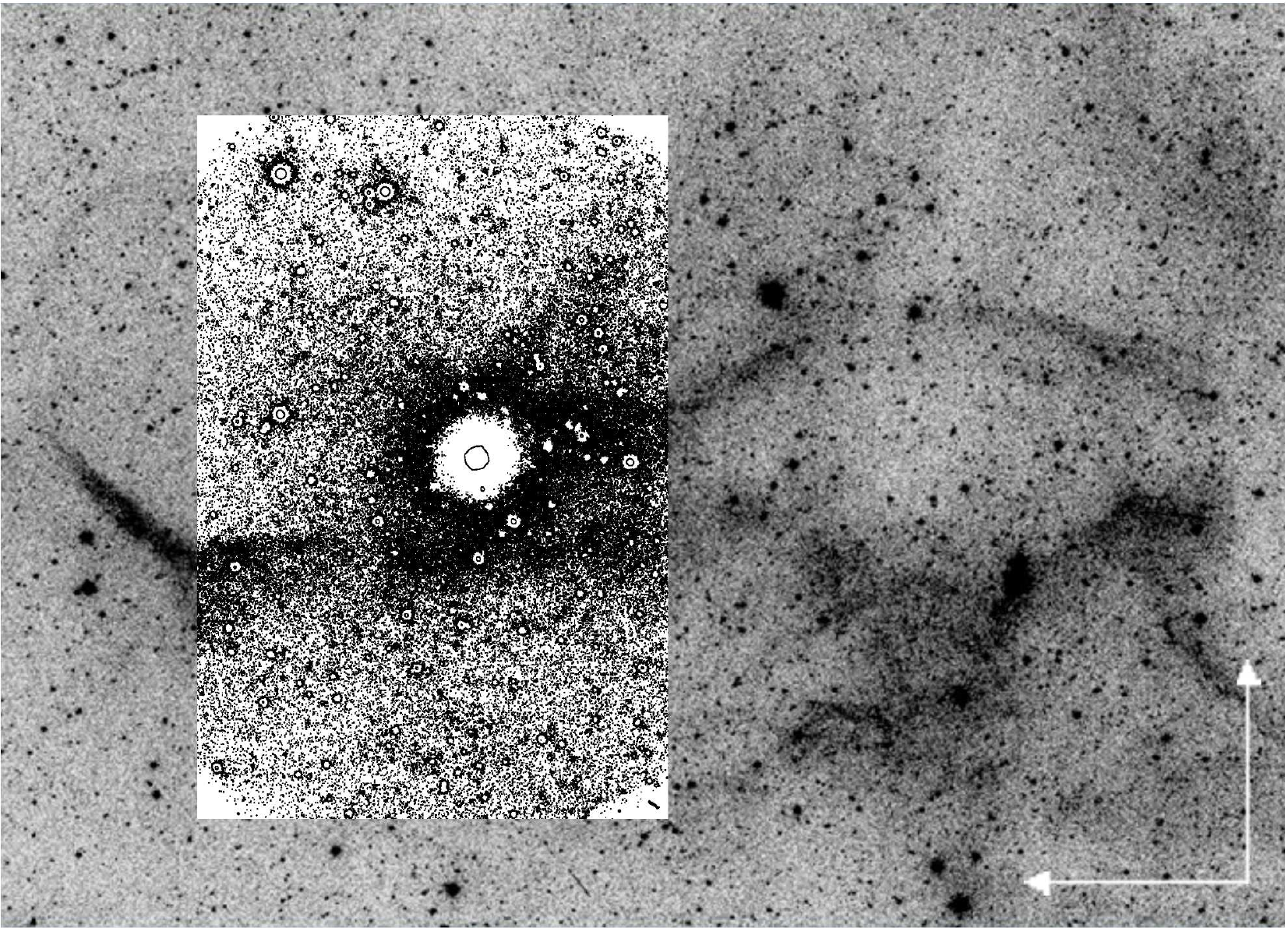}
\caption{Narrow-band H$\alpha$ image of the extended bipolar nebula of MWC~314 ({\em adapted from
\citet{2008A&A...477..193M}}). { The image is 12.5\arcmin vertically}. 
 The inset shows the Mercator-MEROPE intensity contour image with faint H$\alpha$
nebulosity surrounding
the central star. North is up, and east to the left.
\label{fig_1}}
\end{sidewaysfigure}

\begin{figure}
\centering
\includegraphics[width=11.3cm,angle=-90]{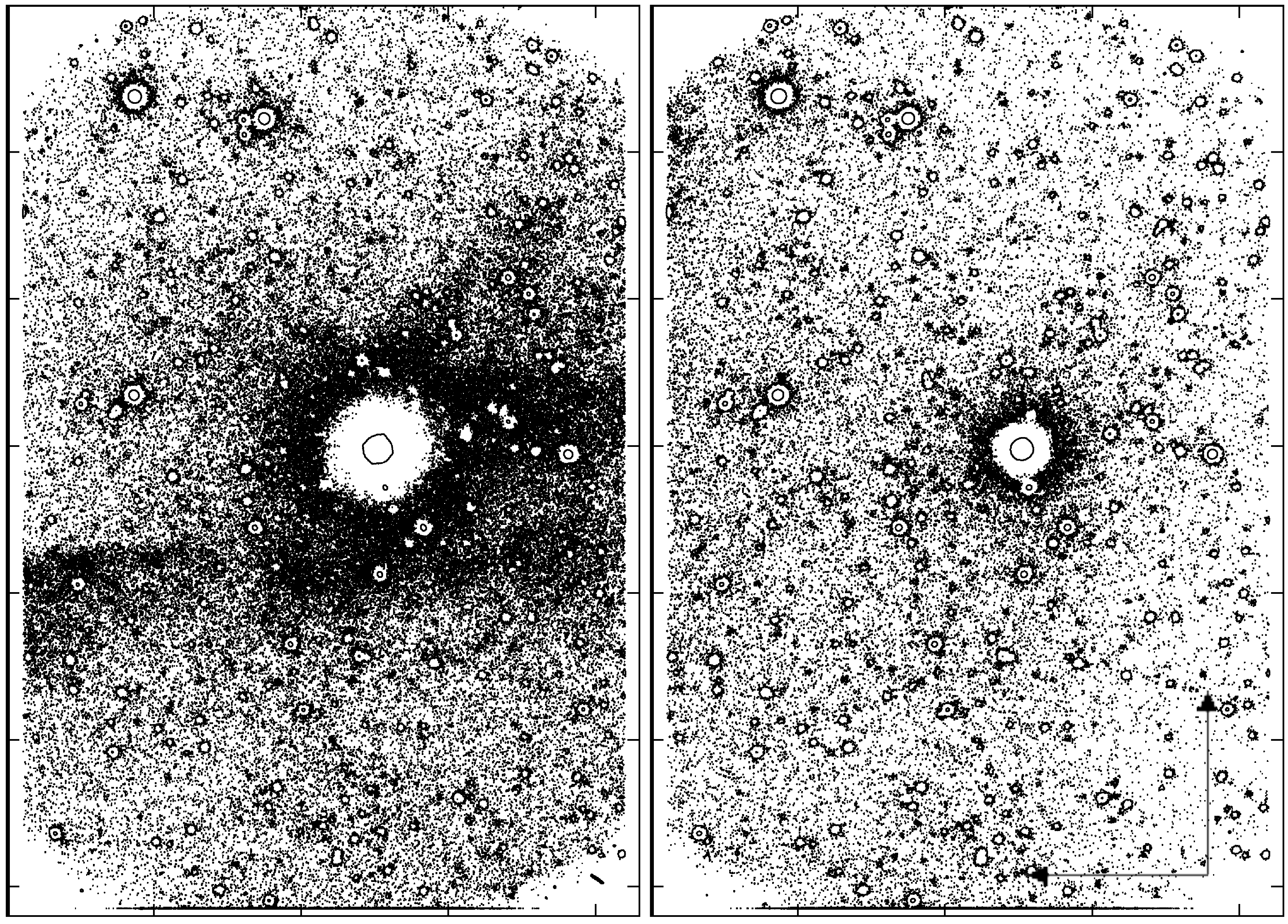}
\caption{ Comparison of the (intensity contour) narrow-band H$\alpha$ emission image ({\it left-hand panel}) and the
continuum image ({\it right-hand panel}) of MWC~314 observed in March 2011 with Mercator-MEROPE. Line emission is absent
in the continuum image where extended east-west filament structures are observed in the H$\alpha$ emission image ({\it
see Sect. 3}). { Both panels are 6.9\arcmin by 10\arcmin}. North is up, and east to the left.
\label{fig_2}}
\end{figure}

\begin{figure}
\centering
\includegraphics[width=7.9cm,angle=-90]{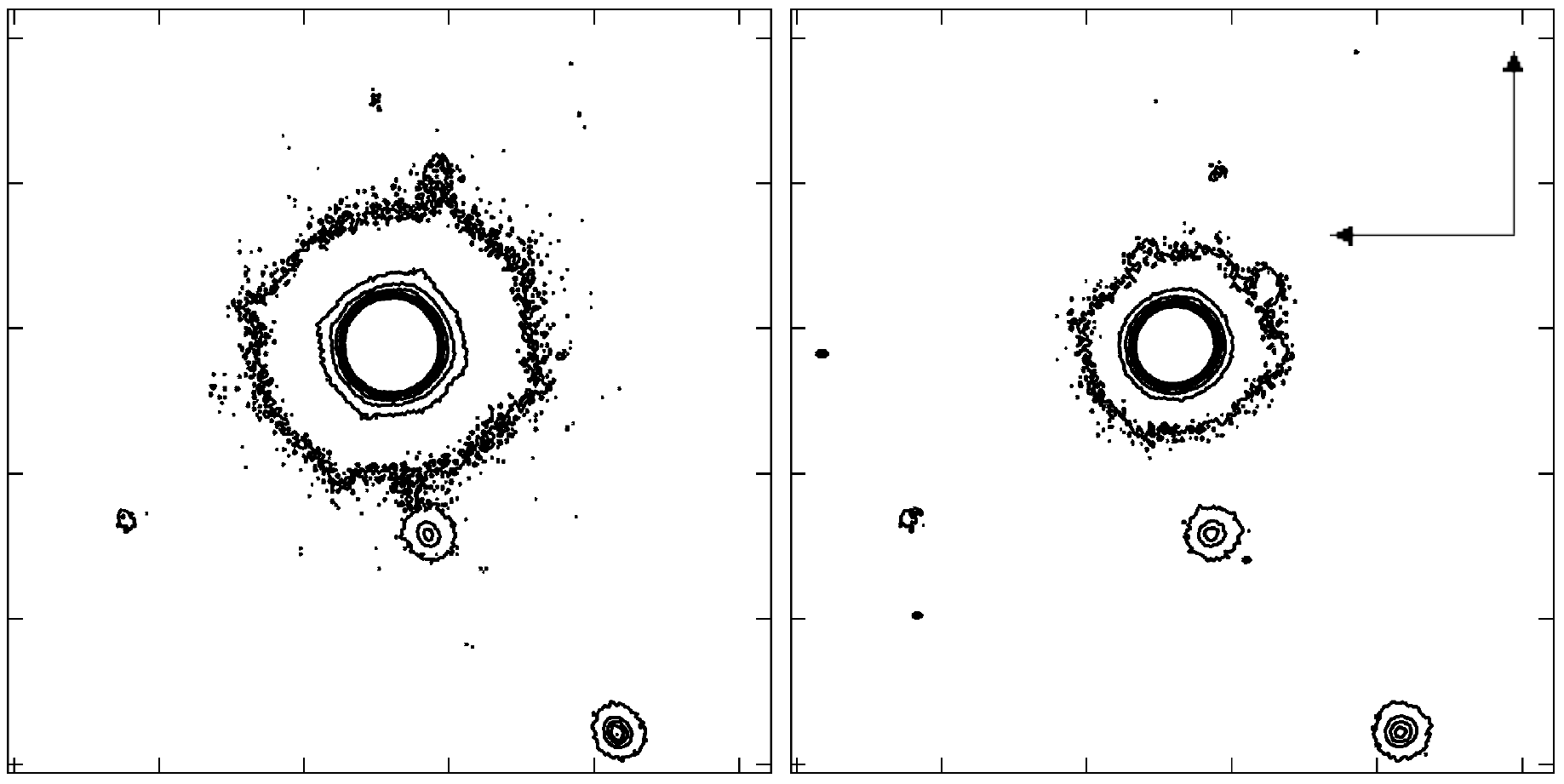}
\caption{ Comparison of 100\arcsec\, $\times$ 100\arcsec\, H$\alpha$ ({\em left-hand panel}) and continuum ({\em
right-hand panel}) intensity contour images of MWC~314. The outer intensity contour signals an extended 
H$\alpha$ envelope that is almost circular symmetric within 20\arcsec\, of the central star.    
\label{fig_3}}
\end{figure}

\begin{sidewaysfigure}
\vspace*{18cm}
\centering
\includegraphics[width=20cm,angle=0]{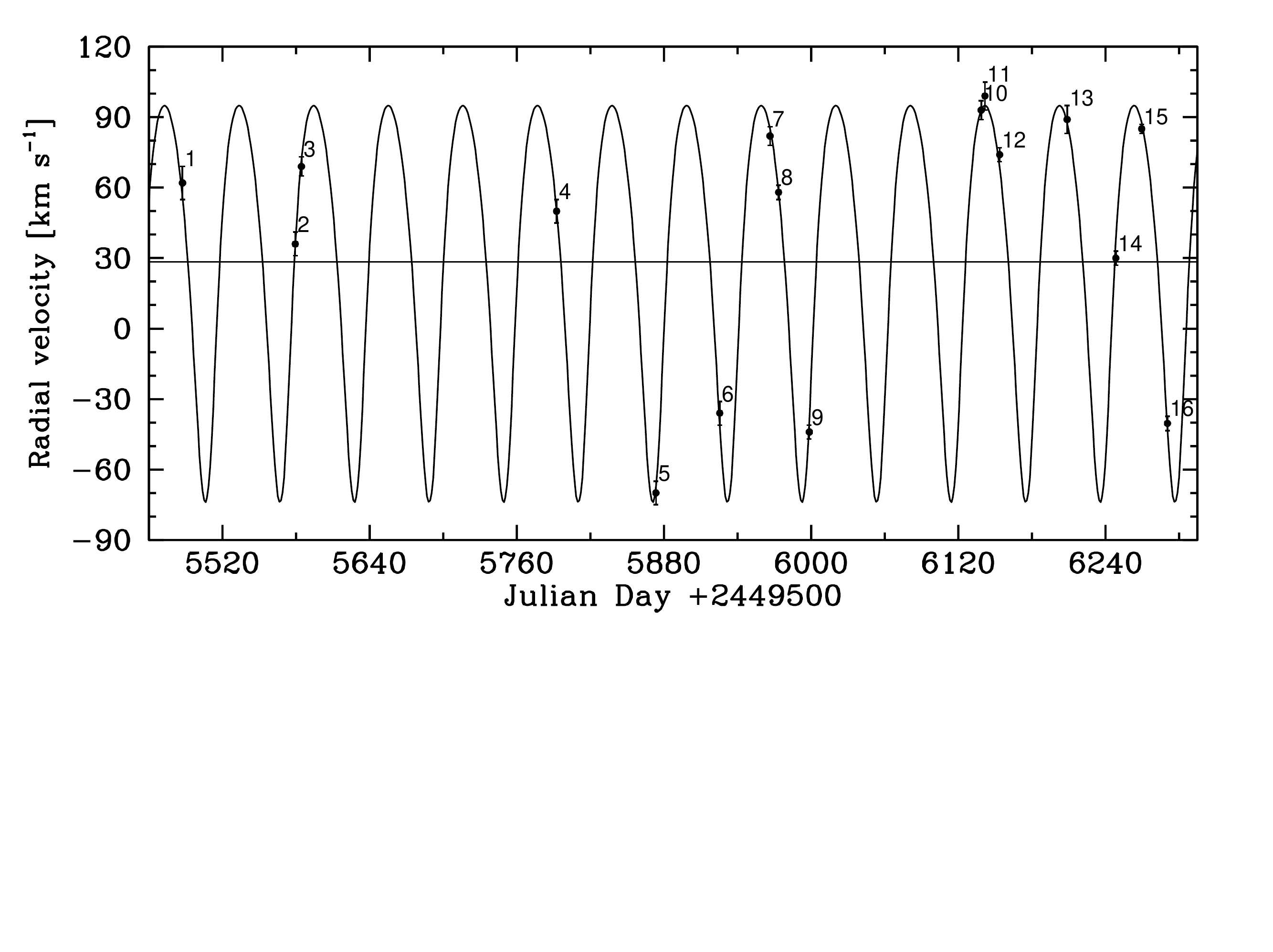}
\caption{ Best fit orbital solution to 16 radial velocity measurements in MWC~314. 
The RV observation dates during 22 m are given in Table 1. The solid horizontal line is 
drawn at the $\gamma$-velocity for the best fit orbital solution.
\label{fig_4}}
\end{sidewaysfigure}

\begin{figure}
\begin{minipage}{8cm}
\centering
\includegraphics[width=16cm,angle=0]{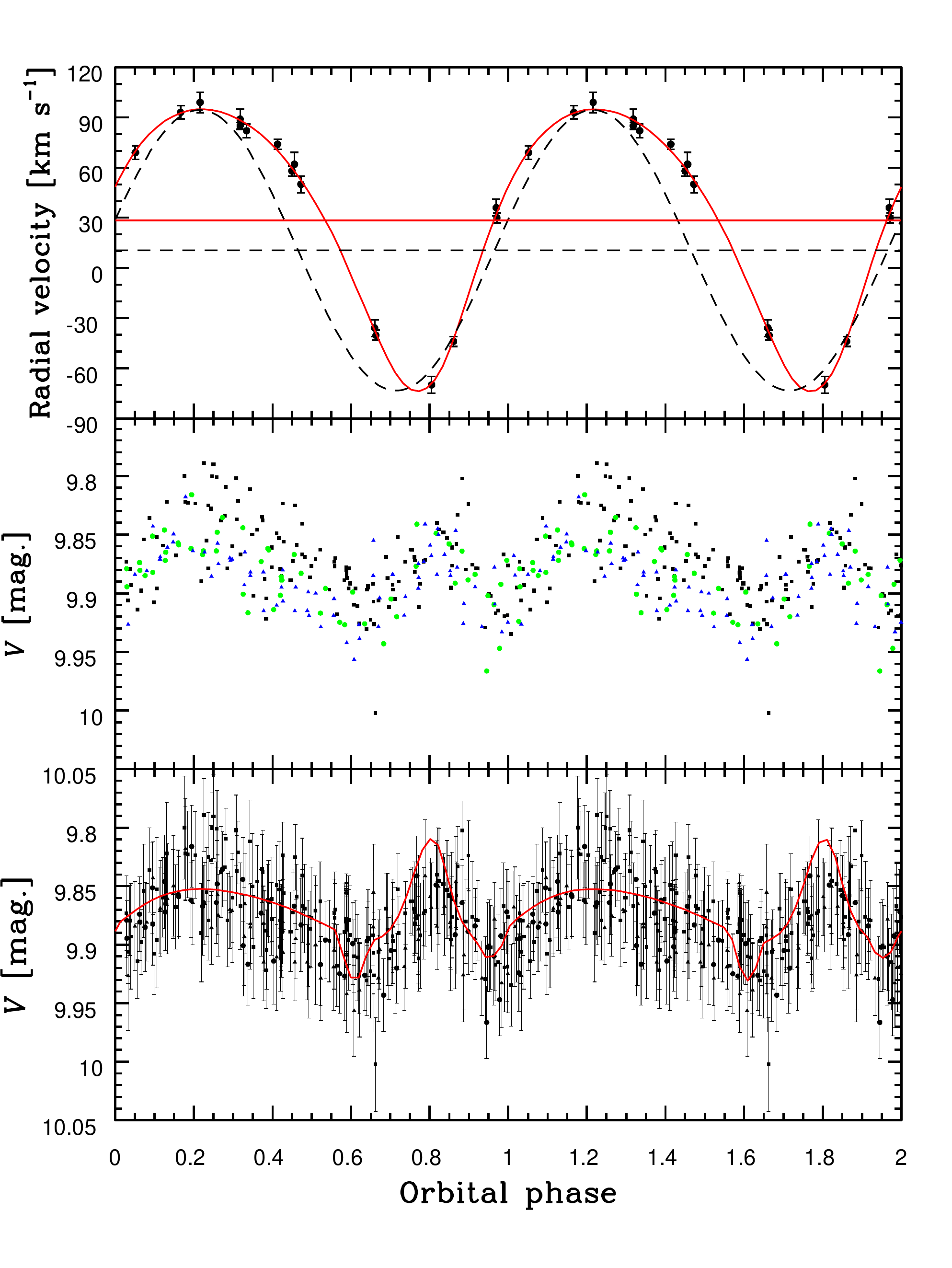}
\end{minipage}
\caption{ {\em Top panel:} Best fit radial velocity curve ({\em curved solid line}) 
as a function of orbital phase. The orbital radial velocity curve signals an
eccentric orbit of $e$=0.23 and $P$=60.8~d. A circular orbit ($e$=0) does not fit 
({\em curved dashed line}) the observations ({\em solid dots}).
The $\gamma$-velocity of $+$28.4~$\rm km\,s^{-1}$ ({\em solid horizontal line}) is
larger than 12.3~$\rm km\,s^{-1}$  ({\em dashed horizontal line}) when $e$ is set equal to 0.
{\em Middle panel:} The ASAS $V$-brightness curve shows two minima signaling an eclipsing binary system.
{ {\em Bottom panel:} The symbols mark the $V$-brightness values
in the middle panel with errorbars. The model best fit is 
drawn with a smooth curve ({\em see Sect. 4.1})}.
\label{fig_5}}
\end{figure}

\begin{sidewaysfigure}
\vspace*{18cm}
\centering
\includegraphics[width=22cm,height=16cm,angle=0]{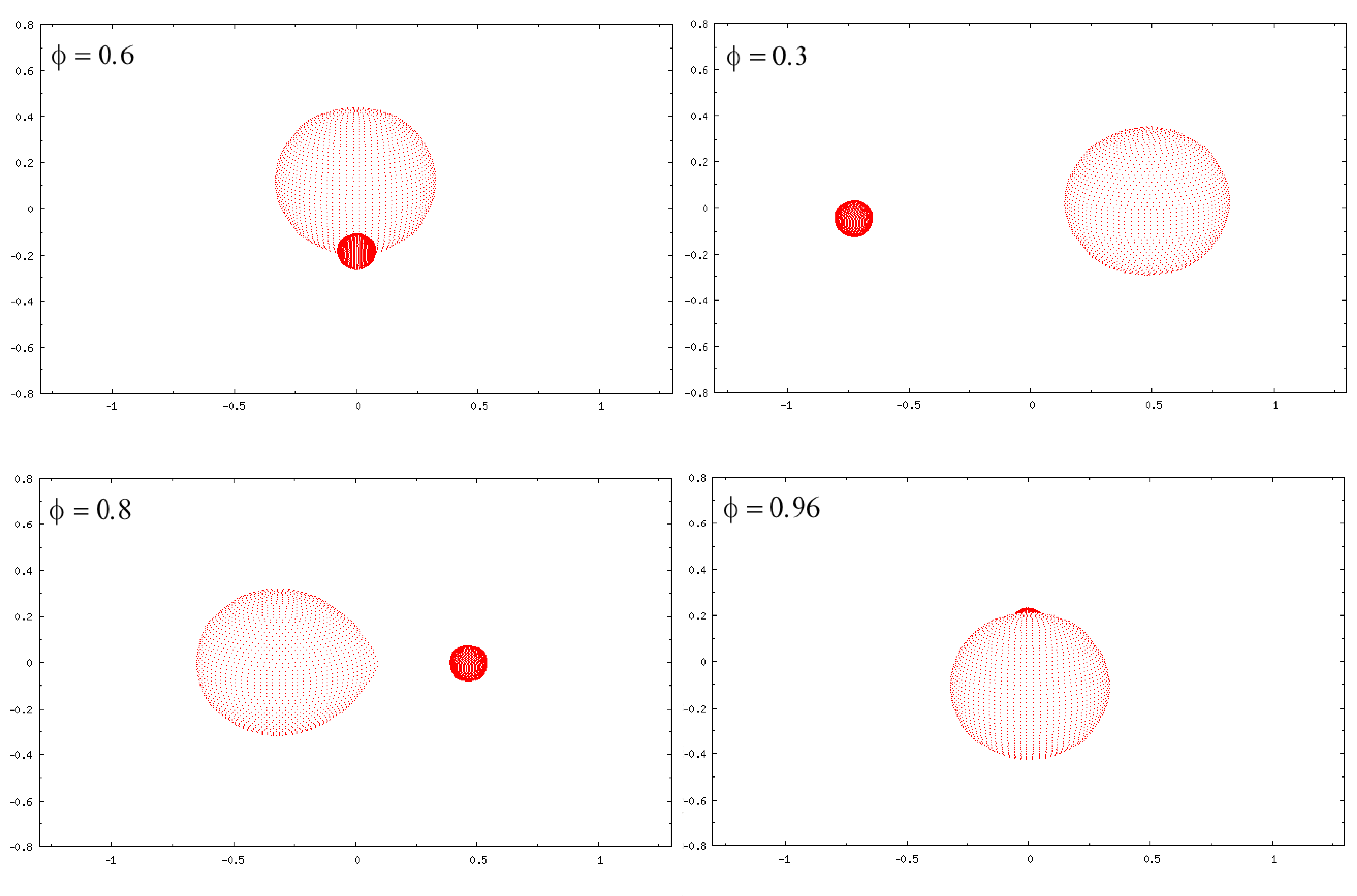}
\caption{Schematic representation of four orbital phases ({\em counter-clockwise}) in MWC 314 computed with the {\sc
Phoebe} code. 
\label{fig_6}}
\end{sidewaysfigure}

\begin{figure}
\centering
\vspace*{-5cm}
\includegraphics[width=20cm,angle=0]{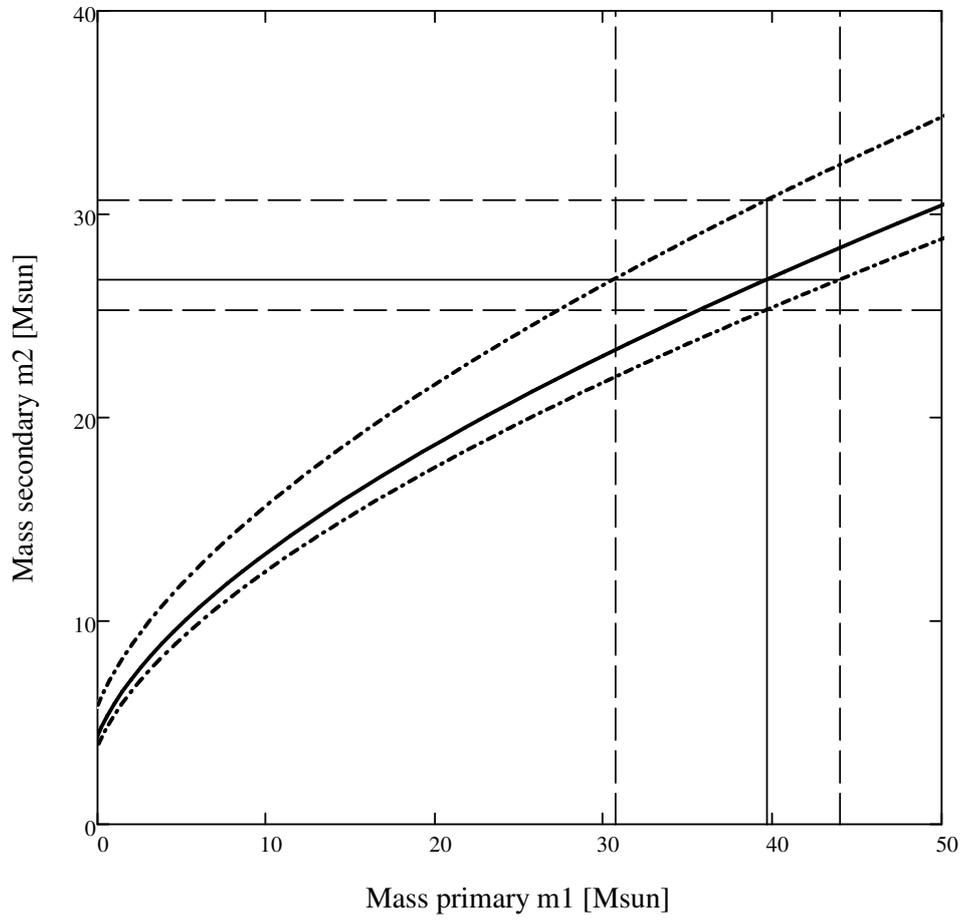}
\caption{The mass function of MWC~314 computed for $i$=72.8$^{\circ}$ ({\em thick drawn solid line}), 
which is also shown for inclination angles of $i-13^{\circ}$ ({\em top dash-dotted line}), and $i+13^{\circ}$
({\em bottom dash-dotted line}). The solid drawn vertical and horizontal lines mark the 
values of $m_{1}$=39.66~$\rm M_{\odot}$ and $m_{2}$=26.26~$\rm M_{\odot}$ computed with {\sc Phoebe}, 
while the dashed drawn lines 
mark the boundaries of 30.7~$\rm M_{\odot}$ $\leq$ $m_{1}$ $\leq$ 44.2~$\rm M_{\odot}$ and 25.3~$\rm M_{\odot}$ $\leq$
$m_{2}$ $\leq$ 30.7~$\rm M_{\odot}$ corresponding to inclination angles ranging 
from $i-13^{\circ}$ to $i+13^{\circ}$. 
\label{fig_7}}
\end{figure}

\begin{figure}
\centering
\includegraphics[width=18cm,angle=0]{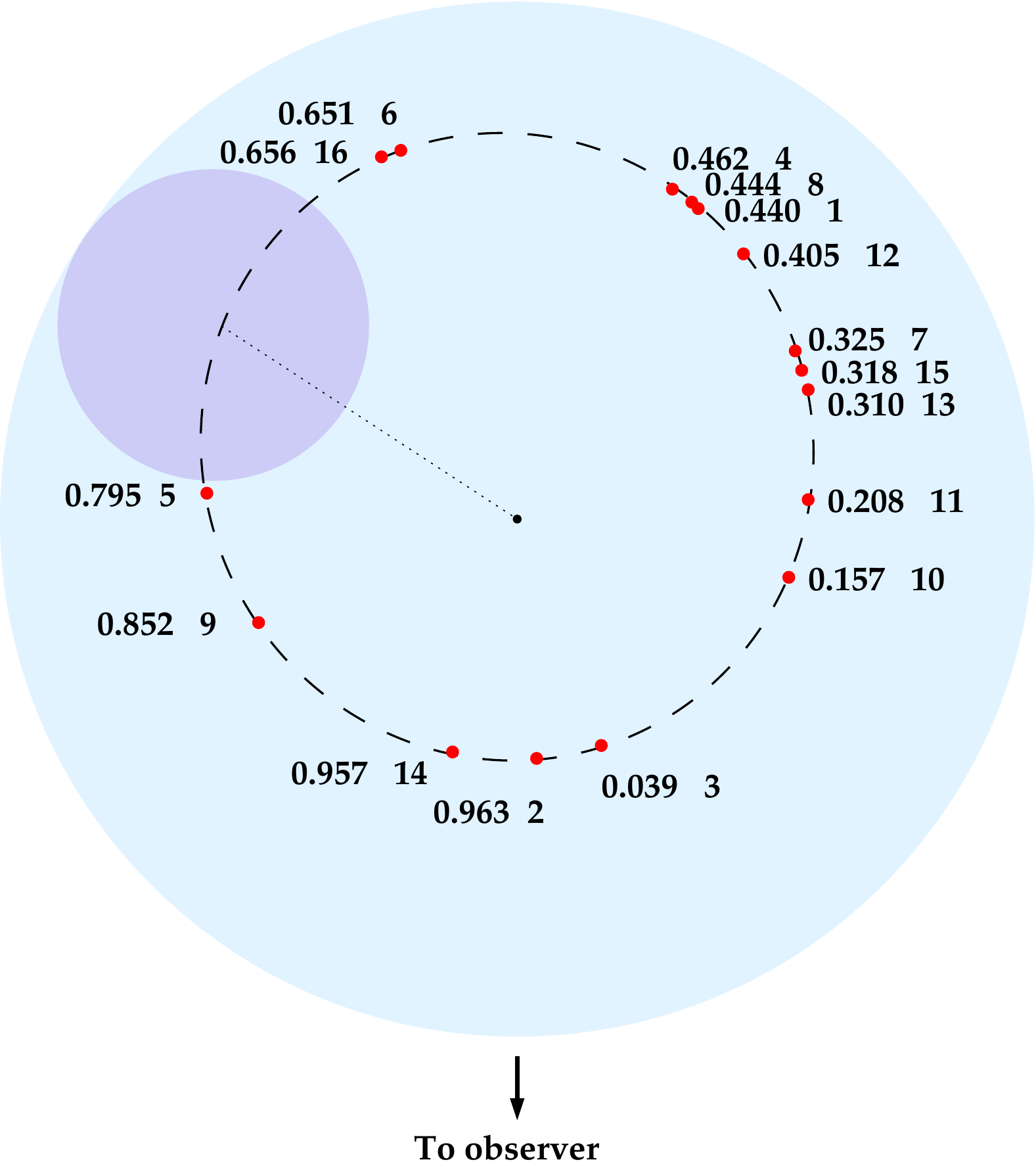}
\caption{ The orbital path ({\em dashed line}) of the primary of MWC~314 
shown around the centre of gravity of the binary ({\em central black dot})
and marked for the 16 observed orbital phases ({\em solid dots}). 
The labels provide the $\phi$-values with spectrum numbers in Table 1. 
The shape and orientation of the orbit are set by $e$ and $\omega$. 
The disc to the left shows the orbital position of the primary star at $\phi$$\sim$0.7. 
The large outer disc represents the circumbinary wind region extending 
further away from the gravity centre.  
\label{fig_8}}
\end{figure}

\begin{sidewaysfigure}
\vspace*{18cm}
\centering
\includegraphics[width=22cm,angle=0]{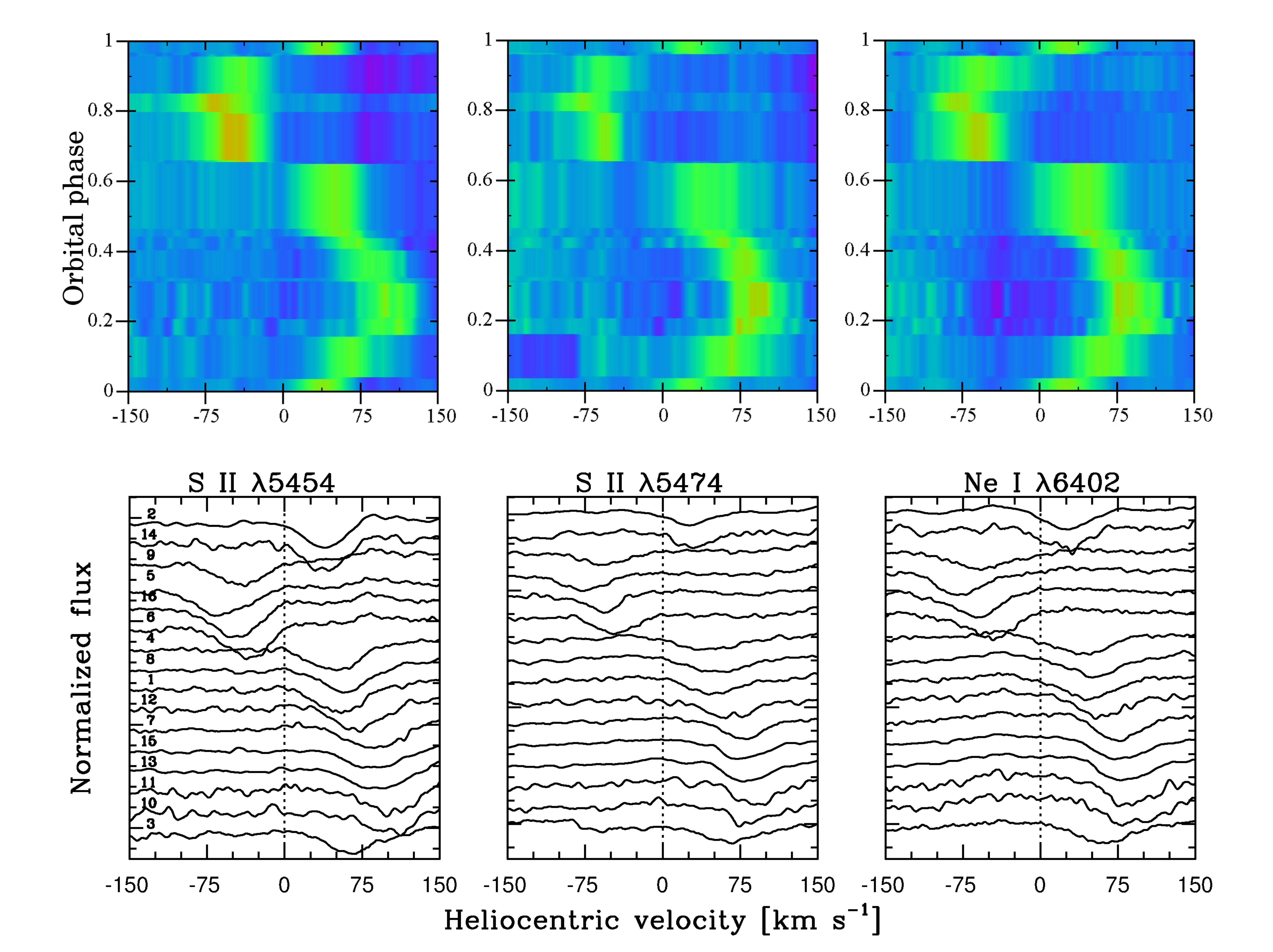}
\caption{ The top panels show dynamic spectra of three absorption lines of MWC~314 
in the heliocentric velocity scale. Blue colours
correspond to normalised stellar continuum flux levels. The bottom panels show the 
corresponding normalised line profiles shifted upwards with orbital phase 
(spectrum numbers are labelled in the left-hand panel). 
The (S-shaped) Doppler shifts of the lines is due to the orbital motion of 
the primary star. We observe significant enhancements in the absorption 
line depth at orbital phases with large Doppler blue-shift of $\phi$=0.65-0.85. 
\label{fig_9}}
\end{sidewaysfigure}

\begin{sidewaysfigure}
\vspace*{18cm}
\centering
\includegraphics[width=16cm,angle=-90]{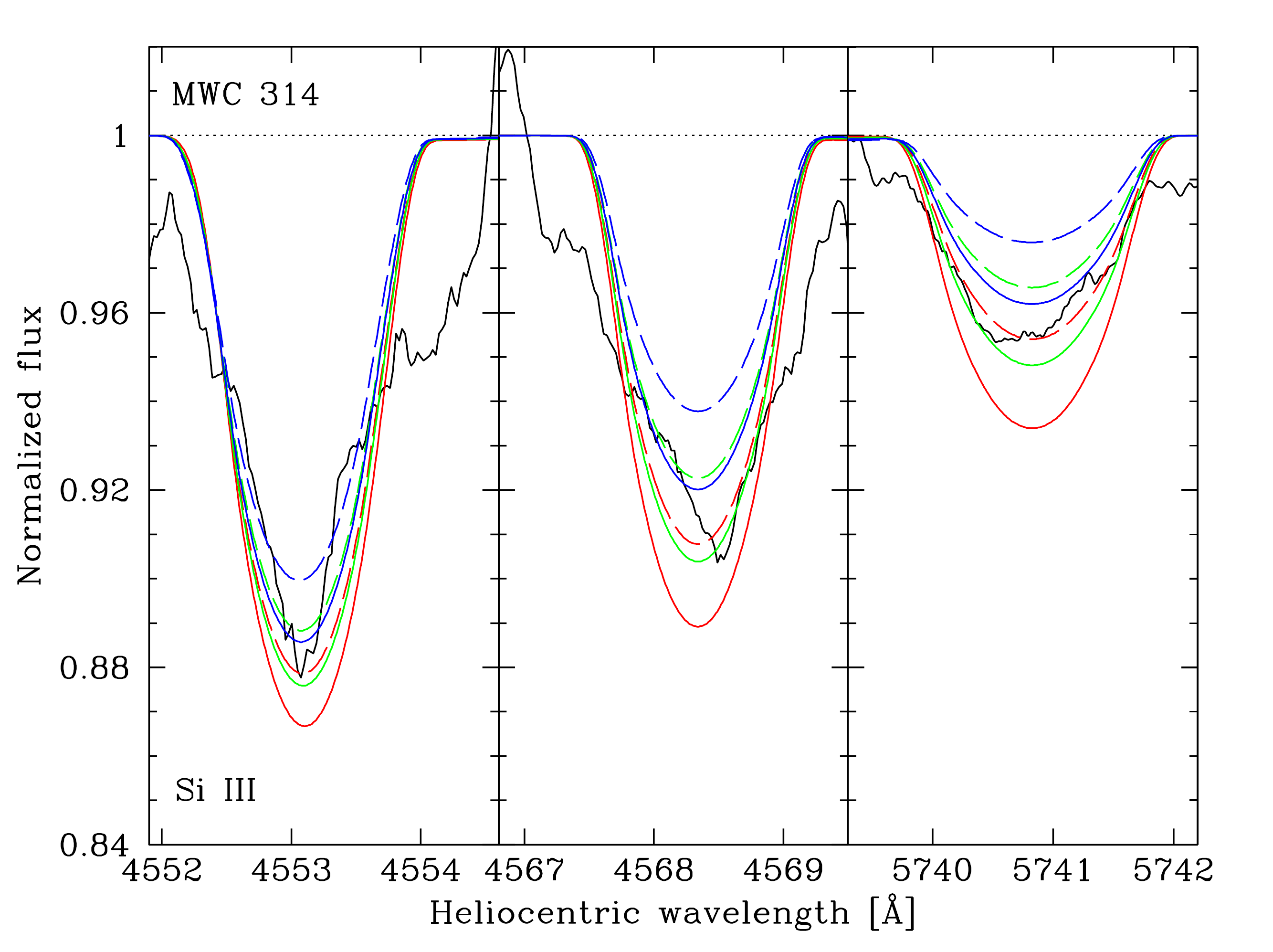}
\caption{Three Si~{\sc iii} absorption lines observed in MWC 314 ({\em black solid lines}) are overplotted 
with theoretical line profiles computed in non-LTE with Multi. The line transfer calculations use
atmosphere models of $T_{\rm eff}$=17 kK ({\em dashed drawn lines}), 18 kK ({\em solid drawn lines}), and 19 kK
({\em dash-dotted lines}). Boldly drawn lines are computed with log $g$=2.5, and thin drawn lines for log
$g$=3.0. Atmosphere models with $T_{\rm eff}$ $>$ 19 kK yield Si~{\sc iii}  line equivalent widths incompatible 
with observed values. The detailed profile fits require $V_{\rm rot}$sin$i$$\simeq$50~$\rm km\,s^{-1}$.
\label{fig_10}}
\end{sidewaysfigure}

\begin{sidewaysfigure}
\vspace*{18cm}
\centering
\includegraphics[width=16cm,angle=-90]{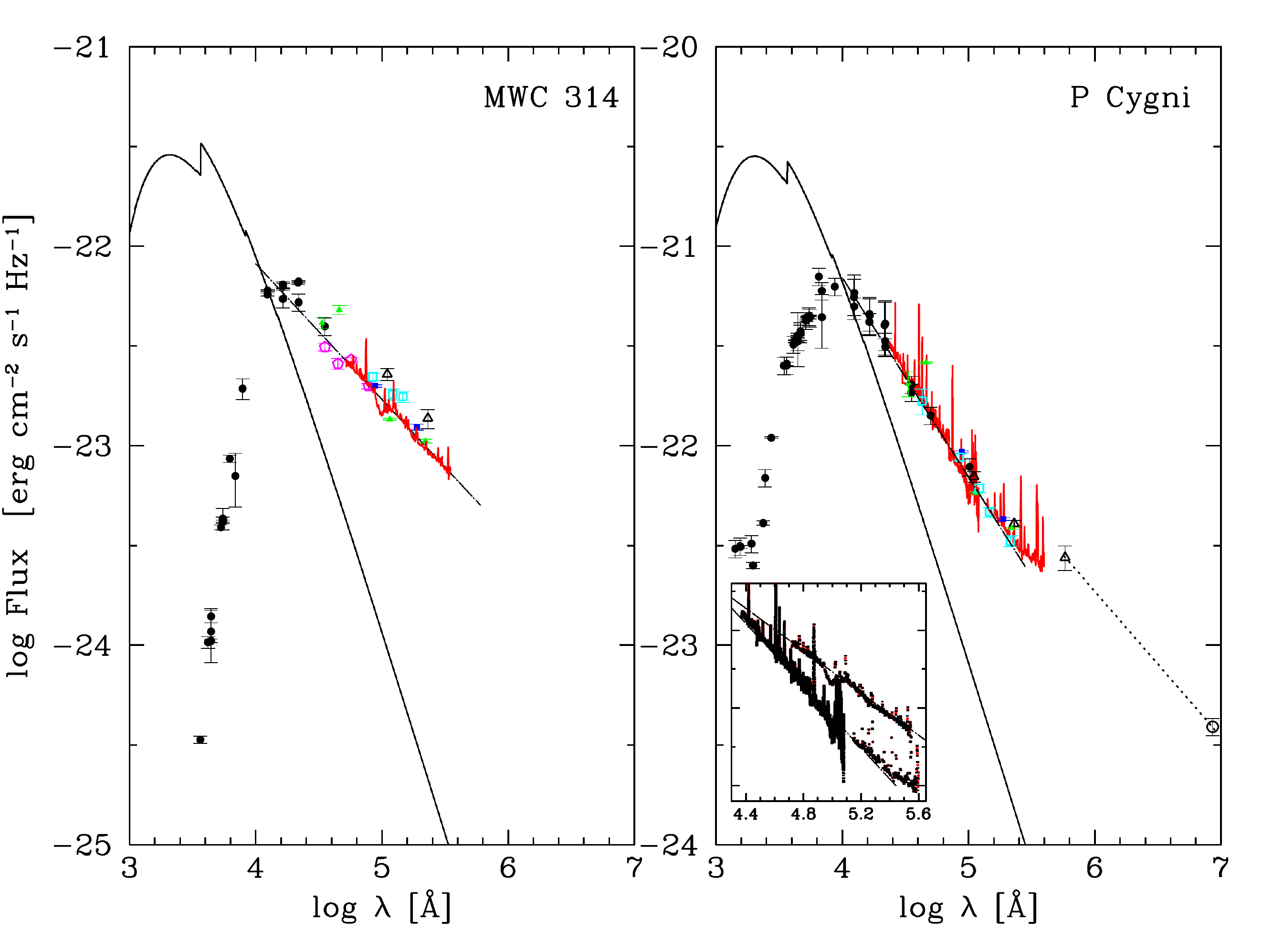}
\caption{A comparison of the SED of MWC 314 and LBV P Cygni reveals very 
similar slopes ({\em dash-dotted and dotted lines}) of the
near-IR and IR flux density in these early B-type Ia supergiants.
Broadband photometric observations ({\em solid black dots}), including 
from five space missions ({\em open \& solid triangles,
open \& solid boxes, and open pentagonal symbols}), are compared to 
{\it Spitzer} and ISO spectra ({\em solid drawn line with spikes, red in online 
version}). { The sub-panel in the right-hand
panel compares the slopes of the {\it Spitzer} spectrum of MWC 314 ({\em upper 
curve}) with the ISO and {\it Spitzer} spectra of P Cyg ({\em see text})}. 
The observed IR flux excesses compared to the photospheric continuum 
fluxes ({\em smoothly drawn solid lines}) are caused by
free-free emission in an expanding stellar wind.
\label{fig_11}}
\end{sidewaysfigure}

\begin{sidewaysfigure}
\vspace*{18cm}
\centering
\includegraphics[width=22cm,angle=0]{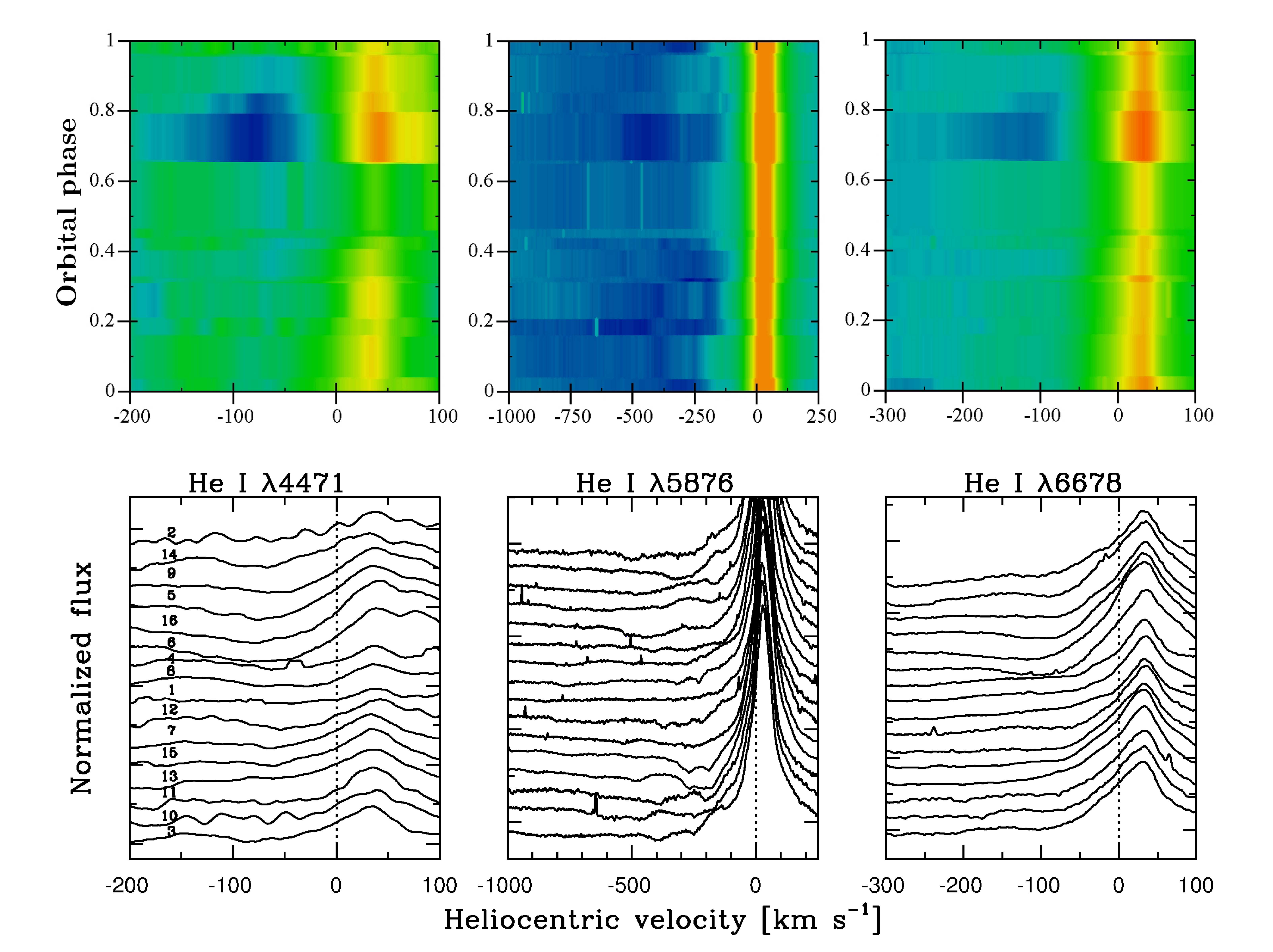}
\caption{Three strong He~{\sc i} lines reveal remarkable 
variability of shape with orbital phase. They show single-peaked 
emission around the systemic velocity. The blue-shifted and variable 
absorption is strongest at $\phi$=0.65-0.85. The lines are 
variable P Cyg-type line profiles. The peculiar P Cyg line shape is 
always observed in He~{\sc i} $\lambda$5876.
\label{fig_12}}
\end{sidewaysfigure}

\begin{figure}
\centering
\includegraphics[width=16cm,angle=0]{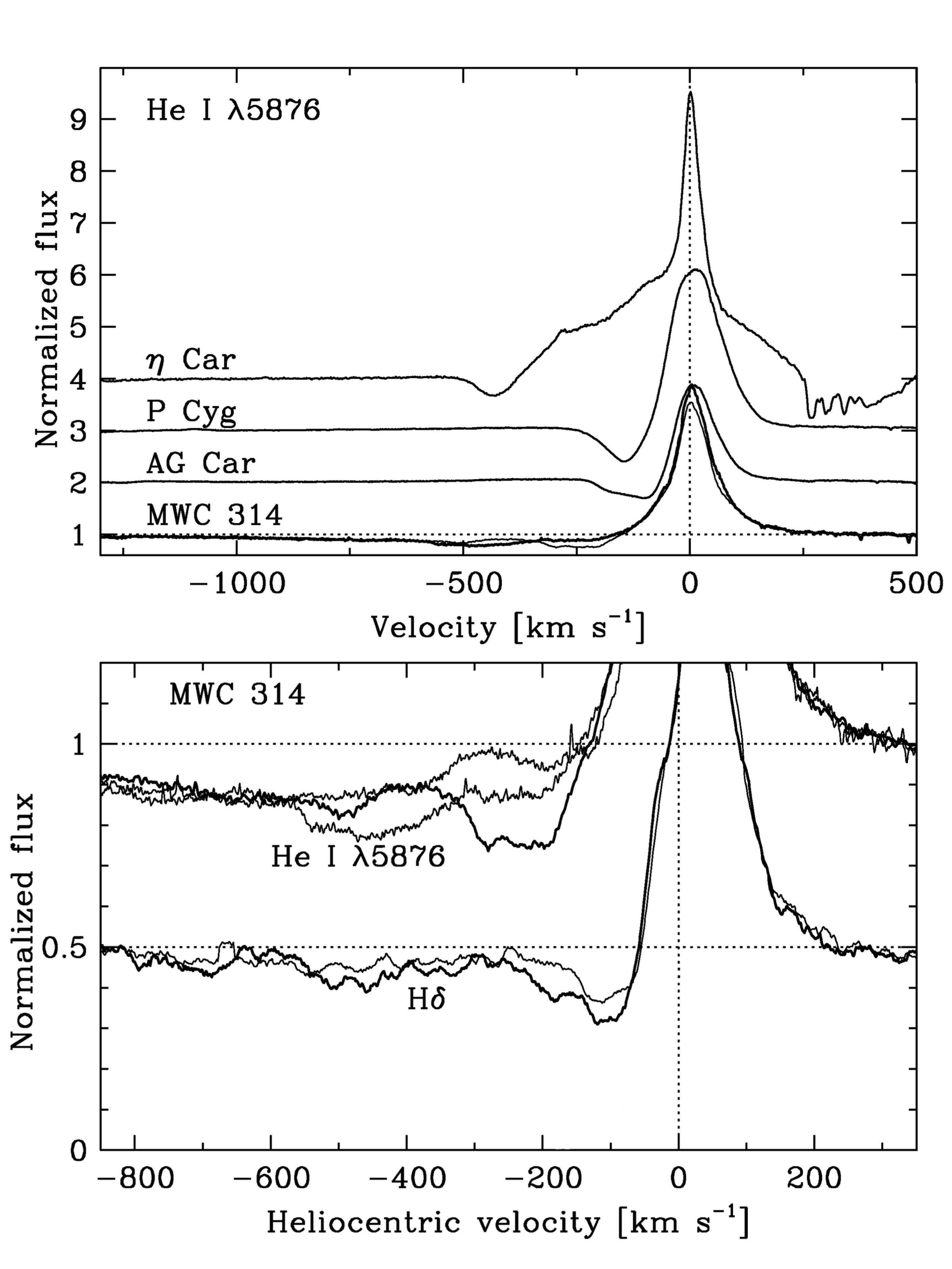}
\caption{ {\em Top panel:} The He~{\sc i} $\lambda$5876
lines in three LBVs, $\eta$ Car, P Cyg, and AG Car are compared 
to MWC~314. We observe blue-shifted wind absorption beyond 1000~$\rm km\,s^{-1}$
in MWC~314. {\em Bottom panel:} The blue-shifted absorption reveals
strong variability below 600~$\rm km\,s^{-1}$ ({\em solid thick and 
thin lines}), which we also observe in H$\delta$.
\label{fig_13}}
\end{figure}

\begin{figure}
\centering
\includegraphics[width=16cm,angle=0]{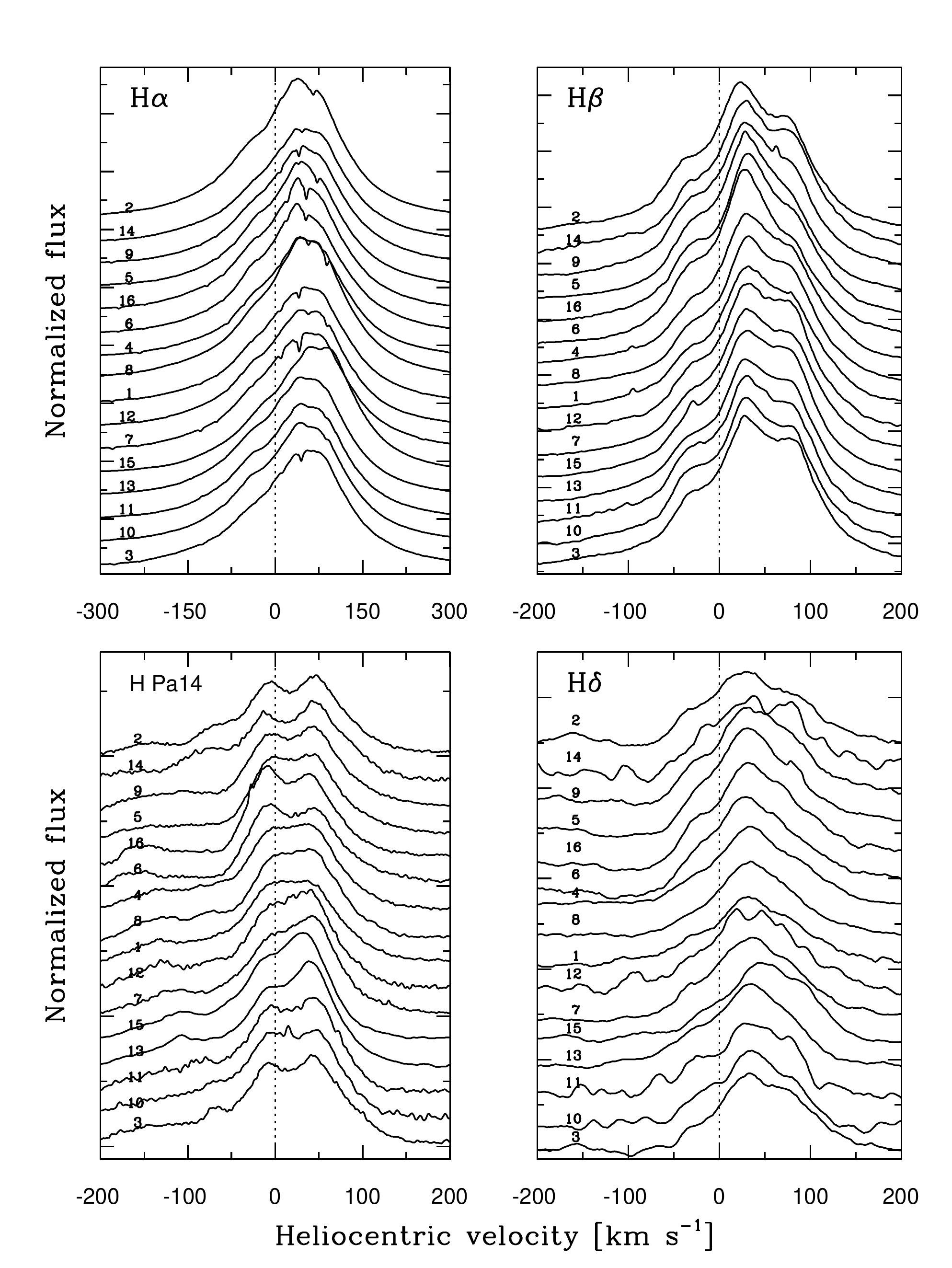}
\caption{ The normalised H$\alpha$, H$\beta$, H$\delta$, and H Pa 14 
lines are plotted with $\phi$ ({\em clockwise}). The line FWHM decreases towards the 
higher H Balmer series lines. The H$\alpha$ line wings extend 
beyond $\pm$300~$\rm km\,s^{-1}$. The higher Balmer series lines also 
show stronger absorption  in the violet 
line wing to $\sim$300~$\rm km\,s^{-1}$. The H$\delta$ and Pa14 lines ({\em bottom panels})  
show enhanced blue-shifted absorption around $\phi$=0.65-0.85, similar to 
the He~{\sc i} lines.    
 \label{fig_14}}
\end{figure}

\begin{sidewaysfigure}
\vspace*{18cm}
\centering
\includegraphics[width=20cm,angle=0]{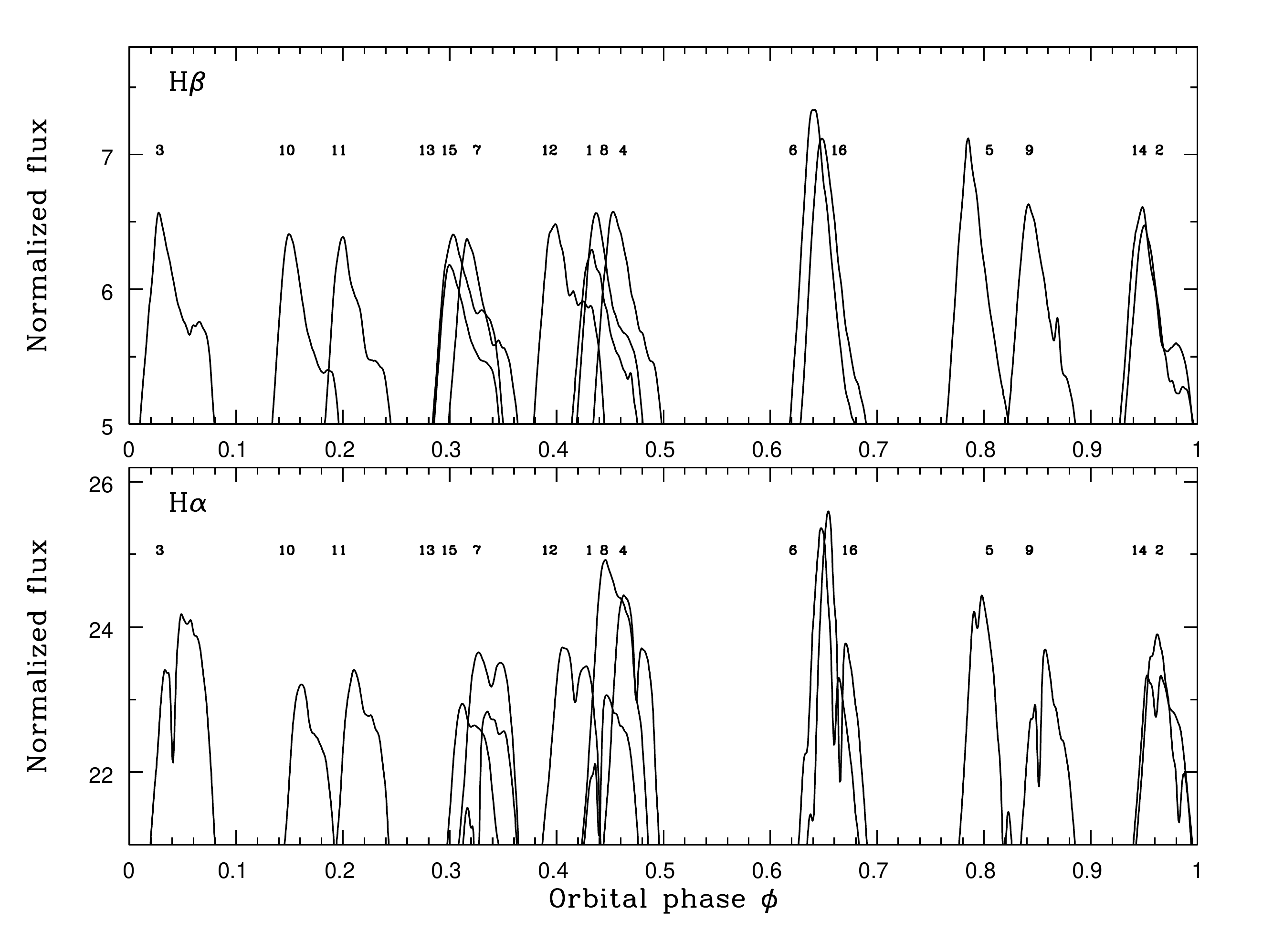}
\caption{ The H$\alpha$ ({\em bottom panel}) and H$\beta$ 
({\em top panel}) emission lines of MWC~314 show largest 
normalised flux maxima at $\phi$$\simeq$0.65 (spectrum Nos. 6 \& 16), 
and smallest flux maxima at $\phi$$\simeq$0.32 (Nos. 13, 15, \& 7). 
The absolute emission line fluxes determined with 
the $V$-brightness curve are however almost invariable, indicating 
that the extended circumbinary H$\alpha$ and H$\beta$ emission line formation 
regions are little influenced by the orbital motion.
\label{fig_15}}
\end{sidewaysfigure}

\begin{sidewaysfigure}
\vspace*{18cm}
\centering
\includegraphics[width=16cm,angle=-90]{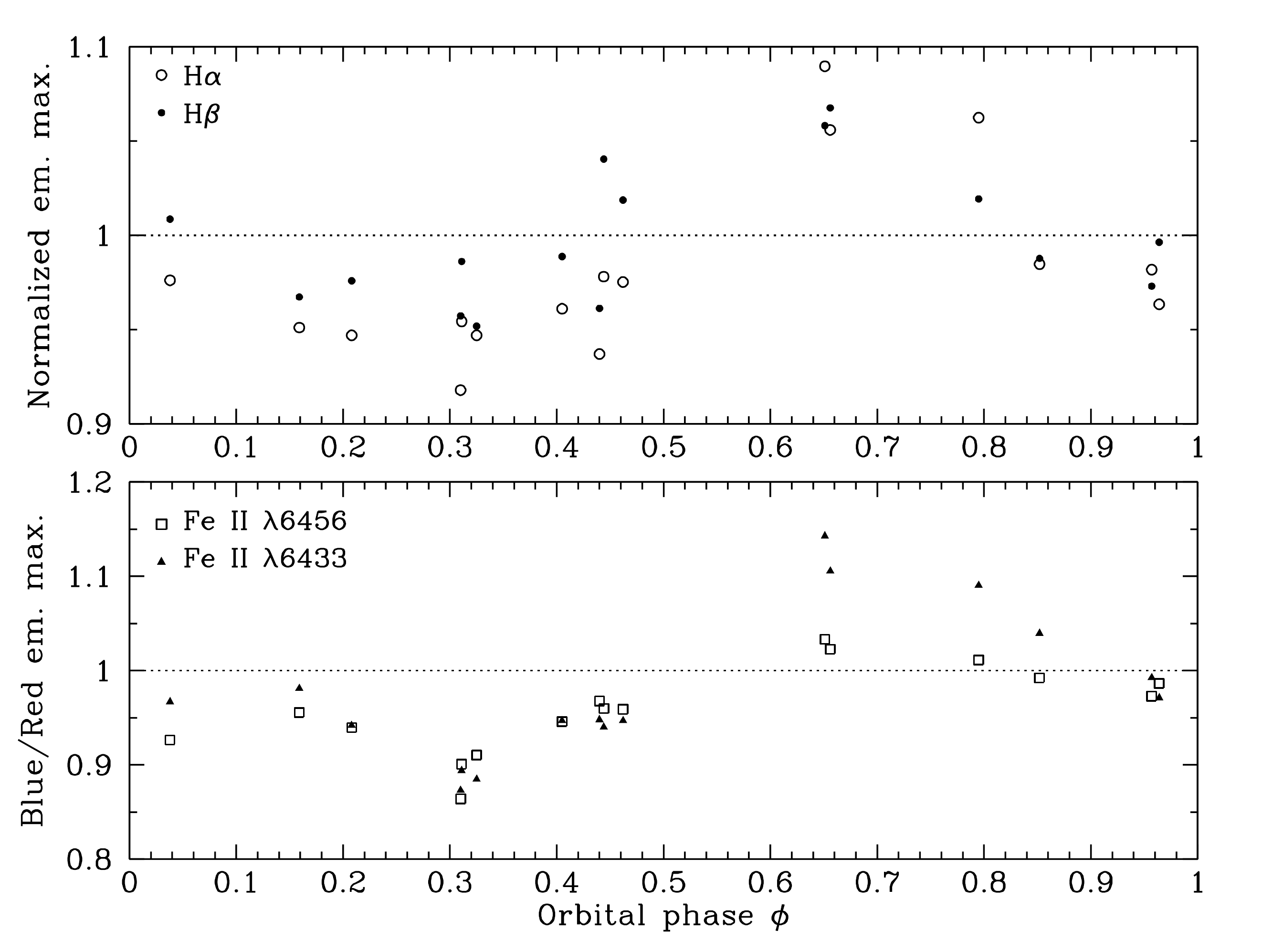}
\caption{ The top panel shows continuum normalised flux maxima observed 
with $\phi$ in strong H$\alpha$ ({\em open symbols}) and H$\beta$ ({\em dots}) 
emission lines of MWC~314. The H$\alpha$ maxima are divided by 24, and 
H$\beta$ maxima by 6.75 for comparison. The bottom panel shows the Blue/Red 
emission peak ratio we observe in two permitted Fe~{\sc ii} 
emission lines formed in the circumbinary disc. The B/R variations of 
$\pm$10~\% are due to orbital modulation of Doppler shifting line 
opacity, whereas the H$\alpha$ and H$\beta$ flux
changes result from variability of the stellar continuum flux with $\phi$. 
\label{fig_16}}
\end{sidewaysfigure}

\begin{sidewaysfigure}
\vspace*{18cm}
\centering
\includegraphics[width=20.cm,angle=0]{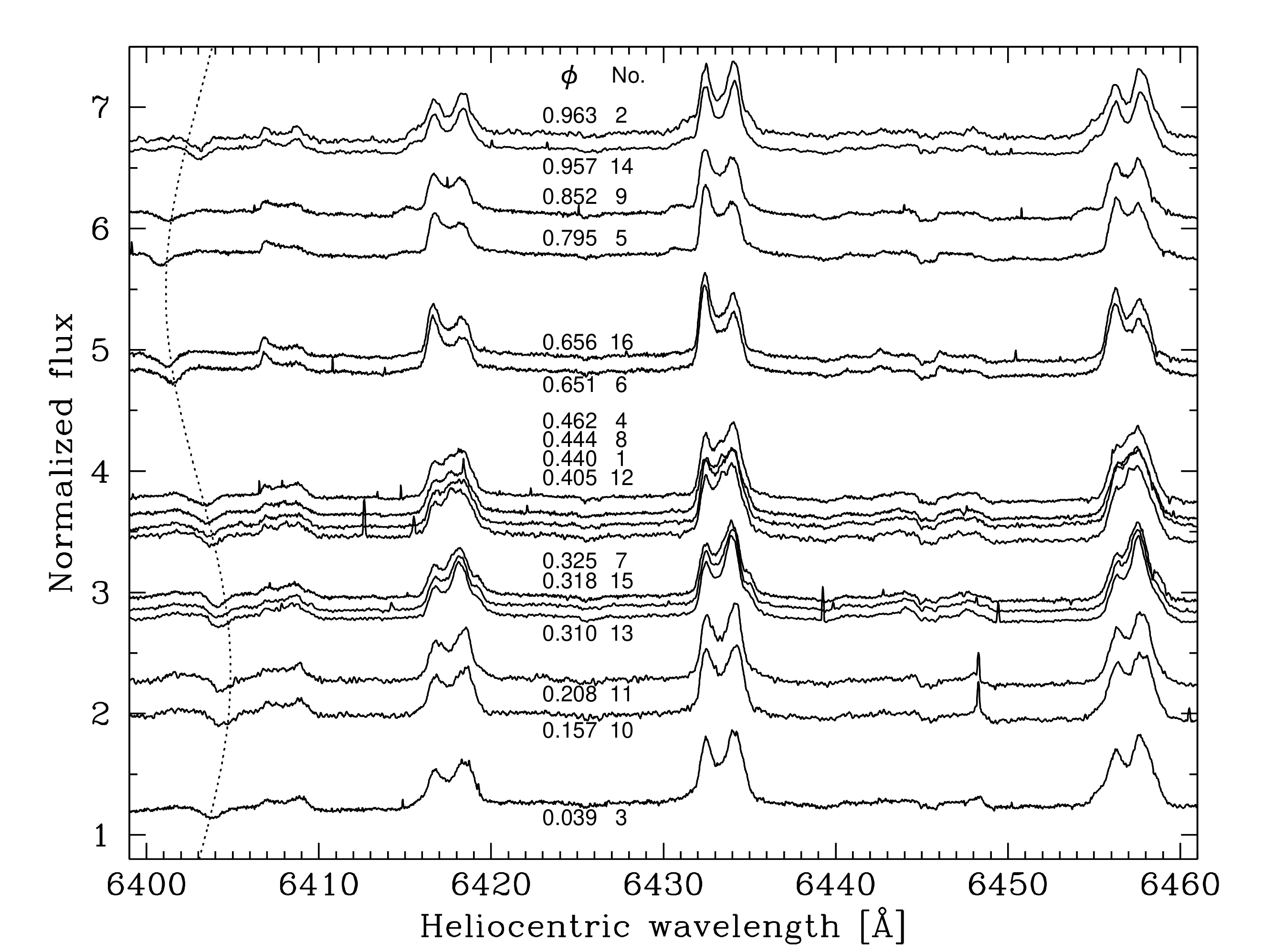}
\caption{ A portion of the MWC~314 spectrum showing 
permitted Fe~{\sc ii} $\lambda$6416.9, $\lambda$6432.7, and $\lambda$6456.4 
emission lines. The spectra are shifted upwards according the orbital phase 
(the labels show $\phi$ and spectrum Nos.). The orbital Doppler shift of the Ne~{\sc i} 
$\lambda$6402.2 absorption line is shown by the vertical curved line ({\em drawn dotted}). 
We show in this paper that the detailed shapes of the emission lines 
precisely depend on $\phi$, or are mainly caused by orbital modulation. 
For example, No. 1 ($\phi$=0.440) and No. 8 ($\phi$=0.444) reveal almost
identical spectra, we observe however 16 m apart, or separated by 
$\sim$8 $\times$ $P$.
\label{fig_17}}
\end{sidewaysfigure}

\begin{sidewaysfigure}
\vspace*{18cm}
\centering
\includegraphics[width=22cm,angle=0]{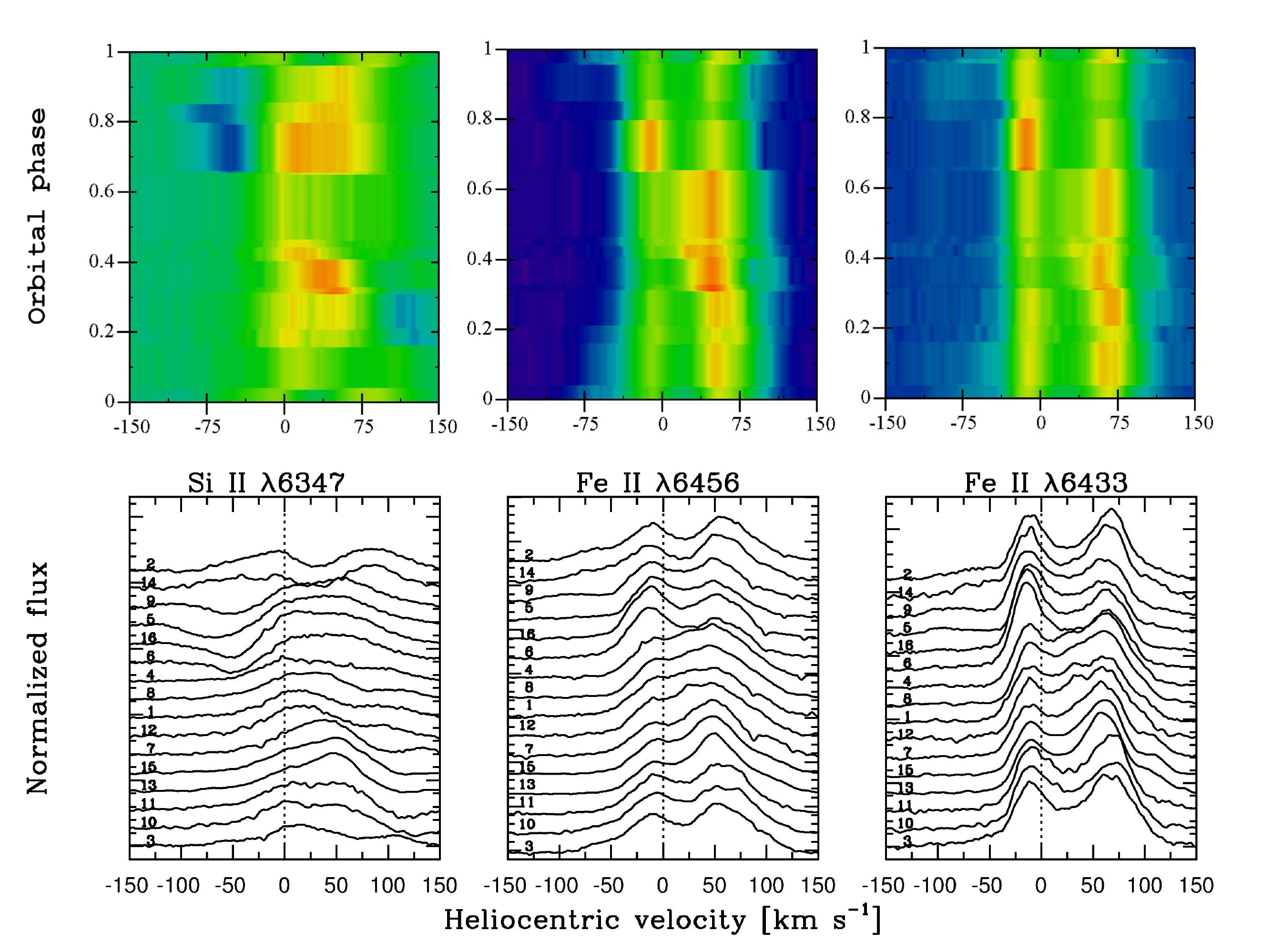}
\caption{ The detailed profiles of permitted Fe~{\sc ii}
lines ({\em middle and right-hand panels}) strongly depend on $\phi$.
The prominent emission lines form in a circumbinary disc and stay 
static (within $\pm$3~$\rm km\,s^{-1}$) around the $\gamma$-velocity.
The B/R emission peak ratios ({\em see bottom panel of Fig. 14})
periodically vary due to Doppler shifting lines formed in the 
orbiting primary star. It causes the triple-peaked emission lines around $\phi$=0.4-0.5.
We observe in Si~{\sc ii} $\lambda$6347 static line emission from the disc
modulated by Doppler shifting absorption in the primary ({\em left-hand panels}).
\label{fig_18}}
\end{sidewaysfigure}

\FloatBarrier

\begin{figure}
\centering
\includegraphics[width=16cm,angle=0]{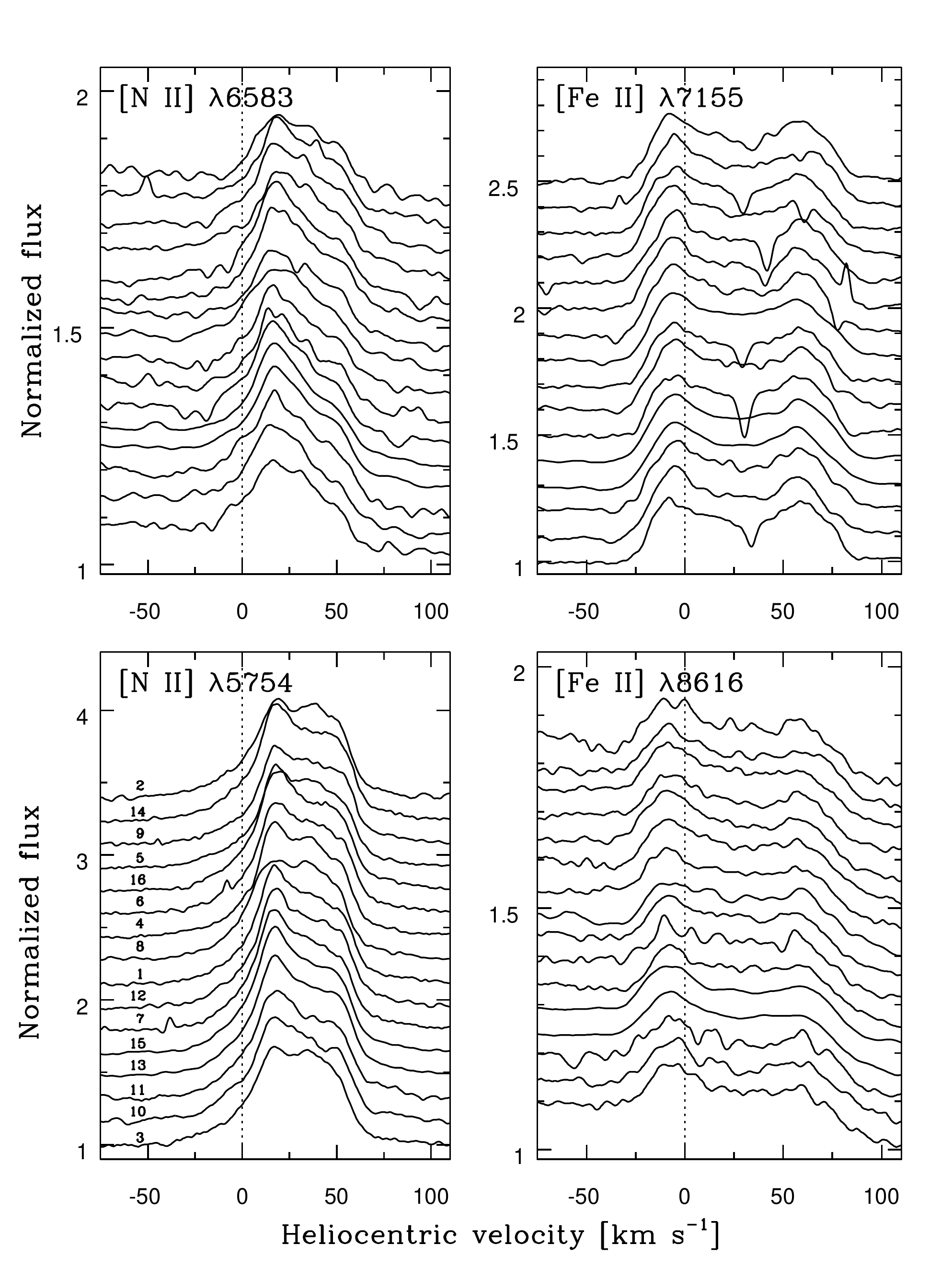}
\caption{ Normalised fluxes plotted with $\phi$ (shifted upwards) 
of two forbidden [N~{\sc ii}] ({\em left-hand panels}) and two [Fe~{\sc ii}] 
({\em right-hand panels}) emission lines in MWC~314.
The FWHM$\simeq$50~$\rm km\,s^{-1}$ of the double-peaked [Fe~{\sc ii}] 
emission lines is comparable to the permitted Fe~{\sc ii} emission lines ({\em see Fig. 16}), 
indicating a common line formation region in a circumbinary disc. 
The [N~{\sc ii}] lines show smaller FWHM$\simeq$25~$\rm km\,s^{-1}$.
\label{fig_19}}
\end{figure}

\begin{sidewaysfigure}
\vspace*{18cm}
\centering
\includegraphics[width=24cm,angle=0]{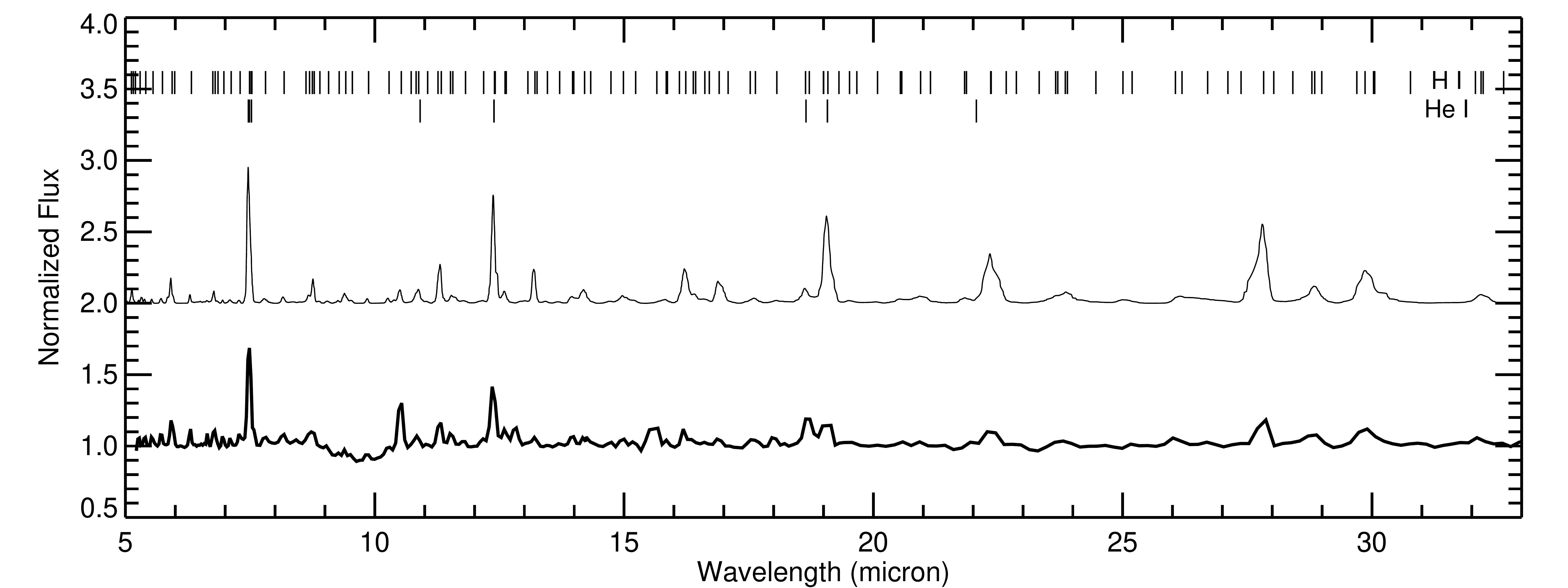}
\caption{Comparison of the {\it Spitzer} spectrum of MWC~314 ({\em bottom spectrum})
and the theoretical model spectrum ({\em top spectrum}). The vertical lines 
mark the wavelength positions of H~{\sc i} and He~{\sc ii} lines used in the 
spectrum calculations. The prominent emission lines computed at 22.38~$\mu$m 
and 27.8~$\mu$m are due to H~{\sc i} and He~{\sc i} lines ({\em see text}).
\label{fig_20}}
\end{sidewaysfigure}

\begin{figure}
\hspace*{-2cm}
\centering
\includegraphics[width=21cm,angle=0]{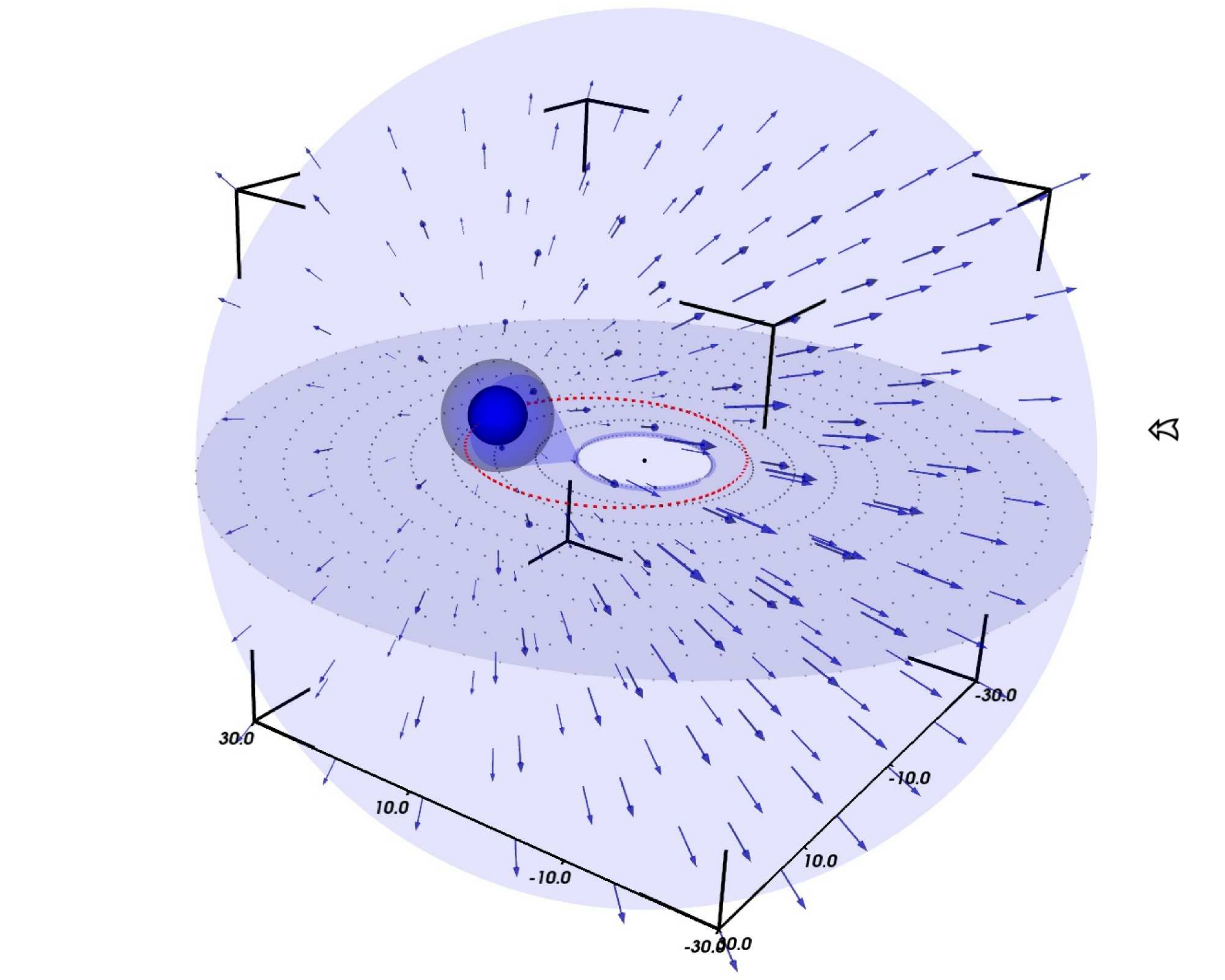}
\caption{3-D representation of the wind geometry in MWC 314. 
The size of the vectors mark the
velocity of the asymmetric wind from the primary due to orbital motion.
The open right-hand arrow marks the observer line of sight.
At longer distances the radiatively driven wind becomes symmetric around 
the centre of gravity ({\em see Sect. 5.1}).
\label{fig_21}}
\end{figure}

\begin{sidewaysfigure}
\vspace*{18cm}
\centering
\includegraphics[width=20cm,angle=0]{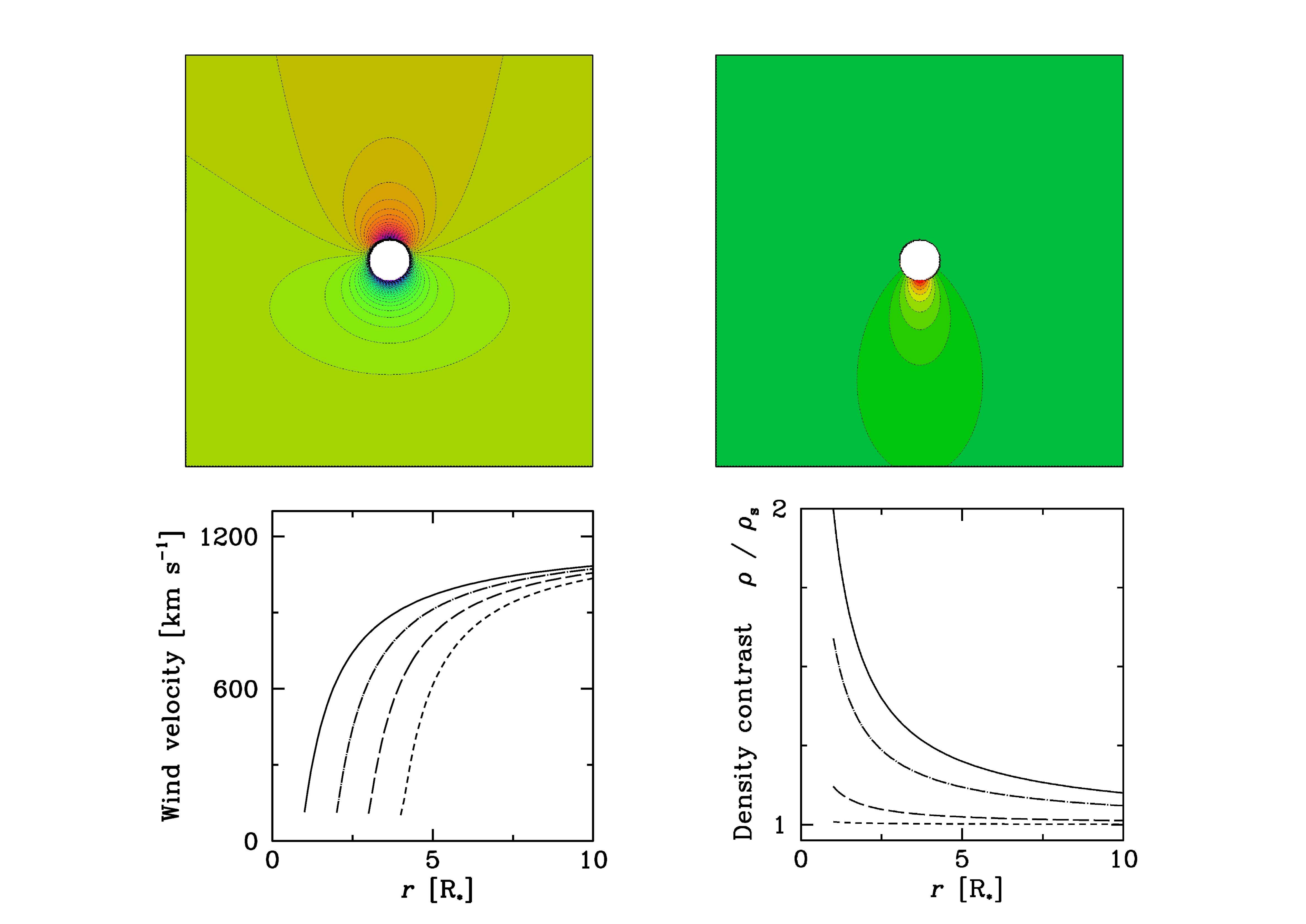}
\caption{Top panels show the changes in
velocity ({\em left-hand panel}) and density ({\em right-hand panel})
to 10~$R_{1}$ in the equatorial plane of the 3-D wind model 
around the primary star due to orbital motion towards the south. 
In front of the star ({\em blue colour in online figure}) the wind velocity {\em difference} 
in the observer's frame decreases from $V_{r}$=$-$85~$\rm km\,s^{-1}$ at 
the surface to $>$$-$10~$\rm km\,s^{-1}$ for $r>$3 $R_{1}$. Behind the star 
({\em red colour in online figure}) it increases from $-$85~$\rm km\,s^{-1}$  
to $<$10~$\rm km\,s^{-1}$ for $r>$3 $R_{1}$. 
At larger distances from the surface the difference of
wind velocity with the stationary wind ($V_{r}$=0) model 
is $<$10~$\rm km\,s^{-1}$ and becomes symmetric in all 
directions around the binary. The bottom left-hand panel 
shows the wind velocity for $\psi$=0 ({\em solid drawn line}), 
$\pi$/3 ({\em dash-dotted line}), 2$\pi$/3 ({\em dashed line}), 
and $\pi$ ({\em short-dashed line}). The curves are shifted 
to the right by $+$1~$R_{1}$ for clarity. The bottom right-hand
panel shows the radial wind density contrast at these $\psi$ angles 
for $f$=1 ({\em see Sect. 5.2}).
\label{fig_22}}
\end{sidewaysfigure}

\begin{sidewaysfigure}
\vspace*{18cm}
\centering
\includegraphics[width=19cm,angle=0]{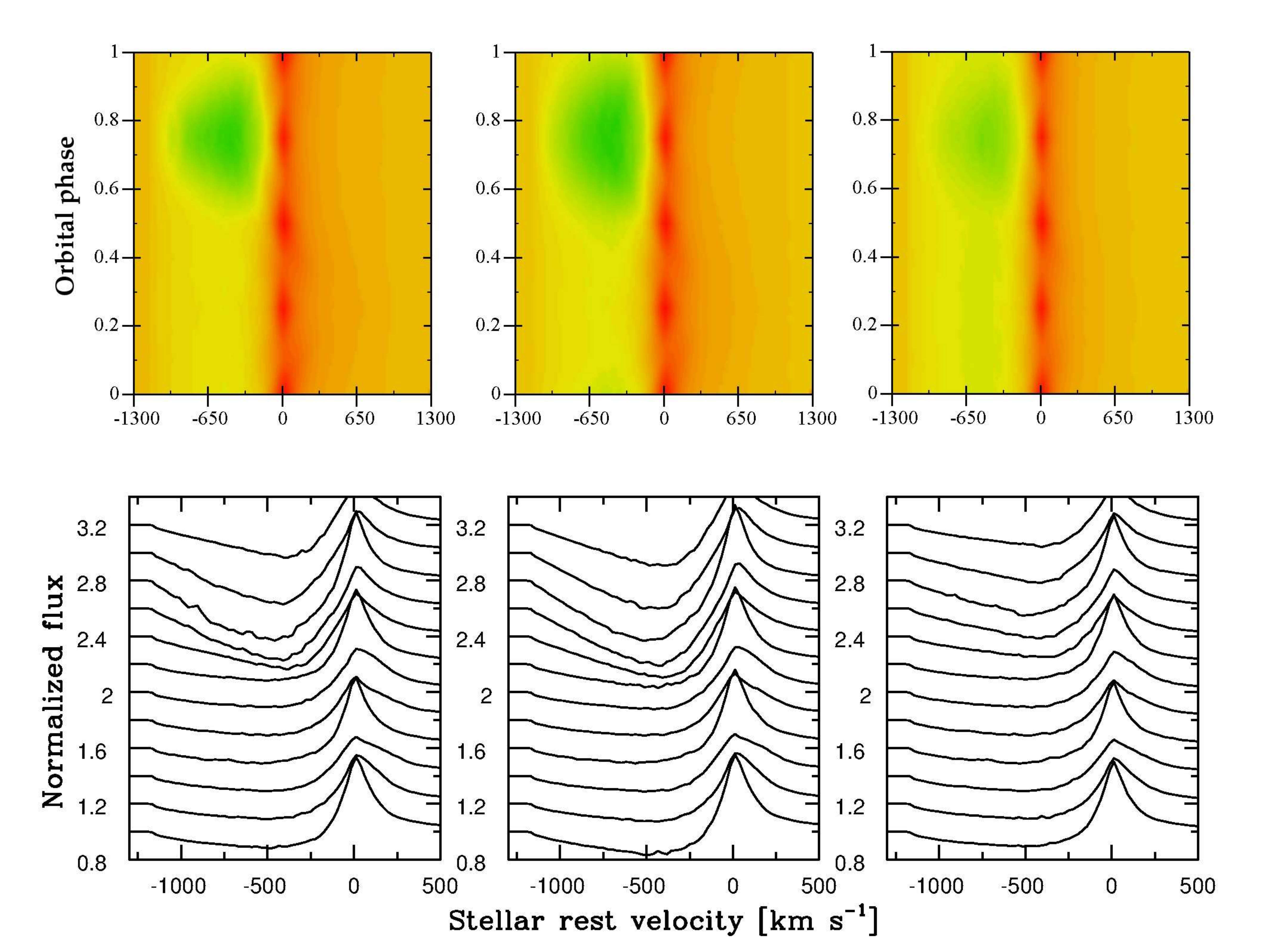}
\caption{ The top panels show dynamic spectra with orbital phase $\phi$ of 
He~{\sc i} $\lambda$5876 computed with the 3-D asymmetric wind model we 
develop for MWC~314 in this paper. The bottom panels show the corresponding normalised 
flux spectra for 12 $\phi$-values. The left-hand panels are computed 
for $f$=10 and $\varsigma_{0}$=$\vartheta_{0}$=$\pi$/4 in the density 
structure of Eq. (6). For comparison the middle panels are computed 
for $f$=10, $\varsigma_{0}$=$\vartheta_{0}$=$\pi$/3, with the radial
velocity amplitude of orbital motion set to 4$\times$ the observed value.
The right-hand panel is computed for $f$=3.3, 
$\varsigma_{0}$=$\vartheta_{0}$=$\pi$/4, and the observed $V_{r}$-curve.
The latter values provide the best fit to the P Cyg 
absorption line profile changes we observe with $\phi$ {    shown in Fig. 12} ({\em see Sect. 5.3}).
\label{fig_23}}
\end{sidewaysfigure}

\begin{sidewaysfigure}
\vspace*{18cm}
\centering
\includegraphics[width=16cm,angle=-90]{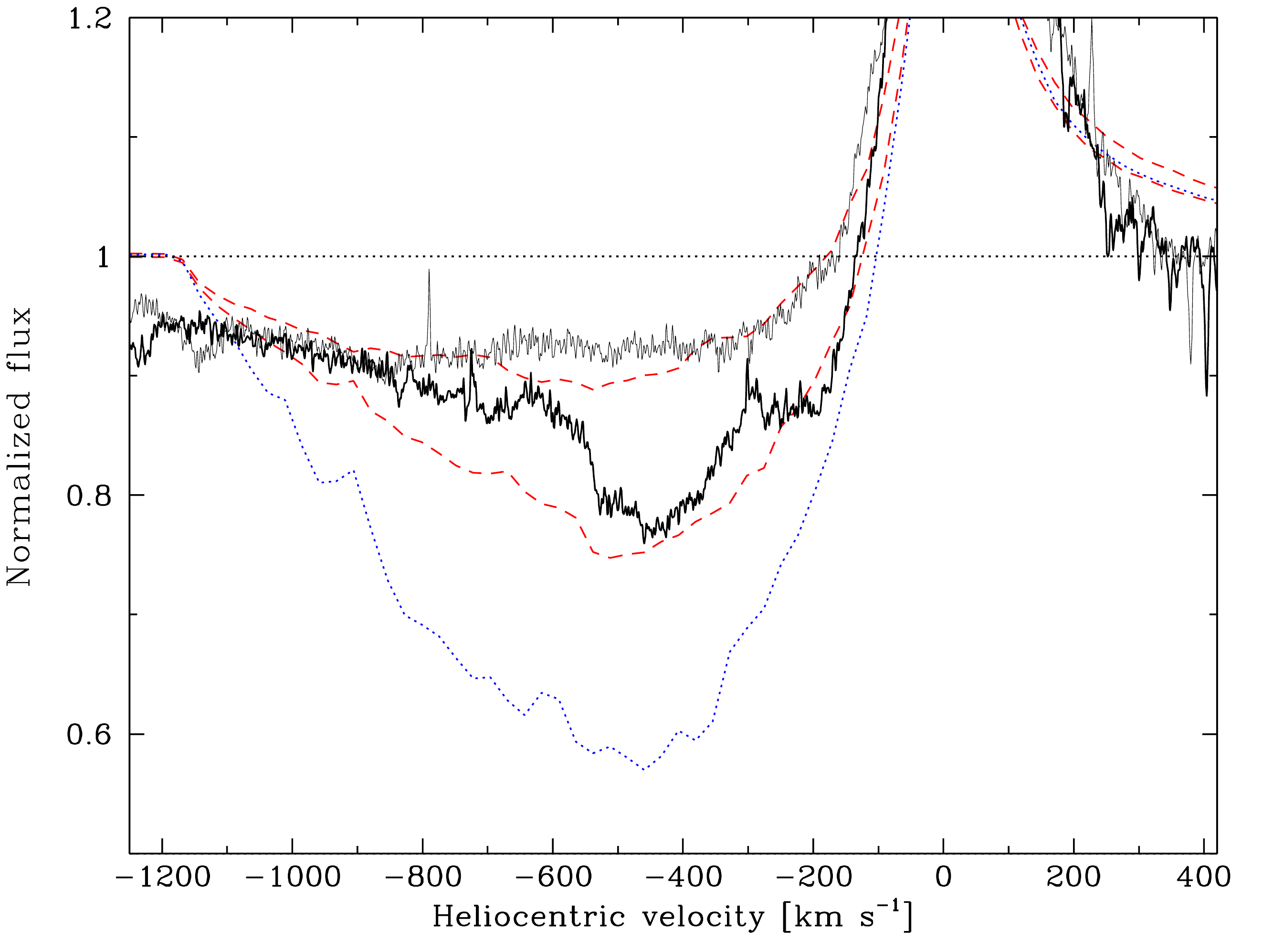}
\caption{Detailed 3-D RT modelling of the orbitally modulated He~{\sc i} $\lambda$5876 line we observe 
in MWC~314 at $\phi$=0.444 ({\em thin drawn solid line}) and at $\phi$=0.795 ({\em boldly drawn solid line}).
The best fit to the P Cyg absorption at both orbital phases ({\em dashed lines}) 
requires an asymmetric distribution of wind density around the primary star with the density 
enhancement factor $f$=3.3 in Eq. (6). The absorption line computed with the model of 
$f$=10 is too strong at $\phi$=0.795 ({\em dotted line}).
\label{fig_24}}
\end{sidewaysfigure}

\begin{figure}
\hspace*{-2cm}
\centering
\includegraphics[width=17cm,angle=0]{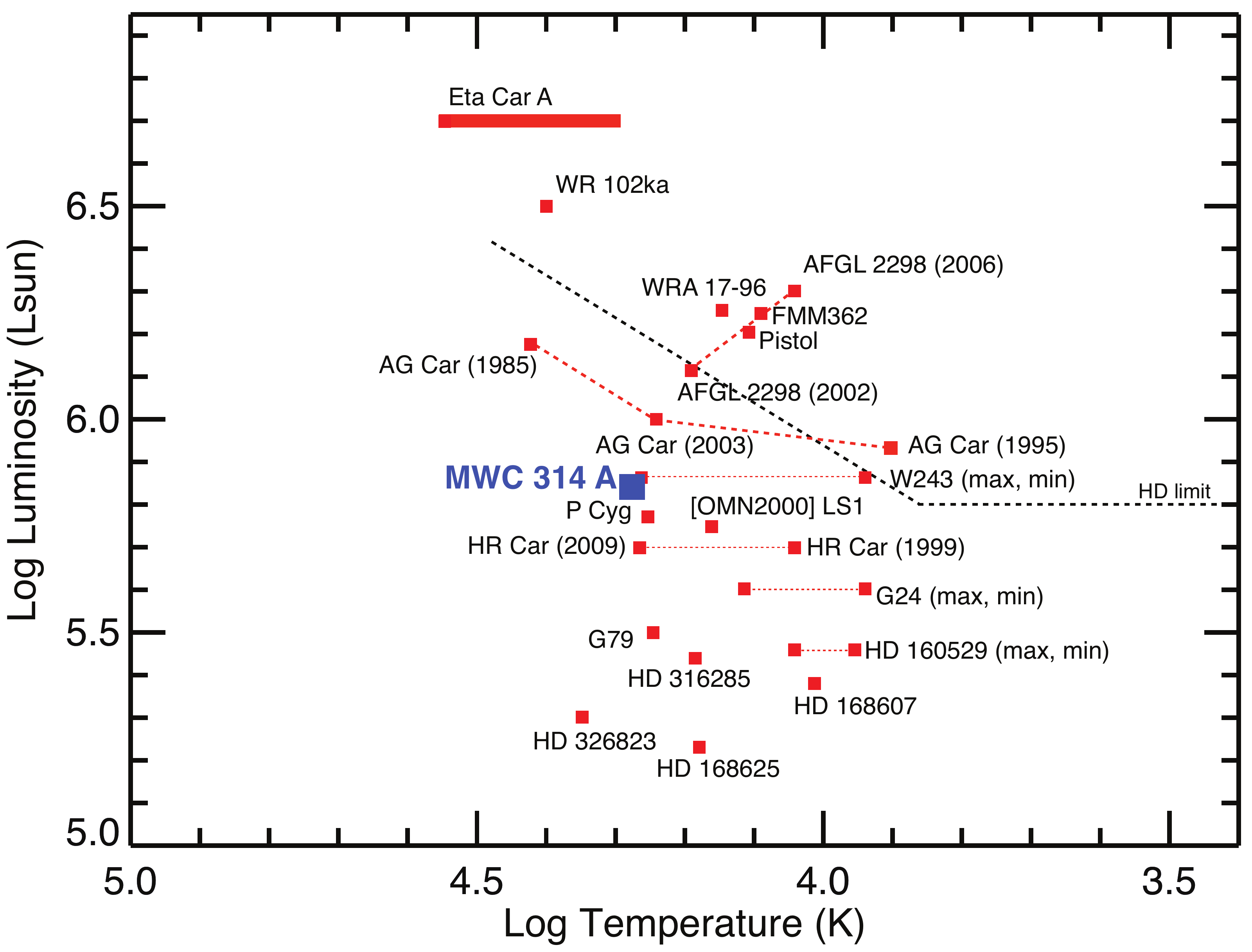}
\caption{
Hertzprung-Russell diagram showing the location of the primary star in MWC 314 (labelled MWC 314 A) 
and other selected LBVs. The sources of the stellar parameters of the LBVs are compiled 
by \citet{GrohAAsubm}.
\label{fig_25}}
\end{figure}
\end{document}